%% file: OptimalCompressionOfAPolylineWithSegmentsAndArcs.tex
\documentclass[conference]{IEEEtran}
\usepackage[table,xcdraw]{xcolor}
\usepackage[noadjust]{cite}
\usepackage{rotating}
\usepackage[cmex10]{amsmath}
\usepackage[caption=false,font=footnotesize]{subfig}
\usepackage{url}
\usepackage{graphicx}
\usepackage{amssymb}
\everymath{\displaystyle}
\usepackage{comment}

\usepackage[shortlabels]{enumitem}
\usepackage{nameref}
\usepackage{slashbox}
\usepackage{textcomp}
\usepackage{gensymb}
\usepackage[hypertexnames=false]{hyperref}
\newcommand{\myref}[1]{\hyperref[#1]{\ref*{#1} ``\nameref*{#1}''}}
\usepackage[all]{hypcap}
\usepackage{tikz}
\usepackage{tikz-qtree}
\usetikzlibrary{patterns}
\usetikzlibrary{shapes}
\usepackage{tkz-euclide}
\usepackage{algpseudocode}
\usepackage[outline]{contour}
\contourlength{1.2pt}
\usepackage[normalem]{ulem}
\usepackage{mathtools}
\usepackage{array}
\usepackage{tabu}

\DeclareMathOperator{\arctan2}{arctan2}

\newcommand{\ImaginaryUnit}{\textrm{i}}

\newcommand{\mytranspose}[1]{{#1}^\top}

\newcommand\norm[1]{\left\lVert#1\right\rVert}

\newcommand{\InversiveGeometryTransformation}[2]{ \Phi_{ #1 }{ \left( { #2 } \right) } }
\newcommand{\Inside}[1]{ Inside{ \left( { #1 } \right) } }
\newcommand{\Border}[1]{ #1 }
\newcommand{\Outside}[1]{ Outside{ \left( { #1 } \right) } }
\newcommand{\InsideAndBorder}[1]{ \Inside{ #1 } \vee \Border{ #1 } }
\newcommand{\OutsideAndBorder}[1]{ \Outside{ #1 } \vee \Border{ #1 } }

\definecolor{MyGreenColor}{HTML}{009901}

\def\addcircle#1{\tikz\node[draw, shape = circle]{#1};}
\def\addsquare#1{\tikz\node[draw, fill = green]{#1};}

\onecolumn

\def\MyGraphicsMode{1}

\begin{document}
\title
{
  Optimal Compression of a Polyline\\
  with Segments and Arcs
}
\author
{
\IEEEauthorblockN{Alexander Gribov}
\IEEEauthorblockA
{
  Esri\\
  380 New York Street\\
  Redlands, CA 92373\\
  Email: agribov@esri.com}
}

\maketitle

\begin{abstract}
\boldmath
This paper describes an efficient approach to constructing a resultant polyline with a minimum number of segments and arcs. While fitting an arc can be done with complexity $O{\left( 1 \right)}$ (see \cite{FittingOfCircularArcsWithO1Complexity} and \cite{EfficientFittingOfCircularArcs}), the main complexity is in checking that the resultant arc is within the specified tolerance. There are additional tests to check for the ends and for changes in direction (see \cite[section~3]{PaperArcFitting} and \cite[sections II.C and II.D]{OptimalCompression}). However, the most important part in reducing complexity is the ability to subdivide the polyline in order to limit the number of arc fittings \cite{EfficientFittingOfCircularArcs}. The approach described in this paper finds a compressed polyline with a minimum number of segments and arcs.
\end{abstract}



\section{Introduction}

Compression of a polyline with a minimum number of segments, within a specified tolerance, is described in \cite{CombinatorialMinimumNumberSegments}. \cite{OptimalCompressionWithArcs} finds a compressed polyline with the minimum number of arcs with complexity $ O{\left( N^3 \right)} $, where $ N $~is the number of vertices in the source polyline. Another algorithm described in \cite{OptimalComressionWithMinimumComplexity} has lower complexity $ O{\left( N^2 \log{ \left( N \right) } \right)} $; however, it might not be applicable for all cases, for example, when there is a backward movement in the source polyline along the true unknown arc. These algorithms construct a resultant polyline from the subset of the vertices of the source polyline. It is possible to achieve a higher compression ratio and better error filtering when vertices of the resultant polyline are not required to be from the subset of the vertices of the source polyline \cite{OptimalCompression}.

To find the optimal solution, the dynamic programming approach is applied (see \cite{CombinatorialMinimumNumberSegments}, \cite{OptimalCompressionWithArcs}, \cite{PolylineGeneralizationCombinatorical}, \cite{DynamicCompressionWithArcs}, \cite{ReconstructionOfOrthogonalPolygonalLines}, and \cite{OptimalCompression}). In the case of straight segments, the reduction of the search in the dynamic programming approach is done by using the convex hull \cite{OptimizedCompressionAlgorithm}. Parts of the polyline having a minimum width of the convex hull more than twice that of the specified tolerance are excluded from the analysis. This paper describes an approach to reduce the search in the dynamic programming approach for arcs. If part of the polyline cannot be fitted with an arc, then any part of the polyline containing that part can be skipped during analysis. The example in \cite{EfficientFittingOfCircularArcs} uses this approach; however, the test is approximate.

\section
{
  \label{sec:EfficientDivisionOfPolyline}
  Finding a Compressed Polyline with a~Minimum Number of Segments and Arcs
}

The task to find a compressed polyline with a minimum number of segments and arcs is solved by dynamic programming \cite{OptimalCompressionWithArcs}, \cite{DynamicCompressionWithArcs}, and \cite{OptimalCompression}. Allowing the resultant vertices to be different from the vertices of the source polyline improves compression \cite{OptimalCompression}. The algorithm in \cite{OptimalCompression} places a finite number of points around each vertex of the source polyline and searches for the resultant polyline connecting these points.

Table~\ref{table:ComparisonComplexitySegmentsArcs} compares the differences between segments and arcs when vertices of the resultant polyline are the subset of the vertices of the source polyline or when using the approach from \cite{OptimalCompression}.

\begin{table*}
  \caption[caption]
  {
    Complexity comparison for evaluation segments and arcs\\
    when end points are known and $ n $ points are between the end points.
  }
  \centering
  \begin{tabular}{ l | l | l }
    \hline
     & \multicolumn{1}{ c | }{Segment} & \multicolumn{1}{ c }{Arc}\\
     \hline
    Fitting by least squares approach                                   & Complexity $ O{ \left( 1 \right) } $, connect start and end points & $ O{ \left( 1 \right) } $, see \cite{EfficientFittingOfCircularArcs}\\
                                                                        &                                                                    & for approximate solution\\
    \hline
    Fitting by least deviation                                          & Complexity $ O{ \left( 1 \right) } $, connect start and end points & Complexity $ O{ \left( n \log{ \left( n \right) } \right) } $,\\
                                                                        &                                                                    & see \myref{appendix:FittingArcByTolerance}\\
    \hline
    Check if the geometric primitive                                    & Complexity $ O{ \left( \log{ \left( n \right) } \right) } $, using convex hull, see \cite{OptimizedCompressionAlgorithm} & Complexity $ O{ \left( n \right) } $,\\
    is within the specified tolerance                                   & (this paper will use the algorithm with complexity $ O{ \left( \log^2{ \left( n \right) } \right) } $,                   & need to check every vertex\\
                                                                        & see \myref{appendix:ConvexHullTree})\\
    \hline
    Testing end points and direction,                                   & Complexity $ O{ \left( \log{ \left( n \right) } \right) } $ (this paper will use the algorithm & Complexity $ O{ \left( n \right) } $,\\
    see \cite{OptimalCompression} and \cite[section 3]{PaperArcFitting} & with complexity $ O{ \left( \log^2{ \left( n \right) } \right) } $, see                        & need to check every vertex\\
                                                                        & \myref{appendix:ConvexHullTree})\\
                                                                        & and $ O{ \left( 1 \right) } $, respectively, see \cite{OptimalCompression}\\
    \hline
  \end{tabular}
  \label{table:ComparisonComplexitySegmentsArcs}
\end{table*}

For fitting arcs by least deviation, it is possible to reduce complexity by using polygons obtained as the intersection of the closest and the farthest Voronoi diagrams (see section~\myref{sec:FindingASmallestWidthAnnulus}). The test for geometric primitives to be within the specified tolerance requires finding the closest and the farthest points from the center. Note that the closest point should belong to the alpha shape with the generalized disk radius equal to or less than the radius of the checked arc, and the farthest point should belong to the convex hull.

Let $ P_{i, j} $, $ 0 \leq i \leq j < N $ be parts of the source polyline from vertex $ i $ to $ j $, where $ N $ is the total number of vertices. The optimal solution is found by induction. First, define the solution for polyline $ P_{0, 0} $. Second, for $ k = \overline{1..N - 1} $, construct the optimal solution for $ P_{0, k} $ from optimal solutions for $ P_{0, k'} $, $ k' = \overline{0..k-1} $. However, it is not always necessary to search over all $ k' $ because, for some $ i_{segment}{\left( k \right)} $ and $ i_{arc}{\left( k \right)} $, either there are no segments for $ P_{i_{segment}{\left( k \right)}, k} $ or no arcs for $ P_{i_{arc}{\left( k \right)}, k} $ exist within the specified tolerance. It is sufficient to search only for segments for $ k' = \overline{i_{segment}{\left( k \right)} + 1..k-1} $ and for arcs for $ k' = \overline{i_{arc}{\left( k \right)} + 1..k-3} $\footnote{At least three points are necessary in order to fit an arc. However, any three points can be fitted with an arc passing through them. Therefore, to avoid generation of too many arcs, this algorithm will require at least four points.}. The task of finding $ i_{arc}{\left( k \right)} $ is described in \myref{appendix:AppendixTestingArcs}. A similar algorithm can be used to find $ i_{segment}{\left( k \right)} $.

\subsection
{
  Finding a Smallest-Width Annulus
  \label{sec:FindingASmallestWidthAnnulus}
}

The task of checking if an arc exists that covers all points within the specified tolerance has an exact solution, described in \cite[section 7.4]{ComputationalGeometry1}. From \cite[section VII]{EfficientFittingOfCircularArcs}:
\textit
{
  The solution is found using the closest and the farthest point Voronoi diagrams. The center of an arc corresponding to the minimum width covering all vertices is either a vertex of the closest or the farthest point Voronoi diagram, or a point on the edge of the closest and the farthest point Voronoi diagrams \cite[p. 167]{ComputationalGeometry1}. The closest and the farthest point Voronoi diagrams are dual with the closest and the farthest point Delaunay triangulations, respectively \cite{FarthestDelaunayTriangulation}. Note that the farthest point Voronoi diagram includes only vertices on the convex hull.\footnote{Delaunay triangulation has an exact solution in integer arithmetic. There is also an exact solution in real arithmetic, because any real number can be exactly represented as a rational number. Both implementations require the use of extended precision. The Voronoi diagram also has an exact solution; however, the vertices of the Voronoi diagram are rational numbers. In the case of Delaunay triangulation on real or even integer numbers, interval arithmetic can be used to decide if extended precision is necessary.}
}

In \myref{appendix:DualityDelaunayDiagramsInversiveGeometry}, it will be shown that the farthest Delaunay triangulation is dual to the closest Delaunay triangulation for the inverted set of points. This gives the possibility to reuse the existing implementation of the closest Delaunay triangulation to construct the farthest Delaunay triangulation and the farthest Voronoi diagram. It also shows that the algorithms for calculation of the closest Delaunay triangulation can be modified to calculate the farthest Delaunay triangulation.

From \cite[section VII]{EfficientFittingOfCircularArcs}:
\textit
{
  Another algorithm to construct the closest and the farthest point Delaunay triangulations for vertices on the convex hull is by mapping each vertex $\left( x, y \right)$ to a vertex in three-dimensional space $\left( x, y, x^2 + y^2 \right)$ and constructing a convex hull for them \cite{FarthestDelaunayTriangulation}.
}

Construction of the closest and the farthest Delaunay triangulation can be implemented by the \textbf{divide-and-conquer} algorithm described in \cite{TwoAlgorithmsForConstructingADelaunayTriangulation} and \cite{VoronoiDiagramDivideAndConquer1985}. While modification of this algorithm described in \cite{Dwyer1987} has complexity $ O{ \left( N \log{ \left( \log{ \left( N \right) } \right) } \right) } $ for uniformly distributed data, there is no algorithm available for the non-uniformly distributed data. However, alternating horizontal and vertical directions (whichever is longer) when dividing data into two almost equal parts by recursively using the \textbf{quick sort} algorithm produces a faster algorithm with average complexity $ O{ \left( N \log{ \left( N \right) } \right) } $ \cite{VoronoiDiagramsAndDelaunayTriangulation}. To achieve worst-case complexity $ O{ \left( N \log{ \left( N \right) } \right) } $, it is necessary to achieve worst-case complexity $ O{ \left( N \right) } $ in finding the median, which is achievable by using the \textbf{median of medians} algorithm, see \cite[section 9.3]{IntroductionToAlgorithms}. Note that construction of the closest or the farthest Delaunay triangulation by use of the \textbf{flipping} algorithm results in the worst-case complexity of $ O{ \left( N^2 \right) } $, see proof in \cite{TwoAlgorithmsForConstructingADelaunayTriangulation}, \cite{VoronoiDiagramsAndDelaunayTriangulation}, and \cite{DelaunayDiagonalFlips}.

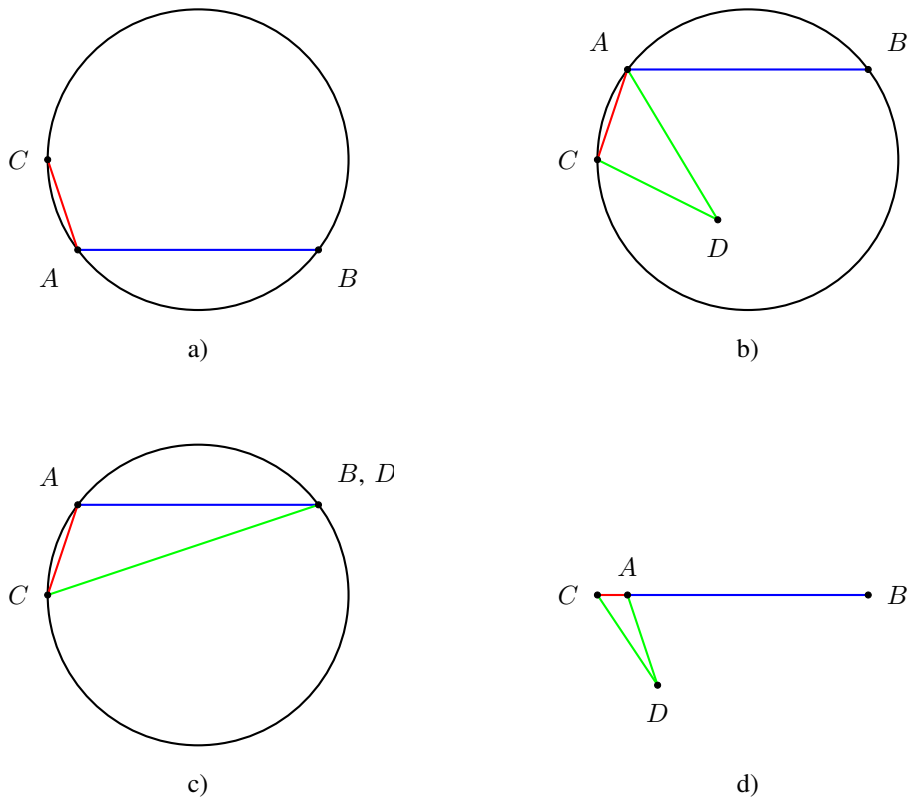
\begin{figure} [htb]
  \centering
  \begin{tabular}{c c c c c c}
    \begin{tikzpicture} [scale = 0.4]
      \begin{scope}
        \clip (-6.5, -5.5) -- (-6.5, 5.5) -- (6.5, 5.5) -- (6.5, -5.5) -- (-6.5, -5.5);
          \draw [thick, black] (0, 0) circle (5);

          \draw [thick, blue] (-4, -3) -- (4, -3);
          \draw [thick, red] (-4, -3) -- (-5, 0);

          \draw [fill = black] (-4, -3) circle [radius = 0.1];
          \node at (-4, -3) [ label = south west : {\scalebox{1}{$A$}} ] {};

          \draw [fill = black] (4, -3) circle [radius = 0.1];
          \node at (4, -3) [ label = south east : {\scalebox{1}{$B$}} ] {};

          \draw [fill = black] (-5, 0) circle [radius = 0.1];
          \node at (-5, 0) [ label = west : {\scalebox{1}{$C$}} ] {};
      \end{scope}
    \end{tikzpicture}
    &&&&&
    \begin{tikzpicture} [scale = 0.4]
      \begin{scope}
        \clip (-6.5, -5.5) -- (-6.5, 5.5) -- (6.5, 5.5) -- (6.5, -5.5) -- (-6.5, -5.5);
          \draw [thick, black] (0, 0) circle (5);

          \draw [thick, blue] (-4, 3) -- (4, 3);
          \draw [thick, red] (-4, 3) -- (-5, 0);
          \draw [thick, green] (-5, 0) -- (-1, -2);
          \draw [thick, green] (-4, 3) -- (-1, -2);

          \draw [fill = black] (-4, 3) circle [radius = 0.1];
          \node at (-4, 3) [ label = north west : {\scalebox{1}{$A$}} ] {};

          \draw [fill = black] (4, 3) circle [radius = 0.1];
          \node at (4, 3) [ label = north east : {\scalebox{1}{$B$}} ] {};

          \draw [fill = black] (-5, 0) circle [radius = 0.1];
          \node at (-5, 0) [ label = west : {\scalebox{1}{$C$}} ] {};

          \draw [fill = black] (-1, -2) circle [radius = 0.1];
          \node at (-1, -2) [ label = south : {\scalebox{1}{$D$}} ] {};
      \end{scope}
    \end{tikzpicture}
    \\
    a)
    &&&&&
    b)
    \\
    \\
    \\
    \begin{tikzpicture} [scale = 0.4]
      \begin{scope}
        \clip (-6.5, -5.5) -- (-6.5, 5.5) -- (6.5, 5.5) -- (6.5, -5.5) -- (-6.5, -5.5);
          \draw [thick, black] (0, 0) circle (5);

          \draw [thick, blue] (-4, 3) -- (4, 3);
          \draw [thick, red] (-4, 3) -- (-5, 0);
          \draw [thick, green] (-5, 0) -- (4, 3);

          \draw [fill = black] (-4, 3) circle [radius = 0.1];
          \node at (-4, 3) [ label = north west : {\scalebox{1}{$A$}} ] {};

          \draw [fill = black] (4, 3) circle [radius = 0.1];
          \node at (4, 3) [ label = north east : {\scalebox{1}{$B$, $D$}} ] {};

          \draw [fill = black] (-5, 0) circle [radius = 0.1];
          \node at (-5, 0) [ label = west : {\scalebox{1}{$C$}} ] {};
      \end{scope}
    \end{tikzpicture}
    &&&&&
    \begin{tikzpicture} [scale = 0.4]
      \begin{scope}
        \clip (-6.5, -5.5) -- (-6.5, 5.5) -- (6.5, 5.5) -- (6.5, -5.5) -- (-6.5, -5.5);
          \draw [thick, blue] (-4, 0) -- (4, 0);
          \draw [thick, red] (-4, 0) -- (-5, 0);
          \draw [thick, green] (-4, 0) -- (-3, -3);
          \draw [thick, green] (-5, 0) -- (-3, -3);

          \draw [fill = black] (-4, 0) circle [radius = 0.1];
          \node at (-4, 0) [ label = north : {\scalebox{1}{$A$}} ] {};
      
          \draw [fill = black] (4, 0) circle [radius = 0.1];
          \node at (4, 0) [ label = east : {\scalebox{1}{$B$}} ] {};

          \draw [fill = black] (-5, 0) circle [radius = 0.1];
          \node at (-5, 0) [ label = west : {\scalebox{1}{$C$}} ] {};

          \draw [fill = black] (-3, -3) circle [radius = 0.1];
          \node at (-3, -3) [ label = south : {\scalebox{1}{$D$}} ] {};
      \end{scope}
    \end{tikzpicture}
    \\
    c)
    &&&&&
    d)
  \end{tabular}
  \caption
  {
    Four cases of the merge process. $ AB $ (blue) is the base edge. $ C $ is the next left neighbor point. The edge $ AC $ (red) is tested if it is part of the resultant Delaunay triangulation. a)~$ C $ is above the base edge. b)~$ C $ is below the base edge, and $ ACD $ is a triangle adjacent to $ AC $. c)~$ D $ is the same point as $ B $. d)~$ A $, $ B $, and $ C $ are collinear.
  }
  \label{fig:DelaunayTriangulationDivideAndConqueTest}
\end{figure}

Adding an \textit{infinite} point to the Delaunay triangulation makes the number of edges equal to $ 3 \left( n - 1 \right) $, for $ n > 1 $ and one edge for one point, where $ n $ is the number of points not counting the \textit{infinite} point. This construction simplifies the \textbf{divide-and-conquer} algorithm, as there is no requirement to have at least two points in both Delaunay triangulations as required in \cite{VoronoiDiagramDivideAndConquer1985}. Note that neighboring convex hull edges are always separated by an edge to the \textit{infinite} point. That \textit{infinite} point is considered to be always outside of any circle in Delaunay triangulation. This paper will follow the \textbf{divide-and-conquer} algorithm described in \cite{TwoAlgorithmsForConstructingADelaunayTriangulation}. The first step in merging two Delaunay triangulations is to find lower and upper common tangents as in \cite{TwoAlgorithmsForConstructingADelaunayTriangulation}. There is a special case where the lower and upper common tangents are the same. This happens when all points are on the line. In such a case, an extra connection to the \textit{infinite} point is added for each Delaunay triangulation, unless it has only one point, and one edge is added to connect two Delaunay triangulations. The second step in merging two Delaunay triangulations is finding first the left and right points to be \textit{hit}, see \cite{TwoAlgorithmsForConstructingADelaunayTriangulation} and \cite{VoronoiDiagramDivideAndConquer1985}. Note that in this step, there is no need to perform the test to check if the point is above the base edge, because the test to check if a point is inside a circle (performed by finding a sign of the determinant \eqref{eq:InvertedDeterminant}, see \cite{VoronoiDiagramDivideAndConquer1985}), will eliminate all cases where a point is not above the base edge. By proving that all cases will be processed in the same way without this test, it follows that this test can be skipped. There are four possible cases shown in Figure~\ref{fig:DelaunayTriangulationDivideAndConqueTest}. All cases except a) are eliminated by the test to check if $ C $ is above the base edge $ AB $; however, all other cases will be eliminated by the test to check if a point is inside the circumscribed circle for the triangle $ ABC $. In case b), $ D $ cannot be outside of the circumscribed circle for the triangle $ ABC $; otherwise, the merging of the two Delaunay triangulations performed up to this step is not valid. This case is eliminated by the sign of the determinant \eqref{eq:InvertedDeterminant}, because $ C $ is below $ AB $ (swapping rows in \eqref{eq:InvertedDeterminant} corresponding to points $ A $ and $ B $ to make the orientation for the triangle $ ABC $ counterclockwise will change the sign of the determinant). Case c) will produce zero of the determinant \eqref{eq:InvertedDeterminant} because two rows of the matrix \eqref{eq:InvertedDeterminant} are the same. Case d) is the same as b). When moving point $ C $ from above $ AB $ to below (see a), d), and b)), it can be thought of as the interior of the circumscribed circle for the triangle $ ABC $, first becoming a half plane and then becoming an inverted circle. When the first left and right points to be \textit{hit} are found, only in the case when they are on the circle passing through ends of the base edge (the determinant is zero), it is necessary to check if one of the points is above the base edge. If not, another point is guaranteed to be above the base edge. Therefore, performing this test only in the case when four points are on the circle produces a valid and slightly faster implementation of the \textbf{divide-and-conquer} algorithm.

For the closest or the farthest Delaunay triangulation when points form a convex hull and are in clockwise or counter-clockwise order, the \textbf{divide-and-conquer} algorithm \cite{TwoAlgorithmsForConstructingADelaunayTriangulation} and \cite{VoronoiDiagramDivideAndConquer1985} does not need to find the division of the points, because any parts of the convex hull are already separated by some line from another part of the convex hull \cite{OnComputingVoronoiDiagramsForSortedPointSets}. Any subset of vertices of the convex hull is a set of vertices of the convex hull; therefore, the farthest Delaunay triangulation can be constructed for any subset of vertices of the convex hull. Therefore, the algorithm only has to recursively merge partial solutions. Note that the case where lower and upper tangents are the same is not possible, and they are known right away. This makes implementation of the \textbf{divide-and-conquer} algorithm for the set of points from a convex hull simpler than for the randomly located points.

\textbf{Divide-and-conquer} algorithm is easily parallelized, as a divided set of points can be processed independently. Dividing a set of points by a median can also be parallelized.

\begin{figure*} [p]
  \input{FigureExampleVoronoiDiagrams.tex}
  \caption
  {
    The closest (blue) and the farthest (red) Voronoi diagrams for $ 100 $ uniformly distributed random points (black) inside annulus with minimum radius $ 0.9 $ and maximum radius $ 1.0 $.
  }
  \label{fig:ExampleVoronoiDiagrams}
\end{figure*}
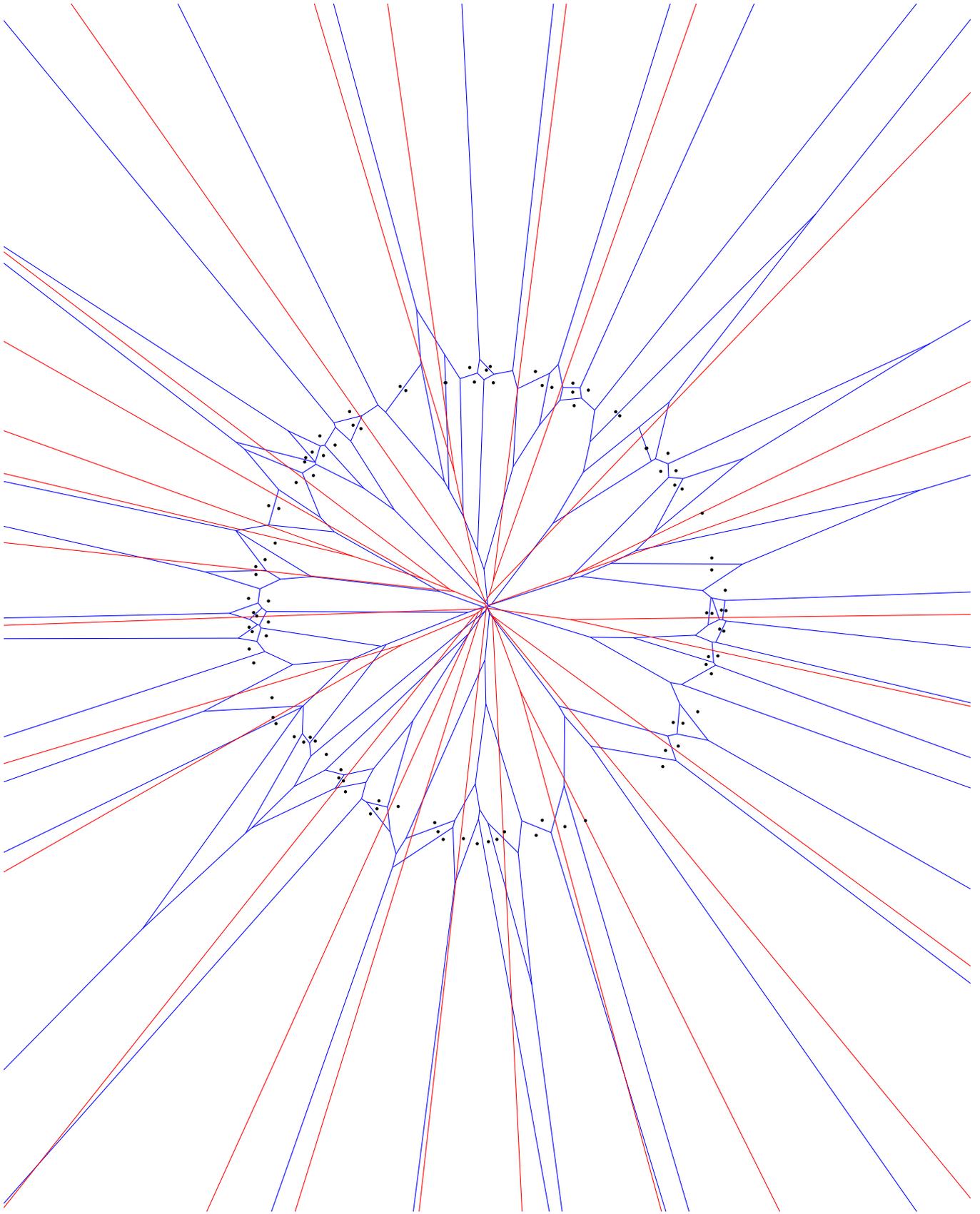

\begin{figure*} [p]
  \input{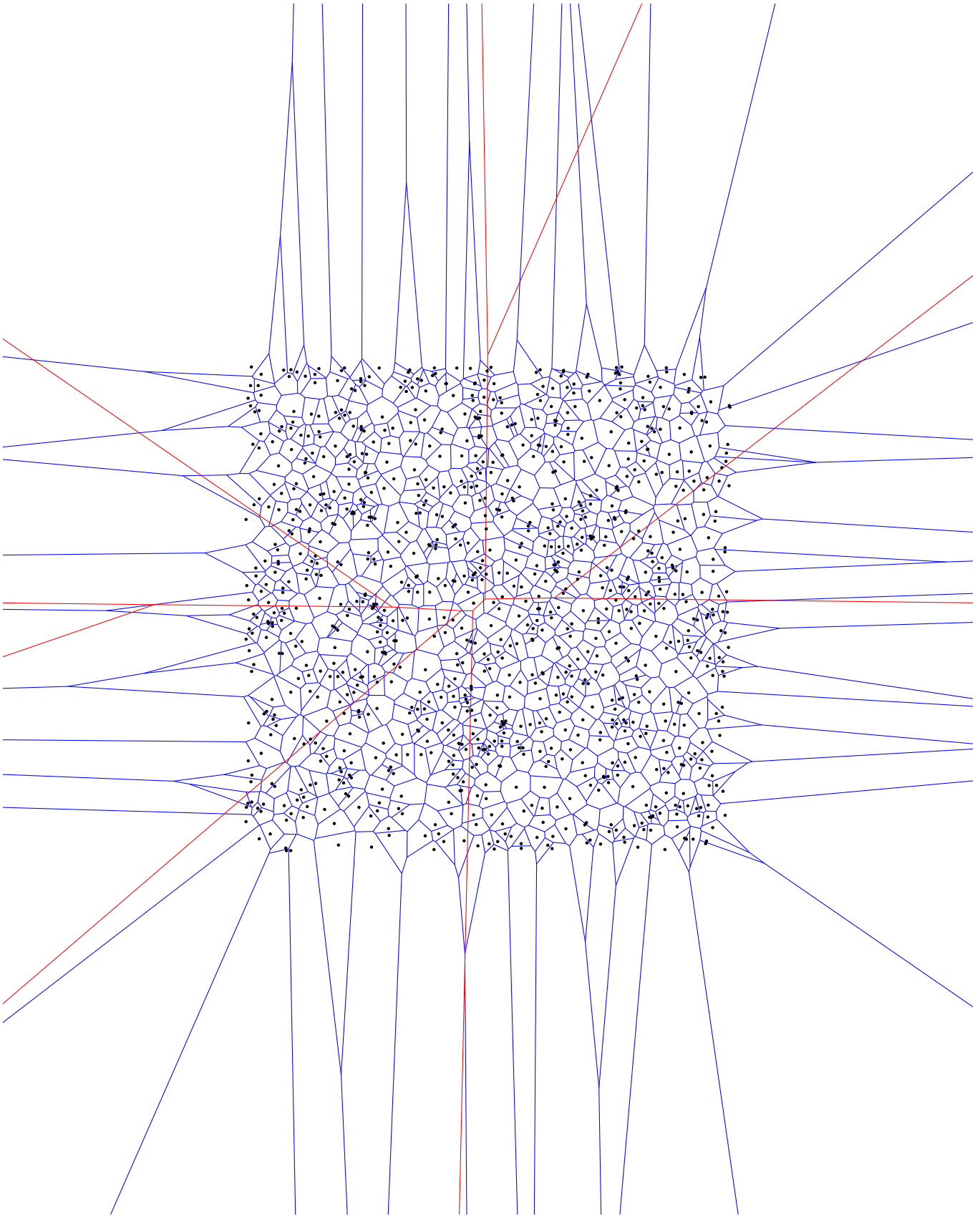}
  \caption
  {
    The closest (blue) and the farthest (red) Voronoi diagrams for $ 1,000 $ uniformly distributed random points (black) inside square.
  }
  \label{fig:ExampleVoronoiDiagramsSquare}
\end{figure*}

Examples of Voronoi diagrams are shown in Figures~\ref{fig:ExampleVoronoiDiagrams} and \ref{fig:ExampleVoronoiDiagramsSquare}.\footnote{When all points are exactly on the line, all edges of the Voronoi diagrams are lines; otherwise, infinite edges can be represented as a vertex and a direction.}


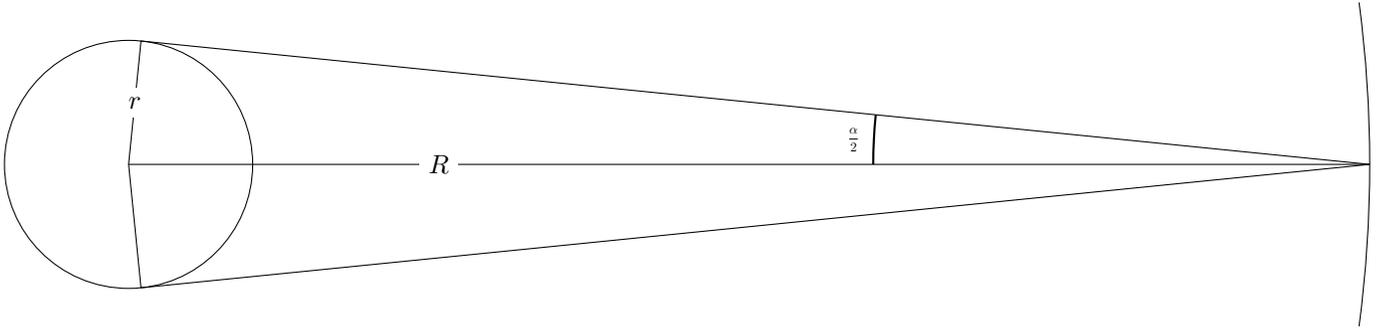
\begin{figure} [htb]
  \centering
  \begin{tikzpicture} [scale = 1.65] 
    \draw (0, 0) circle [radius = 1];
    \draw (10, 0) arc (0 : 7.5 : 10);
    \draw (10, 0) arc (0 : -7.5 : 10);

    \draw (0, 0) -- node[fill = white, pos = 0.25] {$ R $} (10, 0);
    \draw (0, 0) -- node[fill = white, midway] {$ r $} (0.1, 0.99498743710661995473447982100121) -- (10, 0);
    \draw (0, 0) -- (0.1, -0.99498743710661995473447982100121) -- (10, 0);
    \draw [thick] (6, 0) arc (180 : 180 - 5.7391704772667863125149089039304 / 2 : 4) node [left] { \resizebox{0.5\width}{0.5\height}{ $ \dfrac{\alpha}{2} $ } } arc (180 - 5.7391704772667863125149089039304 / 2 : 180 - 5.7391704772667863125149089039304 : 4);
  \end{tikzpicture}
  \caption
  {
    The relationship between the radius of the clipping circle $ R $ when all data is inside the circle of the radius $ r $ from the arc with the smallest angle $ \alpha $, see \eqref{eq:ClippingVoronoiRadius}.
  }
  \label{fig:FormulaForClippingRadius}
\end{figure}

Note that any Voronoi diagram is infinite. However, if we search for an arc with an angle no smaller than some predefined value $ \alpha $, the Voronoi diagram can be clipped. Let all vertices of the source polyline be inside a circle of radius $ r $, then the center of the arc is inside the circle of radius
\begin{equation}
  R = \dfrac{r}{\sin{\dfrac{\alpha}{2}}}
  ,
  \label{eq:ClippingVoronoiRadius}
\end{equation}
see Figure~\ref{fig:FormulaForClippingRadius}.
For an angle of $ 0.1 \degree $ and $ r = 1 $, $ R \approx 1146 $.\footnote{The increase of the domain will consume about $ 10 $ additional bits for each coordinate.}

Tables~\ref{table:TimeDelaunayTriangulation} and \ref{table:TimeDelaunayTriangulationConvexHull} compare calculation time for the \textbf{divide-and-conquer} algorithm when division of the points is performed versus when the order of the points in the convex hull is known.\footnote{In parallel implementation, the part of the algorithm that divides data into two equal parts is not parallelized.} The comparison is performed using extended precision integer arithmetic with coordinates using $ 64 $-bit integers. Algorithms to generate a random convex hull are described in \myref{appendix:RandomConvexHull}. Time was measured on the processor Intel Xeon CPU $\text{E5-2670}$. The algorithm, which uses a known order of points in the convex hull, performs similarly or faster when compared to the algorithm that divides points. There is also a comparison with parallel implementation but only for the cases where division of points are performed. Parallel implementation performs faster when there are more than a few hundred points.

\begin{table*} [htb]
  \caption[caption]
  {
    Time to calculate Delaunay triangulation versus the number of points\\
    with unknown division in advance (all time is measured in seconds).
  }
  \centering
  \begin{tabular}{ | l | c | c | c | c | c | c | c | c | }
    \hline
    Number of points & 10 & 100 & 1,000 & 10,000 & 100,000 & 1,000,000 & 10,000,000 & 100,000,000\\
    \hline
    Uniformly distributed points & & & & & & & &\\
    The closest Delaunay triangulation & 6.76e-05 & 7.41e-04 & 7.74e-03 & 8.27e-02 & 8.40e-01 & 8.76e+00 & 8.88e+01 & 9.49e+02\\
    Calculated in parallel with $ 16 $ threads & 3.94e-04 & 1.03e-03 & 1.93e-03 & 1.31e-02 & 1.21e-01 & 1.16e+00 & 1.03e+01 & 1.04e+02\\
    \hline
    \multicolumn{9}{|c|}{Algorithm ``\nameref{sec:RandomSetOfDirections}'' described in \myref{appendix:RandomConvexHull}}\\
    \hline
    The closest Delaunay triangulation & 4.33e-05 & 4.06e-04 & 4.91e-03 & 5.15e-02 & 5.39e-01 & 5.24e+00 & 5.40e+01 & 5.03e+02\\
    Calculated in parallel with $ 16 $ threads & 1.02e-03 & 2.53e-03 & 1.73e-03 & 9.21e-03 & 7.05e-02 & 5.76e-01 & 5.22e+00 & 4.95e+01\\
    The farthest Delaunay triangulation & 4.01e-05 & 5.71e-04 & 4.53e-03 & 4.93e-02 & 4.95e-01 & 4.97e+00 & 5.00e+01 & 4.90e+02\\
    Calculated in parallel with $ 16 $ threads & 1.11e-03 & 2.43e-03 & 1.75e-03 & 9.47e-03 & 6.92e-02 & 5.16e-01 & 4.88e+00 & 4.60e+01\\
    \hline
    \multicolumn{9}{|c|}{Algorithm ``\nameref{sec:RandomConvexHullFromTheFarthestDelaunayTriangulation}'' described in \myref{appendix:RandomConvexHull}}\\
    \hline
    The closest Delaunay triangulation & 2.09e-05 & 4.52e-04 & 6.74e-03 & 8.44e-02 & & & &\\
    Calculated in parallel with $ 16 $ threads & 1.15e-03 & 2.51e-03 & 2.02e-03 & 1.85e-02 & & & &\\
    The farthest Delaunay triangulation & 2.51e-05 & 3.78e-04 & 3.73e-03 & 3.74e-02 & & & &\\
    Calculated in parallel with $ 16 $ threads & 1.04e-03 & 2.45e-03 & 1.19e-03 & 7.17e-03 & & & &\\
    \hline
  \end{tabular}
  \label{table:TimeDelaunayTriangulation}
\end{table*}

\begin{table*} [htb]
  \caption[caption]
  {
    Time to calculate Delaunay triangulation versus the number of points\\
    when the order of points in the convex hull is known (all time is measured in seconds).
  }
  \centering
  \begin{tabular}{ | l | c | c | c | c | c | c | c | c | }
    \hline
    Number of points & 10 & 100 & 1,000 & 10,000 & 100,000 & 1,000,000 & 10,000,000 & 100,000,000\\
    \hline
    \multicolumn{9}{|c|}{Algorithm ``\nameref{sec:RandomSetOfDirections}'' described in \myref{appendix:RandomConvexHull}}\\
    \hline
    The closest Delaunay triangulation & 2.48e-05 & 3.62e-04 & 4.40e-03 & 4.83e-02 & 4.85e-01 & 4.81e+00 & 4.74e+01 & 4.51e+02\\
    The farthest Delaunay triangulation & 2.77e-05 & 3.69e-04 & 4.31e-03 & 4.42e-02 & 4.50e-01 & 4.37e+00 & 4.35e+01 & 4.12e+02\\
    \hline
    \multicolumn{9}{|c|}{Algorithm ``\nameref{sec:RandomConvexHullFromTheFarthestDelaunayTriangulation}'' described in \myref{appendix:RandomConvexHull}}\\
    \hline
    The closest Delaunay triangulation & 1.95e-05 & 4.60e-04 & 6.67e-03 & 8.76e-02 & & & &\\
    The farthest Delaunay triangulation & 1.61e-05 & 2.62e-04 & 2.84e-03 & 2.97e-02 & & & &\\
    \hline
  \end{tabular}
  \label{table:TimeDelaunayTriangulationConvexHull}
\end{table*}

If vertices of the Voronoi diagram are not represented as rational numbers, they are not exact and can lead to improper geometry. This means that some Voronoi diagram cells can have self-intersections, overlap other cells, etc. We are going to treat each Voronoi diagram cell by the ``XOR'' rule in order to be able to work properly with rounded vertices of the Voronoi diagram.

The next step would be to intersect the closest and the farthest Voronoi diagrams:
\begin{itemize}
  \item Find the intersection of the edges of the closest Voronoi diagram with the edges of the farthest Voronoi diagram.
  \item For the vertices of the closest Voronoi diagram, find the corresponding cells of the farthest Voronoi diagram.
  \item For the vertices of the farthest Voronoi diagram, find the corresponding cells of the closest Voronoi diagram.
\end{itemize}

An algorithm to resolve the above tasks follows:
\begin{enumerate}
  \item Put all edges of the closest and the farthest Voronoi diagrams\footnote{For points with equal coordinates, only one point is used in the Voronoi diagrams.} into an array with information for the corresponding Voronoi diagrams' cells. Edges of the Voronoi diagrams are represented as segments with two indices with the exception of the additional edges to close infinite Voronoi cells. They will have only one index.\footnote{The closest Voronoi diagram can be constructed with finite cells by adding extra points around the source data. However, this solution is not appropriate for the farthest Voronoi diagram because the farthest Voronoi diagram has only cells corresponding to the vertices of the convex hull.}
  \item Clip all segments by the square. The algorithm is described in \myref{appendix:ClippingBySquare}.
  \item Put all segments into an integer grid.
  \item Remove overlapping segments. The algorithm is described in \myref{appendix:RemoveOverlappingSegments}.
  \item Apply a plane sweep algorithm (see \cite{ComputationalGeometry2} and \cite{ComputationalGeometry1}) on rational arithmetic without modifying segments, and report all events with at least one index for the closest Voronoi diagram and one index for the farthest Voronoi diagram.\footnote{This will require integer arithmetic with about eight times the number of bits used in the coordinates.}
  \item Find a point corresponding to the center of a minimum width annulus.
\end{enumerate}

If the minimum width annulus has a width larger than two tolerances, then there are no arcs that can be fitted to this set of points.

\subsection{Complexity of the Algorithm ``Finding a Smallest-Width Annulus''}

\begin{figure*} [p]
  \input{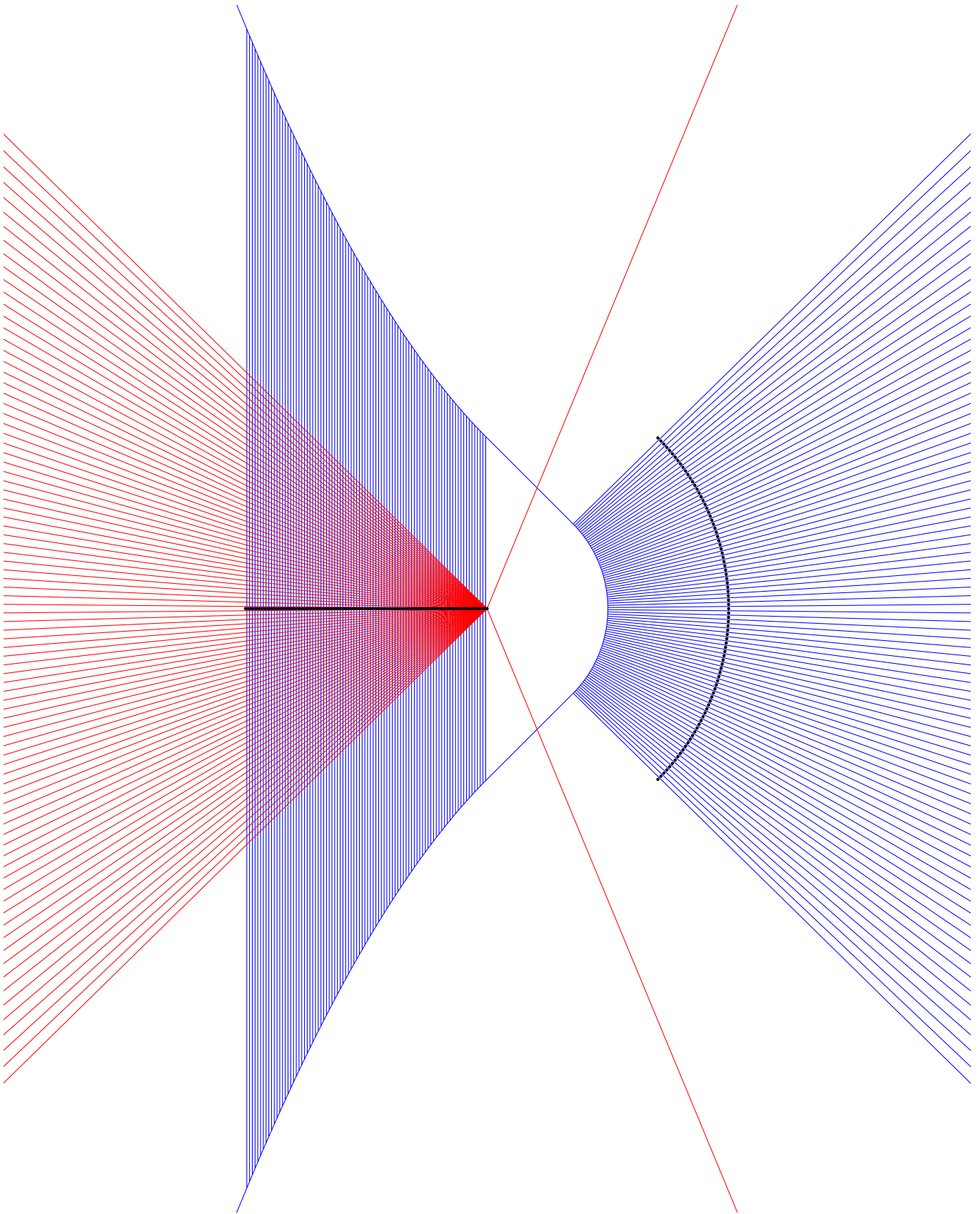}
  \caption
  {
    The closest (blue) and the farthest (red) Voronoi diagrams for points (black).
  }
  \label{fig:BadVoronoi}
\end{figure*}

The algorithm described in the previous section gives an exact answer\footnote{With the precision of floating-point arithmetic. If rational arithmetic is used, the result would be exact.} to the question of whether the arc with the specified tolerance can be fitted to the set of points. The algorithm has the complexity $ O{ \left( \left( n + k \right) \log{ \left( n \right) } \right) } $, where $ n $ is the number of points and $ k $ is the number of intersections between the edges of the closest and the farthest Voronoi diagrams. In the worst case, $ k $ can have order of $ n^2 $ (see, for example, Figure~\ref{fig:BadVoronoi}). Therefore, the worst-case complexity is $ O{ \left( n^2 \log{ \left( n \right) } \right) } $. It is possible to speed up the algorithm if the first test is performed by the direct fitting of an arc to the set of points, for example, by approximation to the least squares approach (see \cite{FittingOfCircularArcsWithO1Complexity} and \cite{EfficientFittingOfCircularArcs}) and then by checking if all points are within the specified tolerance. The complexity of this test is $ O{ \left( n \right) } $. If the points are within the specified tolerance from the estimated center, then there is no need to perform the test described in the previous section.

\subsection{Efficient Restriction of Arc Fittings}

The algorithm described in section~\myref{sec:FindingASmallestWidthAnnulus} cannot be used for all parts of the source polyline, as it will produce an algorithm with the worst-case complexity $ O{ \left( N^4 \log{ \left( N \right) } \right) } $. However, using it on the next set $ P_{ k 2^q, \left( k + 4 \right) 2^q - 1 } $, $ \forall q, k \in \mathbb{N}_0 $ and $ \left( k + 4 \right) 2^q \leq N $, where $ N $ is the number of points in the source polyline, will produce an algorithm with the worst-case complexity $ O{ \left( N^2 \log{ \left( N \right) } \right) } $ (see \myref{appendix:AppendixTestingArcs}). Note that if there is a subset of a tested set of points for which an arc cannot be fitted within the specified tolerance, then no arcs can be fitted to the tested set of points within the same specified tolerance. This algorithm breaks the long polyline into parts that are considered separate by the dynamic programming approach. For certain applications, it is unusual to have arcs with more than a few hundred points. If the limitation that the arc cannot have more than a specified number of points is accepted, then this approach to restrict the fitting of arcs changes the complexity of the algorithm from $ O{ \left( N^3 \log{ \left( N \right) } \right) } $ to $ O{ \left( N \right) } $.

\section
{
  \label{sec:DynamicProgramming}
  Dynamic Programming
}

Similar to \cite[section II.E]{OptimalCompression}, the goal of this algorithm is to find the solution with the minimum number of segments and arcs while satisfying the tolerance restriction, and among them with the minimum sum of square differences. Therefore, minimization is performed in two parts
$
  \left\{
    \begin{aligned}
      & T^{\#}\\
      & T^{\epsilon}
    \end{aligned}
  \right\}
$,
where the first part $ T^{\#} $ is the penalty for the number of segments and arcs, and the second part $ T^{\epsilon} $ is the sum of the squared deviations between points of the source polyline and corresponding resultant segment or arc. The solutions are compared by the penalty for the number of segments and arcs and, if they have the same penalty, by squared deviations between segments and the source polyline. It is reasonable to have a preference for segments versus arcs, because the arc is a more complicated geometric shape and requires one additional number for the curvature. Therefore, the penalty for a segment is $ PENALTY_{segment} = 2 $ (one for each coordinate), and the penalty for an arc is $ PENALTY_{arc} = 3 $ (for coordinates and curvature).

The optimal solution is found by induction:
\begin{enumerate}
  \item
    Define the penalty for the optimal solution for polyline $ P_{0, 0} $ as
    $
      \left\{
        \begin{aligned}
          & T_0^{\#}\\
          & T_0^{\epsilon}
        \end{aligned}
      \right\}
      =
      \left\{
        \begin{aligned}
          0\\
          0
        \end{aligned}
      \right\}
    $.
  \item \label{enum:SubArrayStep}
    For $ k $ from $ 1 $ to $ N - 1 $.
    \begin{enumerate}
      \item
        Set the penalty for the polyline $ P_{0, k} $ as
        $
          \left\{
            \begin{aligned}
              & T_k^{\#}\\
              & T_k^{\epsilon}
            \end{aligned}
          \right\}
          =
          \left\{
            \begin{aligned}
              & T_{k - 1}^{\#} + PENALTY_{segment}\\
              & T_{k - 1}^{\epsilon}
            \end{aligned}
          \right\}
        $.
        This is equivalent to taking the last segment as a solution. The square difference for the last segment is $ 0 $.
      \item
        Find $ i_{segment}{\left( k \right)} $ (see section~\myref{sec:EfficientDivisionOfPolyline}).
      \item
        Process all indices $ k' $ from $ i_{segment}{\left( k \right)} + 1 $ to $ k - 2 $ in ascending order of
        $
          \left\{
            \begin{aligned}
              & T_{k'}^{\#}\\
              & T_{k'}^{\epsilon}
            \end{aligned}
          \right\}
        $ (see \myref{appendix:MinimumSubElement}).

        If
        $
          \left\{
            \begin{aligned}
              & T_k^{\#}\\
              & T_k^{\epsilon}
            \end{aligned}
          \right\}
          \leq
          \left\{
            \begin{aligned}
              & T_{k'}^{\#} + PENALTY_{segment}\\
              & T_{k'}^{\epsilon}
            \end{aligned}
          \right\}
        $,
        then stop processing further indices.

        Evaluate the solution for the segment from vertex $ k $ to $ k' $. If it produces a solution better than
        $
          \left\{
            \begin{aligned}
              & T_k^{\#}\\
              & T_k^{\epsilon}
            \end{aligned}
          \right\}
        $
        and satisfies the tolerance, end points, and directional requirements, then update it and store $ k' $.
      \item
        Find $ i_{arc}{\left( k \right)} $ (see section~\myref{sec:EfficientDivisionOfPolyline}).
      \item
        Process all indices $ k' $ from $ i_{arc}{\left( k \right)} + 1 $ to $ k - 3 $ in ascending order of
        $
          \left\{
            \begin{aligned}
              & T_{k'}^{\#}\\
              & T_{k'}^{\epsilon}
            \end{aligned}
          \right\}
        $ (see \myref{appendix:MinimumSubElement}).

        If
        $
          \left\{
            \begin{aligned}
              & T_k^{\#}\\
              & T_k^{\epsilon}
            \end{aligned}
          \right\}
          \leq
          \left\{
            \begin{aligned}
              & T_{k'}^{\#} + PENALTY_{arc}\\
              & T_{k'}^{\epsilon}
            \end{aligned}
          \right\}
        $,
        then stop processing further indices.

        Fit the arc to the polyline $ P_{k', k} $ passing through vertices $ k' $ and $ k $ by approximation to the least squares approach \cite{EfficientFittingOfCircularArcs}.
        If it produces a solution better than
        $
          \left\{
            \begin{aligned}
              & T_k^{\#}\\
              & T_k^{\epsilon}
            \end{aligned}
          \right\}
        $
        and satisfies tolerance, end points, and directional requirements, then update it and store $ k' $.
        Otherwise, fit the arc by the algorithm described in \myref{appendix:FittingArcByTolerance}.
        If it produces a solution better than
        $
          \left\{
            \begin{aligned}
              & T_k^{\#}\\
              & T_k^{\epsilon}
            \end{aligned}
          \right\}
        $
        and satisfies end points and directional requirements, then update it and store $ k' $.
    \end{enumerate}
\end{enumerate}

The optimal solution is reconstructed by recurrently using stored $ k' $ values. Note that this algorithm finds the optimal solution with the limitation due to approximation in the check for direction (see \cite[section II.D]{OptimalCompression}).

\section{Compression of the polyline by Vertices versus Segments}

The task of polyline compression can be formulated to find a resultant polyline having the vertices of the source polyline within the specified tolerance or having the source polyline (segments of the source polyline) within the specified tolerance. If arcs have been lost due to limitations of the format, projection (performed on vertices instead of segments), and so forth, then the segments between vertices do not represent arcs. Enforcing segments of the source polyline to be within the specified tolerance from the arc does not make sense. Note that for the segments of the resultant polyline, if all vertices of the source polyline are within the specified tolerance, then the same is true for the segments of the source polyline. Now, let's discuss cases where it makes sense to enforce tolerance compliance for the segments of the source polyline. One case would be where vertices of the source polyline were removed by some compression. In that case, we cannot rely only on the vertices of the source polyline, as we do not know all of them. Another case would be where it is a requirement to have the resultant polyline within the specified tolerance from the source polyline.

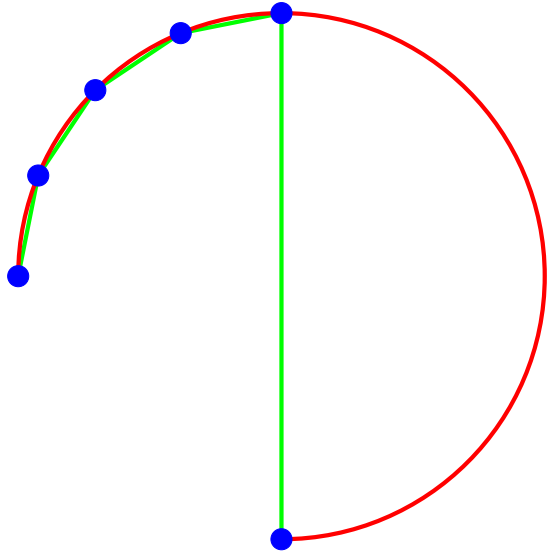
\begin{figure} [htb]
  \centering
  \begin{tikzpicture} [scale = 3.5] 
    \draw [ultra thick, green] (-1, 0) -- (-0.92387953251128675612818318939679, 0.3826834323650897717284599840304) -- (-0.70710678118654752440084436210485, 0.70710678118654752440084436210485) -- (-0.3826834323650897717284599840304, 0.92387953251128675612818318939679) -- (0, 1) -- (0, -1);

    \draw [ultra thick, red] (-1, 0) arc (180 : -90 : 1);

    \draw [blue, fill = blue] (-1, 0) circle [radius = 0.04];
    \draw [blue, fill = blue] (-0.92387953251128675612818318939679, 0.3826834323650897717284599840304) circle [radius = 0.04];
    \draw [blue, fill = blue] (-0.70710678118654752440084436210485, 0.70710678118654752440084436210485) circle [radius = 0.04];
    \draw [blue, fill = blue] (-0.3826834323650897717284599840304, 0.92387953251128675612818318939679) circle [radius = 0.04];
    \draw [blue, fill = blue] (0, 1) circle [radius = 0.04];
    \draw [blue, fill = blue] (0, -1) circle [radius = 0.04];
  \end{tikzpicture}
  \caption
  {
    The circular arc (red) perfectly fits the vertices (blue) of the source polyline (green).
  }
  \label{fig:UndesirableResultForArcFittingFromVertices}
\end{figure}

Requiring only vertices of the source polyline to be within the specified tolerance without additional limitations can produce undesirable results, as shown in Figure~\ref{fig:UndesirableResultForArcFittingFromVertices}. This can be resolved by limiting the angle between neighboring vertices. Another solution would be to require the source polyline to be within the specified tolerance from the resultant polyline. Notice that the effect is similar to densification of the source polyline.

The algorithm described in section~\myref{sec:FindingASmallestWidthAnnulus} can be extended for the segments; however, this will require implementation of the closest segment Voronoi diagram and a plane sweep algorithm for parabolas. An approximate solution can be obtained by densifying segments and using the algorithm described in section~\myref{sec:FindingASmallestWidthAnnulus} without any modifications.

\section{Examples}

\begin{sidewaysfigure} [p]
  \centering
  \includegraphics[width = \columnwidth, keepaspectratio]{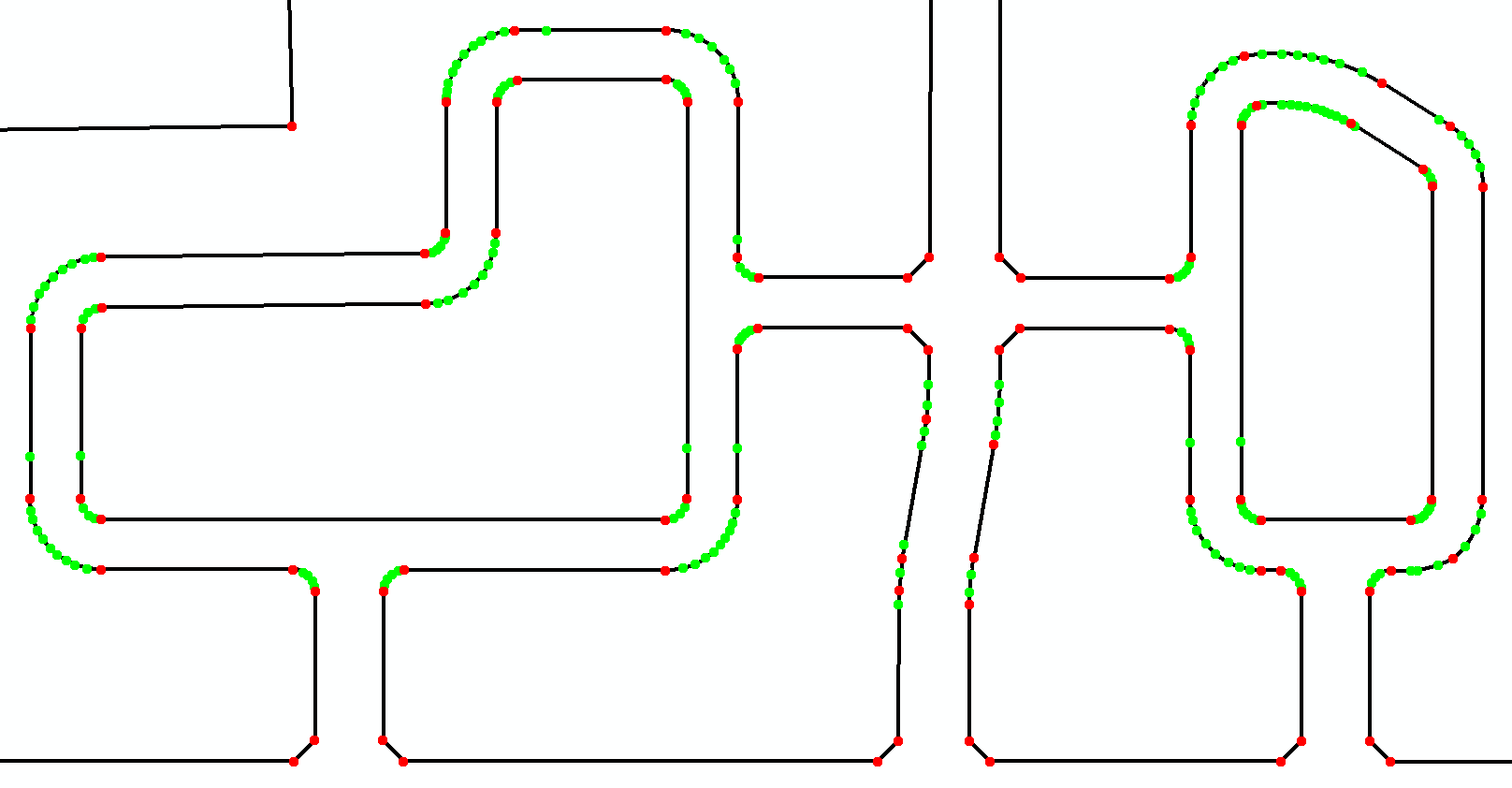} 
  \caption
  {
    Example of parcel data. Source polylines are shown as black lines. Green points are the vertices of the source polylines. Red points are the vertices of the resultant polylines.
  }
  \label{fig:ParcelExampleArcRecognition}
\end{sidewaysfigure}

An example of parcel data is shown in Figure~\ref{fig:ParcelExampleArcRecognition}.

From \cite[section VI]{EfficientFittingOfCircularArcs}:
\textit
{
  The original arcs were lost due to digitization, limitations of the format, projection, and so forth. The restoration of arcs is an important task because it is the original representation. Restoring original arcs creates cleaner databases and simplifies future editing.
}

\begin{figure} [htb]
  \input{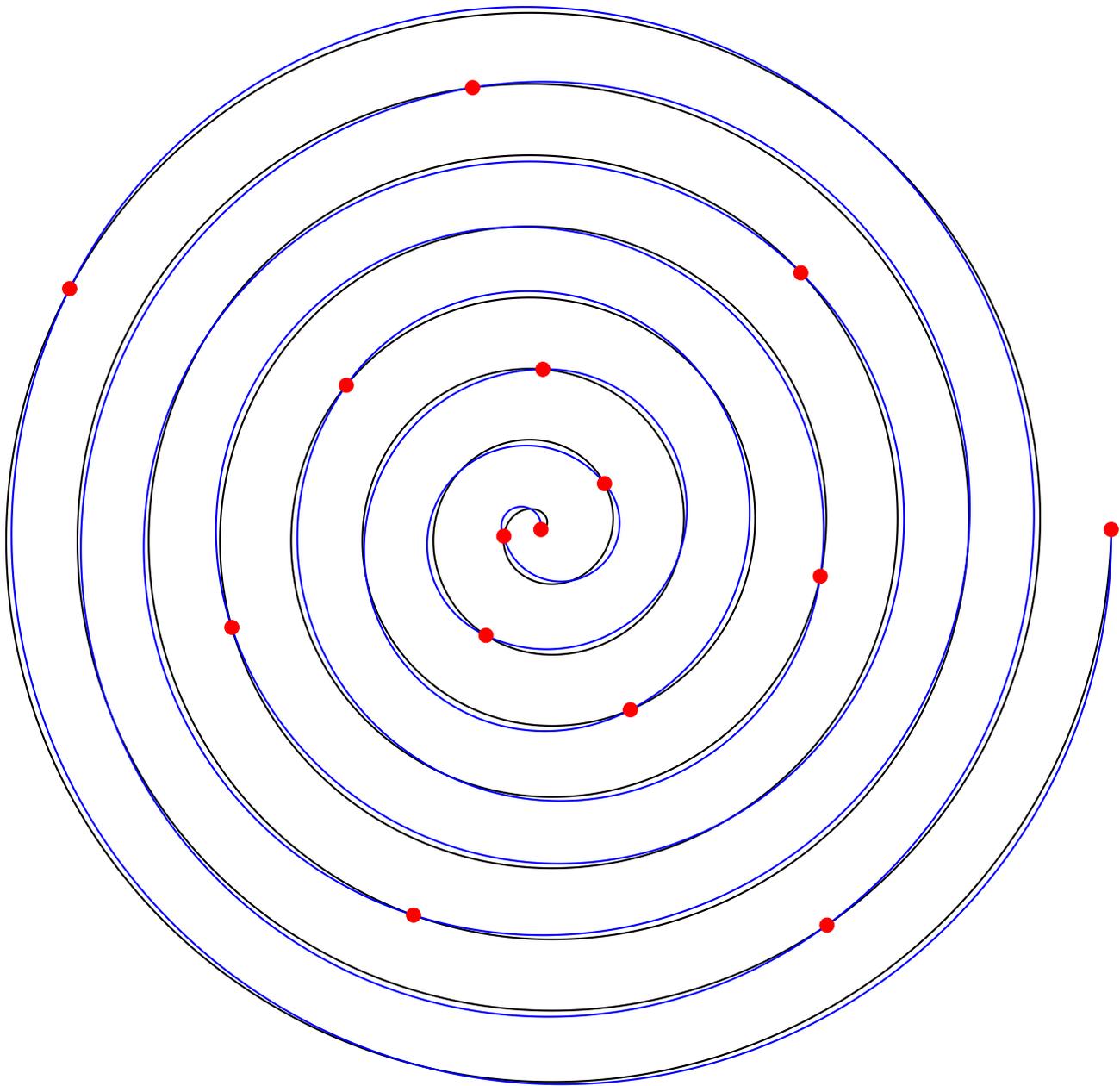}
  \caption
  {
    The source polyline is shown as a black polyline. The resultant polyline is shown as a blue polyline. Vertices of the resultant polyline are shown as red circles.
  }
  \label{fig:SpiralExampleArcRecognition}
\end{figure}

An example of compression of the Archimedean spiral with a constant separation distance between arms equal to $ 1 $ and the tolerance equal to $ 0.1 $ is shown in Figure~\ref{fig:SpiralExampleArcRecognition}.

\section{Conclusion}

This paper describes an efficient algorithm for compressing polylines with segments and arcs. The algorithm guarantees to find the resultant polyline, which minimizes the next penalty criteria. First, minimization is performed by the number of segments and arcs. The penalty for each segment is $ 2 $, and the penalty for each arc is $ 3 $ (other values can be used). Second, among solutions with the same minimum penalty, the solution with the minimum sum of squared deviations (for arcs, this value is approximate; see \cite{FittingOfCircularArcsWithO1Complexity}, \cite{EfficientFittingOfCircularArcs}, and \myref{appendix:FittingArcByTolerance}) is chosen.

Higher compression is achievable when the resultant polyline is not limited by the vertices of the source polyline. The algorithm in \cite{OptimalCompression} places a finite number of points around each vertex of the source polyline and searches for the resultant polyline connecting these points.

The most complicated parts of the algorithm are for arc fittings and arc verifications:
\begin{itemize}
  \item Checking an arc for tolerance (when fitted by the approximation to the least squares approach, see \cite{FittingOfCircularArcsWithO1Complexity} and \cite{EfficientFittingOfCircularArcs}).
  \item Tolerance fitting (see \myref{appendix:FittingArcByTolerance}).
  \item End points and directional check (see \cite[section 3]{PaperArcFitting}).
\end{itemize}

The closest or the farthest Delaunay triangulation for the set of vertices of the convex hull can be constructed in linear time, see \cite{ALinearTimeAlgorithmForComputingTheVoronoiDiagramOfAConvexPolygon} and \cite{OnComputingVoronoiDiagramsForSortedPointSets}. In the future, it would be interesting to compare \cite{ALinearTimeAlgorithmForComputingTheVoronoiDiagramOfAConvexPolygon} and \cite{OnComputingVoronoiDiagramsForSortedPointSets} with the \textbf{divide-and-conquer} approach \cite{TwoAlgorithmsForConstructingADelaunayTriangulation} and \cite{VoronoiDiagramDivideAndConquer1985}.

\section*{Acknowledgment}

The author would like to thank Linda Thomas for proofreading this paper.

\newcounter{CurrentSectionValue}
\setcounter{CurrentSectionValue}{\value{section}}
\setcounter{section}{0}
\renewcommand{\thesection}{Appendix \Roman{section}}
\input{OptimalCompressionOfAPolylineWithSegmentsAndArcsAppendices.tex}
\setcounter{section}{\value{CurrentSectionValue}}
\renewcommand{\thesection}{\Roman{section}}

\clearpage

\newcommand{\doi}[1]{\textsc{doi}: \href{http://dx.doi.org/#1}{\nolinkurl{#1}}}


\bibliographystyle{IEEEtran}
\bibliography{OptimalCompressionOfAPolylineWithSegmentsAndArcs}


\end{document}

%% file: FigureExampleVoronoiDiagrams.tex

  \centering
  \begin{tikzpicture} [scale = 4.5]
    \begin{scope}
      \clip (-2, -2.5) -- (-2, 2.5) -- (2, 2.5) -- (2, -2.5) -- (-2, -2.5);
      \draw [blue] (-0.306774, -0.469122) -- (-0.469249, -0.667499);
      \draw [blue] (-0.501019, -0.724247) -- (-0.627881, -0.748381);
      \draw [blue] (-0.007214, -0.014596) -- (-0.074254, -0.107055);
      \draw [blue] (-0.669373, -0.673695) -- (-0.591645, -0.693946);
      \draw [blue] (-0.627881, -0.748381) -- (-0.591645, -0.693946);
      \draw [blue] (-0.591645, -0.693946) -- (-0.469249, -0.667499);
      \draw [blue] (-0.798411, -0.742126) -- (-0.678575, -0.678575);
      \draw [blue] (-0.669373, -0.673695) -- (-0.678575, -0.678575);
      \draw [blue] (-1.426012, -1.331241) -- (-0.999589, -0.935283);
      \draw [blue] (-0.973057, -0.913029) -- (-0.798411, -0.742126);
      \draw [blue] (-0.764253, -0.522312) -- (-0.734197, -0.562633);
      \draw [blue] (-0.729722, -0.616355) -- (-0.051382, -0.051382);
      \draw [blue] (-0.007214, -0.014596) -- (-0.051382, -0.051382);
      \draw [blue] (-0.734197, -0.562633) -- (-0.729722, -0.616355);
      \draw [blue] (-0.729722, -0.616355) -- (-0.798411, -0.742126);
      \draw [blue] (-0.669373, -0.673695) -- (-0.074254, -0.107055);
      \draw [blue] (0.002968, -0.001685) -- (0.002402, -0.002402);
      \draw [blue] (-0.007214, -0.014596) -- (0.002402, -0.002402);
      \draw [blue] (-0.627881, -0.748381) -- (-0.858281, -0.858281);
      \draw [blue] (-0.973057, -0.913029) -- (-0.858281, -0.858281);
      \draw [blue] (-4.416503, -5.000000) -- (-4.658704, -5.000000);
      \draw [blue] (-4.658704, -5.000000) -- (-4.913202, -5.000000);
      \draw [blue] (-1.757009, -5.000000) -- (-2.934406, -5.000000);
      \draw [blue] (-2.934406, -5.000000) -- (-4.416503, -5.000000);
      \draw [blue] (-0.519621, -0.792863) -- (-0.499928, -0.805605);
      \draw [blue] (-0.499928, -0.805605) -- (-0.400742, -0.931105);
      \draw [blue] (-0.519621, -0.792863) -- (-0.501019, -0.724247);
      \draw [blue] (-0.519621, -0.792863) -- (-4.235788, -5.000000);
      \draw [blue] (-4.416503, -5.000000) -- (-4.235788, -5.000000);
      \draw [blue] (-0.411936, -0.827564) -- (-0.400742, -0.931105);
      \draw [blue] (-0.074254, -0.107055) -- (-0.306774, -0.469122);
      \draw [blue] (-0.078936, -0.020463) -- (-0.417116, -0.153550);
      \draw [blue] (0.009831, -0.024123) -- (-0.008734, -0.217601);
      \draw [blue] (-0.008734, -0.217601) -- (-0.335985, -0.957128);
      \draw [blue] (-0.306774, -0.469122) -- (-0.411936, -0.827564);
      \draw [blue] (-0.411936, -0.827564) -- (-0.499928, -0.805605);
      \draw [blue] (-0.400742, -0.931105) -- (-0.376621, -1.021357);
      \draw [blue] (-0.133225, -0.882094) -- (-0.335985, -0.957128);
      \draw [blue] (-0.469249, -0.667499) -- (-0.501019, -0.724247);
      \draw [blue] (-0.973057, -0.913029) -- (-0.999589, -0.935283);
      \draw [blue] (-0.078936, -0.020463) -- (-0.912806, -0.017110);
      \draw [blue] (-4.913202, -5.000000) -- (-5.000000, -5.000000);
      \draw [blue] (-5.000000, -3.598093) -- (-5.000000, -5.000000);
      \draw [blue] (-5.000000, -3.598093) -- (-5.000000, -2.436215);
      \draw [blue] (-4.913202, -5.000000) -- (-5.000000, -5.000000);
      \draw [blue] (-1.426012, -1.331241) -- (-5.000000, -4.968368);
      \draw [blue] (-5.000000, -5.000000) -- (-5.000000, -4.968368);
      \draw [blue] (-0.999589, -0.935283) -- (-0.764253, -0.522312);
      \draw [blue] (-0.760283, -0.406996) -- (-0.764253, -0.522312);
      \draw [blue] (-0.767806, -0.414249) -- (-5.000000, -2.476355);
      \draw [blue] (-5.000000, -2.436215) -- (-5.000000, -2.476355);
      \draw [blue] (-0.767806, -0.414249) -- (-1.426012, -1.331241);
      \draw [blue] (-0.762229, -0.408107) -- (-0.760283, -0.406996);
      \draw [blue] (-5.000000, -2.436215) -- (-5.000000, -2.094688);
      \draw [blue] (-5.000000, -2.094688) -- (-5.000000, -1.769017);
      \draw [blue] (-0.767806, -0.414249) -- (-0.762229, -0.408107);
      \draw [blue] (-5.000000, -1.769017) -- (-5.000000, -1.701253);
      \draw [blue] (-5.000000, -1.701253) -- (-5.000000, -1.634044);
      \draw [blue] (-1.172620, -0.430180) -- (-0.803851, -0.236866);
      \draw [blue] (-1.172620, -0.430180) -- (-5.000000, -1.784320);
      \draw [blue] (-5.000000, -1.769017) -- (-5.000000, -1.784320);
      \draw [blue] (-1.172620, -0.430180) -- (-0.762229, -0.408107);
      \draw [blue] (-0.760283, -0.406996) -- (-0.561174, -0.214544);
      \draw [blue] (-0.919678, -0.183134) -- (-0.803851, -0.236866);
      \draw [blue] (-0.803851, -0.236866) -- (-0.561174, -0.214544);
      \draw [blue] (-5.000000, -1.634044) -- (-5.000000, -0.829427);
      \draw [blue] (-5.000000, -0.829427) -- (-5.000000, -0.064552);
      \draw [blue] (-1.030109, -0.127537) -- (-5.000000, -0.136185);
      \draw [blue] (-5.000000, -0.106503) -- (-5.000000, -0.136185);
      \draw [blue] (-0.919678, -0.183134) -- (-5.000000, -1.516620);
      \draw [blue] (-5.000000, -1.634044) -- (-5.000000, -1.516620);
      \draw [blue] (-0.935694, -0.085265) -- (-0.952361, -0.140950);
      \draw [blue] (-1.030109, -0.127537) -- (-0.952361, -0.140950);
      \draw [blue] (-0.952361, -0.140950) -- (-0.919678, -0.183134);
      \draw [blue] (-0.942546, -0.073714) -- (-0.935694, -0.085265);
      \draw [blue] (-5.000000, -0.064552) -- (-5.000000, -0.106503);
      \draw [blue] (-1.067016, -0.024250) -- (-5.000000, -0.106933);
      \draw [blue] (-5.000000, -0.106503) -- (-5.000000, -0.106933);
      \draw [blue] (-1.067016, -0.024250) -- (-0.982383, -0.049768);
      \draw [blue] (-0.931796, -0.002243) -- (-0.982383, -0.049768);
      \draw [blue] (-0.931796, -0.002243) -- (-0.947678, 0.018187);
      \draw [blue] (-0.912806, -0.017110) -- (-0.931796, -0.002243);
      \draw [blue] (-0.912806, -0.017110) -- (-0.942546, -0.073714);
      \draw [blue] (-0.982383, -0.049768) -- (-0.949959, -0.071565);
      \draw [blue] (-0.949959, -0.071565) -- (-1.030109, -0.127537);
      \draw [blue] (-0.949959, -0.071565) -- (-0.942546, -0.073714);
      \draw [blue] (-0.935694, -0.085265) -- (-0.437563, -0.160977);
      \draw [blue] (-0.561174, -0.214544) -- (-0.437563, -0.160977);
      \draw [blue] (-0.437563, -0.160977) -- (-0.417116, -0.153550);
      \draw [blue] (-0.734197, -0.562633) -- (-0.417116, -0.153550);
      \draw [blue] (0.002815, 0.001237) -- (0.002968, -0.001685);
      \draw [blue] (-0.335985, -0.957128) -- (-0.376621, -1.021357);
      \draw [blue] (-0.048467, -0.730257) -- (-0.133225, -0.882094);
      \draw [blue] (-0.633129, -5.000000) -- (-1.180949, -5.000000);
      \draw [blue] (-1.180949, -5.000000) -- (-1.757009, -5.000000);
      \draw [blue] (-0.048467, -0.730257) -- (-0.004763, -0.398813);
      \draw [blue] (-0.141259, -0.911369) -- (-0.132858, -1.142289);
      \draw [blue] (-0.133225, -0.882094) -- (-0.141259, -0.911369);
      \draw [blue] (-0.141259, -0.911369) -- (-0.391419, -1.077403);
      \draw [blue] (-0.035204, -0.875974) -- (-0.132858, -1.142289);
      \draw [blue] (0.711668, -5.000000) -- (0.038572, -5.000000);
      \draw [blue] (0.038572, -5.000000) -- (-0.633129, -5.000000);
      \draw [blue] (0.128830, -1.016913) -- (0.185791, -1.565988);
      \draw [blue] (-0.030537, -0.837521) -- (0.003666, -0.892965);
      \draw [blue] (0.003666, -0.892965) -- (0.185791, -1.565988);
      \draw [blue] (-0.030537, -0.837521) -- (-0.048467, -0.730257);
      \draw [blue] (-0.035204, -0.875974) -- (0.463430, -3.645813);
      \draw [blue] (-0.030537, -0.837521) -- (-0.035204, -0.875974);
      \draw [blue] (-0.132858, -1.142289) -- (-0.621344, -5.000000);
      \draw [blue] (-0.633129, -5.000000) -- (-0.621344, -5.000000);
      \draw [blue] (0.185791, -1.565988) -- (0.463430, -3.645813);
      \draw [blue] (-0.008734, -0.217601) -- (-0.004763, -0.398813);
      \draw [blue] (1.493459, -5.000000) -- (1.096172, -5.000000);
      \draw [blue] (1.096172, -5.000000) -- (0.711668, -5.000000);
      \draw [blue] (0.009831, -0.024123) -- (0.003534, -0.003534);
      \draw [blue] (0.002968, -0.001685) -- (0.003534, -0.003534);
      \draw [blue] (0.143527, -0.883911) -- (0.264871, -0.932656);
      \draw [blue] (-0.004763, -0.398813) -- (0.143527, -0.883911);
      \draw [blue] (0.143527, -0.883911) -- (0.128830, -1.016913);
      \draw [blue] (0.318954, -0.738920) -- (0.264871, -0.932656);
      \draw [blue] (0.321144, -0.450745) -- (0.318954, -0.738920);
      \draw [blue] (3.497300, -5.000000) -- (2.417186, -5.000000);
      \draw [blue] (2.417186, -5.000000) -- (1.493459, -5.000000);
      \draw [blue] (5.000000, -3.786783) -- (5.000000, -5.000000);
      \draw [blue] (4.814343, -5.000000) -- (5.000000, -5.000000);
      \draw [blue] (4.814343, -5.000000) -- (3.497300, -5.000000);
      \draw [blue] (0.429261, -0.574055) -- (3.525032, -5.000000);
      \draw [blue] (3.497300, -5.000000) -- (3.525032, -5.000000);
      \draw [blue] (0.321144, -0.450745) -- (0.297327, -0.408705);
      \draw [blue] (0.318954, -0.738920) -- (1.568378, -5.000000);
      \draw [blue] (1.506176, -5.000000) -- (1.568378, -5.000000);
      \draw [blue] (0.264871, -0.932656) -- (1.495629, -5.000000);
      \draw [blue] (1.506176, -5.000000) -- (1.495629, -5.000000);
      \draw [blue] (1.506176, -5.000000) -- (1.493459, -5.000000);
      \draw [blue] (0.003666, -0.892965) -- (0.128830, -1.016913);
      \draw [blue] (0.463430, -3.645813) -- (0.656176, -5.000000);
      \draw [blue] (0.711668, -5.000000) -- (0.656176, -5.000000);
      \draw [blue] (0.321144, -0.450745) -- (0.429261, -0.574055);
      \draw [blue] (0.009831, -0.024123) -- (0.297327, -0.408705);
      \draw [blue] (5.000000, -2.834943) -- (5.000000, -3.290101);
      \draw [blue] (5.000000, -3.290101) -- (5.000000, -3.786783);
      \draw [blue] (0.747509, -0.532653) -- (0.786548, -0.524429);
      \draw [blue] (0.786548, -0.524429) -- (0.915137, -0.550800);
      \draw [blue] (0.747509, -0.532653) -- (0.451022, -0.451022);
      \draw [blue] (0.297327, -0.408705) -- (0.451022, -0.451022);
      \draw [blue] (0.747509, -0.532653) -- (0.782309, -0.634612);
      \draw [blue] (0.797676, -0.399945) -- (0.915137, -0.550800);
      \draw [blue] (0.002815, 0.001237) -- (0.001226, 0.001226);
      \draw [blue] (0.000000, 0.001217) -- (0.001226, 0.001226);
      \draw [blue] (-0.014301, 0.001117) -- (-0.001209, 0.001209);
      \draw [blue] (0.000000, 0.001217) -- (-0.001209, 0.001209);
      \draw [blue] (5.000000, -1.799326) -- (5.000000, -2.296825);
      \draw [blue] (5.000000, -2.296825) -- (5.000000, -2.834943);
      \draw [blue] (0.759080, -0.311313) -- (0.804995, -0.319404);
      \draw [blue] (0.804995, -0.319404) -- (5.000000, -1.822665);
      \draw [blue] (5.000000, -1.797181) -- (5.000000, -1.822665);
      \draw [blue] (0.759080, -0.311313) -- (0.427939, -0.123730);
      \draw [blue] (0.759080, -0.311313) -- (0.797676, -0.399945);
      \draw [blue] (0.786548, -0.524429) -- (0.797676, -0.399945);
      \draw [blue] (0.915137, -0.550800) -- (5.000000, -2.866871);
      \draw [blue] (5.000000, -2.834943) -- (5.000000, -2.866871);
      \draw [blue] (0.946033, -0.239463) -- (5.000000, -1.711478);
      \draw [blue] (5.000000, -1.797181) -- (5.000000, -1.711478);
      \draw [blue] (0.006614, 0.004926) -- (0.427939, -0.123730);
      \draw [blue] (0.427939, -0.123730) -- (0.603895, -0.125828);
      \draw [blue] (5.000000, -1.187103) -- (5.000000, -1.488092);
      \draw [blue] (5.000000, -1.488092) -- (5.000000, -1.799326);
      \draw [blue] (0.931437, -0.143705) -- (0.937893, -0.228222);
      \draw [blue] (0.603895, -0.125828) -- (0.937893, -0.228222);
      \draw [blue] (0.937893, -0.228222) -- (0.946033, -0.239463);
      \draw [blue] (0.861800, -0.115381) -- (0.931437, -0.143705);
      \draw [blue] (0.603895, -0.125828) -- (0.861800, -0.115381);
      \draw [blue] (5.000000, -0.543567) -- (5.000000, -0.861869);
      \draw [blue] (5.000000, -0.861869) -- (5.000000, -1.187103);
      \draw [blue] (0.989674, -0.056551) -- (0.946641, -0.144741);
      \draw [blue] (0.914950, -0.073125) -- (0.961731, -0.050233);
      \draw [blue] (0.005004, 0.004423) -- (0.006614, 0.004926);
      \draw [blue] (0.892237, 0.069618) -- (0.926800, 0.041266);
      \draw [blue] (0.926800, 0.041266) -- (0.914950, -0.073125);
      \draw [blue] (0.861800, -0.115381) -- (0.914950, -0.073125);
      \draw [blue] (0.931185, 0.038748) -- (0.961731, -0.050233);
      \draw [blue] (0.931437, -0.143705) -- (0.946641, -0.144741);
      \draw [blue] (0.946641, -0.144741) -- (5.000000, -1.107091);
      \draw [blue] (5.000000, -1.187103) -- (5.000000, -1.107091);
      \draw [blue] (0.804995, -0.319404) -- (0.946033, -0.239463);
      \draw [blue] (5.000000, -1.797181) -- (5.000000, -1.799326);
      \draw [blue] (0.429261, -0.574055) -- (0.604033, -0.604033);
      \draw [blue] (0.782309, -0.634612) -- (0.604033, -0.604033);
      \draw [blue] (0.782309, -0.634612) -- (5.000000, -3.828908);
      \draw [blue] (5.000000, -3.786783) -- (5.000000, -3.828908);
      \draw [blue] (-0.376621, -1.021357) -- (-0.391419, -1.077403);
      \draw [blue] (-0.391419, -1.077403) -- (-1.769827, -5.000000);
      \draw [blue] (-1.757009, -5.000000) -- (-1.769827, -5.000000);
      \draw [blue] (0.974805, -0.051489) -- (0.989674, -0.056551);
      \draw [blue] (-1.067016, -0.024250) -- (-0.947678, 0.018187);
      \draw [blue] (-5.000000, -0.064552) -- (-5.000000, 0.503193);
      \draw [blue] (-5.000000, 0.503193) -- (-5.000000, 1.084080);
      \draw [blue] (-1.166137, 0.146972) -- (-5.000000, 1.015183);
      \draw [blue] (-5.000000, 1.092356) -- (-5.000000, 1.015183);
      \draw [blue] (-1.166137, 0.146972) -- (-0.940009, 0.076856);
      \draw [blue] (-0.947678, 0.018187) -- (-0.940009, 0.076856);
      \draw [blue] (-0.940009, 0.076856) -- (-0.855055, 0.116882);
      \draw [blue] (-0.915558, 0.153664) -- (-0.855055, 0.116882);
      \draw [blue] (-5.000000, 1.084080) -- (-5.000000, 1.092356);
      \draw [blue] (-1.166137, 0.146972) -- (-0.915558, 0.153664);
      \draw [blue] (-1.040074, 0.318313) -- (-1.038551, 0.318211);
      \draw [blue] (-1.038551, 0.318211) -- (-0.726690, 0.127496);
      \draw [blue] (-1.040074, 0.318313) -- (-5.000000, 1.156963);
      \draw [blue] (-5.000000, 1.092356) -- (-5.000000, 1.156963);
      \draw [blue] (-1.040074, 0.318313) -- (-0.915558, 0.153664);
      \draw [blue] (-0.855055, 0.116882) -- (-0.726690, 0.127496);
      \draw [blue] (-0.726690, 0.127496) -- (-0.196489, 0.066307);
      \draw [blue] (-0.632564, 0.312978) -- (-0.196489, 0.066307);
      \draw [blue] (-5.000000, 1.084080) -- (-5.000000, 2.251103);
      \draw [blue] (-5.000000, 2.251103) -- (-5.000000, 3.665242);
      \draw [blue] (-1.038551, 0.318211) -- (-0.903730, 0.341582);
      \draw [blue] (-1.038703, 0.684108) -- (-1.004003, 0.655595);
      \draw [blue] (-0.862408, 0.487743) -- (-0.689333, 0.372368);
      \draw [blue] (-1.004003, 0.655595) -- (-0.862408, 0.487743);
      \draw [blue] (-0.862408, 0.487743) -- (-0.903730, 0.341582);
      \draw [blue] (-0.764416, 0.556781) -- (-0.689333, 0.372368);
      \draw [blue] (-0.710791, 0.600967) -- (-0.708072, 0.592950);
      \draw [blue] (-2.559448, 1.857232) -- (-1.038703, 0.684108);
      \draw [blue] (-2.559448, 1.857232) -- (-5.000000, 3.646274);
      \draw [blue] (-5.000000, 3.665242) -- (-5.000000, 3.646274);
      \draw [blue] (-1.038703, 0.684108) -- (-0.710791, 0.600967);
      \draw [blue] (-0.764416, 0.556781) -- (-0.708072, 0.592950);
      \draw [blue] (-0.710896, 0.601847) -- (-0.710791, 0.600967);
      \draw [blue] (-1.004003, 0.655595) -- (-0.764416, 0.556781);
      \draw [blue] (-0.689333, 0.372368) -- (-0.632564, 0.312978);
      \draw [blue] (-0.903730, 0.341582) -- (-0.632564, 0.312978);
      \draw [blue] (-0.196489, 0.066307) -- (-0.014301, 0.001117);
      \draw [blue] (-0.383204, 0.403182) -- (0.003057, 0.011683);
      \draw [blue] (-2.559448, 1.857232) -- (-0.823403, 0.731747);
      \draw [blue] (-0.823403, 0.731747) -- (-0.710896, 0.601847);
      \draw [blue] (-0.517600, 0.794355) -- (-0.564197, 0.688842);
      \draw [blue] (-0.672142, 0.672042) -- (-0.509884, 0.493150);
      \draw [blue] (-0.690299, 0.668716) -- (-0.823403, 0.731747);
      \draw [blue] (-0.690299, 0.668716) -- (-0.710896, 0.601847);
      \draw [blue] (-0.690299, 0.668716) -- (-0.672142, 0.672042);
      \draw [blue] (-5.000000, 3.665242) -- (-5.000000, 4.735406);
      \draw [blue] (-5.000000, 4.735406) -- (-5.000000, 5.000000);
      \draw [blue] (-4.100749, 5.000000) -- (-5.000000, 5.000000);
      \draw [blue] (-0.451690, 0.838082) -- (-0.517600, 0.794355);
      \draw [blue] (-4.100749, 5.000000) -- (-3.236243, 5.000000);
      \draw [blue] (-3.236243, 5.000000) -- (-2.489523, 5.000000);
      \draw [blue] (-0.633560, 0.764754) -- (-0.517600, 0.794355);
      \draw [blue] (-0.633560, 0.764754) -- (-4.107097, 5.000000);
      \draw [blue] (-4.100749, 5.000000) -- (-4.107097, 5.000000);
      \draw [blue] (-0.633560, 0.764754) -- (-0.627576, 0.746243);
      \draw [blue] (-0.672142, 0.672042) -- (-0.672104, 0.672104);
      \draw [blue] (-0.627576, 0.746243) -- (-0.672104, 0.672104);
      \draw [blue] (-0.627576, 0.746243) -- (-0.564197, 0.688842);
      \draw [blue] (-0.564197, 0.688842) -- (-0.383204, 0.403182);
      \draw [blue] (-0.451690, 0.838082) -- (-2.523929, 5.000000);
      \draw [blue] (-2.489523, 5.000000) -- (-2.523929, 5.000000);
      \draw [blue] (-2.489523, 5.000000) -- (-1.901581, 5.000000);
      \draw [blue] (-1.901581, 5.000000) -- (-1.356250, 5.000000);
      \draw [blue] (-0.291775, 1.235055) -- (-0.273396, 1.008473);
      \draw [blue] (-0.273396, 1.008473) -- (-0.419970, 0.808675);
      \draw [blue] (-0.451690, 0.838082) -- (-0.419970, 0.808675);
      \draw [blue] (-0.419970, 0.808675) -- (-0.176862, 0.522454);
      \draw [blue] (-0.273396, 1.008473) -- (-0.176862, 0.522454);
      \draw [blue] (-0.291775, 1.235055) -- (-1.313016, 5.000000);
      \draw [blue] (-1.356250, 5.000000) -- (-1.313016, 5.000000);
      \draw [blue] (-0.291775, 1.235055) -- (-0.175611, 1.047781);
      \draw [blue] (-0.175611, 1.047781) -- (-0.157573, 0.487080);
      \draw [blue] (-0.111746, 0.947528) -- (-0.098443, 0.377815);
      \draw [blue] (-0.030774, 1.028226) -- (-0.039922, 0.970641);
      \draw [blue] (-1.356250, 5.000000) -- (-0.793071, 5.000000);
      \draw [blue] (-0.793071, 5.000000) -- (-0.248860, 5.000000);
      \draw [blue] (-0.111746, 0.947528) -- (-0.039922, 0.970641);
      \draw [blue] (-0.175611, 1.047781) -- (-0.111746, 0.947528);
      \draw [blue] (-0.157573, 0.487080) -- (-0.176862, 0.522454);
      \draw [blue] (-0.157573, 0.487080) -- (-0.098443, 0.377815);
      \draw [blue] (-0.098443, 0.377815) -- (-0.039281, 0.233844);
      \draw [blue] (-0.509884, 0.493150) -- (-0.708072, 0.592950);
      \draw [blue] (-0.383204, 0.403182) -- (-0.452141, 0.452141);
      \draw [blue] (-0.509884, 0.493150) -- (-0.452141, 0.452141);
      \draw [blue] (-0.012632, 0.157011) -- (0.003057, 0.011683);
      \draw [blue] (-0.012752, 0.943162) -- (-0.039281, 0.233844);
      \draw [blue] (0.005004, 0.004423) -- (0.002815, 0.001237);
      \draw [blue] (5.000000, 0.172708) -- (5.000000, -0.184485);
      \draw [blue] (5.000000, -0.184485) -- (5.000000, -0.543567);
      \draw [blue] (0.017158, 0.011667) -- (0.006614, 0.004926);
      \draw [blue] (5.000000, 1.493618) -- (5.000000, 0.819248);
      \draw [blue] (5.000000, 0.819248) -- (5.000000, 0.172708);
      \draw [blue] (0.931185, 0.038748) -- (0.984814, 0.028617);
      \draw [blue] (0.974805, -0.051489) -- (0.984814, 0.028617);
      \draw [blue] (0.984814, 0.028617) -- (5.000000, 0.167308);
      \draw [blue] (5.000000, 0.172708) -- (5.000000, 0.167308);
      \draw [blue] (0.388711, 0.128426) -- (0.892237, 0.069618);
      \draw [blue] (-0.078936, -0.020463) -- (-0.014301, 0.001117);
      \draw [blue] (0.892237, 0.069618) -- (1.058020, 0.179583);
      \draw [blue] (5.000000, 2.855603) -- (5.000000, 2.141128);
      \draw [blue] (5.000000, 2.141128) -- (5.000000, 1.493618);
      \draw [blue] (0.614877, 0.234456) -- (1.793565, 0.486115);
      \draw [blue] (0.511727, 0.182050) -- (0.388711, 0.128426);
      \draw [blue] (0.511727, 0.182050) -- (1.058020, 0.179583);
      \draw [blue] (1.058020, 0.179583) -- (1.793565, 0.486115);
      \draw [blue] (1.793565, 0.486115) -- (5.000000, 1.443953);
      \draw [blue] (5.000000, 1.493618) -- (5.000000, 1.443953);
      \draw [blue] (0.511727, 0.182050) -- (0.614877, 0.234456);
      \draw [blue] (0.005004, 0.004423) -- (0.004881, 0.004881);
      \draw [blue] (0.003057, 0.011683) -- (0.004881, 0.004881);
      \draw [blue] (0.017158, 0.011667) -- (0.336738, 0.115436);
      \draw [blue] (0.614877, 0.234456) -- (0.689712, 0.307598);
      \draw [blue] (0.811047, 0.533618) -- (0.689712, 0.307598);
      \draw [blue] (0.751619, 0.538792) -- (0.749228, 0.594840);
      \draw [blue] (0.751619, 0.538792) -- (0.336738, 0.115436);
      \draw [blue] (0.751619, 0.538792) -- (0.811047, 0.533618);
      \draw [blue] (0.017158, 0.011667) -- (0.034499, 0.034499);
      \draw [blue] (0.272480, 0.347830) -- (0.034499, 0.034499);
      \draw [blue] (0.689712, 0.307598) -- (1.060930, 0.616727);
      \draw [blue] (3.974692, 5.000000) -- (5.000000, 5.000000);
      \draw [blue] (5.000000, 4.289231) -- (5.000000, 5.000000);
      \draw [blue] (5.000000, 4.289231) -- (5.000000, 2.855603);
      \draw [blue] (0.749228, 0.594840) -- (1.833371, 1.093267);
      \draw [blue] (0.679524, 0.605254) -- (0.659736, 0.659736);
      \draw [blue] (0.628398, 0.746024) -- (0.659736, 0.659736);
      \draw [blue] (0.679524, 0.605254) -- (0.477472, 0.477472);
      \draw [blue] (0.272480, 0.347830) -- (0.477472, 0.477472);
      \draw [blue] (0.696963, 0.615706) -- (0.749228, 0.594840);
      \draw [blue] (0.811047, 0.533618) -- (1.060930, 0.616727);
      \draw [blue] (1.060930, 0.616727) -- (1.833371, 1.093267);
      \draw [blue] (1.833371, 1.093267) -- (5.000000, 2.901793);
      \draw [blue] (5.000000, 2.855603) -- (5.000000, 2.901793);
      \draw [blue] (0.679524, 0.605254) -- (0.696963, 0.615706);
      \draw [blue] (-0.039922, 0.970641) -- (-0.012752, 0.943162);
      \draw [blue] (-0.030774, 1.028226) -- (-0.228458, 5.000000);
      \draw [blue] (-0.248860, 5.000000) -- (-0.228458, 5.000000);
      \draw [blue] (-0.248860, 5.000000) -- (0.152522, 5.000000);
      \draw [blue] (0.152522, 5.000000) -- (0.555878, 5.000000);
      \draw [blue] (-0.030774, 1.028226) -- (0.028596, 0.965855);
      \draw [blue] (-0.012752, 0.943162) -- (0.028596, 0.965855);
      \draw [blue] (0.028596, 0.965855) -- (0.106038, 0.980372);
      \draw [blue] (0.106038, 0.980372) -- (0.552923, 5.000000);
      \draw [blue] (0.555878, 5.000000) -- (0.552923, 5.000000);
      \draw [blue] (0.555878, 5.000000) -- (1.043646, 5.000000);
      \draw [blue] (1.043646, 5.000000) -- (1.551222, 5.000000);
      \draw [blue] (0.106038, 0.980372) -- (0.125987, 0.905657);
      \draw [blue] (0.294594, 1.006688) -- (1.533497, 5.000000);
      \draw [blue] (1.551222, 5.000000) -- (1.533497, 5.000000);
      \draw [blue] (0.258837, 0.969673) -- (0.216695, 0.754665);
      \draw [blue] (0.294594, 1.006688) -- (0.258837, 0.969673);
      \draw [blue] (0.258837, 0.969673) -- (0.125987, 0.905657);
      \draw [blue] (0.125987, 0.905657) -- (0.107979, 0.582503);
      \draw [blue] (0.301652, 0.858136) -- (0.216695, 0.754665);
      \draw [blue] (0.216695, 0.754665) -- (0.107979, 0.582503);
      \draw [blue] (0.313751, 0.910964) -- (0.301652, 0.858136);
      \draw [blue] (0.294594, 1.006688) -- (0.313751, 0.910964);
      \draw [blue] (1.551222, 5.000000) -- (1.899479, 5.000000);
      \draw [blue] (1.899479, 5.000000) -- (2.264623, 5.000000);
      \draw [blue] (0.313751, 0.910964) -- (0.385295, 0.909673);
      \draw [blue] (0.301652, 0.858136) -- (0.390907, 0.866916);
      \draw [blue] (0.385295, 0.909673) -- (2.237905, 5.000000);
      \draw [blue] (2.264623, 5.000000) -- (2.237905, 5.000000);
      \draw [blue] (0.385295, 0.909673) -- (0.390907, 0.866916);
      \draw [blue] (2.264623, 5.000000) -- (3.046845, 5.000000);
      \draw [blue] (3.046845, 5.000000) -- (3.955388, 5.000000);
      \draw [blue] (3.955388, 5.000000) -- (3.965031, 5.000000);
      \draw [blue] (3.965031, 5.000000) -- (3.974692, 5.000000);
      \draw [blue] (0.445564, 0.816799) -- (3.754800, 5.000000);
      \draw [blue] (3.955388, 5.000000) -- (3.754800, 5.000000);
      \draw [blue] (1.360604, 1.630041) -- (0.426385, 0.686374);
      \draw [blue] (0.398311, 0.559860) -- (0.628398, 0.746024);
      \draw [blue] (0.696963, 0.615706) -- (0.722894, 0.722894);
      \draw [blue] (0.753791, 0.850608) -- (0.722894, 0.722894);
      \draw [blue] (0.628398, 0.746024) -- (0.753791, 0.850608);
      \draw [blue] (0.445564, 0.816799) -- (0.390907, 0.866916);
      \draw [blue] (0.426385, 0.686374) -- (0.445564, 0.816799);
      \draw [blue] (0.426385, 0.686374) -- (0.398311, 0.559860);
      \draw [blue] (0.398311, 0.559860) -- (0.272480, 0.347830);
      \draw [blue] (0.107979, 0.582503) -- (-0.012632, 0.157011);
      \draw [blue] (-0.012632, 0.157011) -- (-0.039281, 0.233844);
      \draw [blue] (0.753791, 0.850608) -- (1.360604, 1.630041);
      \draw [blue] (1.360604, 1.630041) -- (4.039514, 5.000000);
      \draw [blue] (3.974692, 5.000000) -- (4.039514, 5.000000);
      \draw [blue] (0.388711, 0.128426) -- (0.336738, 0.115436);
      \draw [blue] (0.931185, 0.038748) -- (0.926800, 0.041266);
      \draw [blue] (0.961731, -0.050233) -- (0.974805, -0.051489);
      \draw [blue] (0.989674, -0.056551) -- (5.000000, -0.492527);
      \draw [blue] (5.000000, -0.543567) -- (5.000000, -0.492527);
      \draw [red] (1.757009, 5.000000) -- (2.934406, 5.000000);
      \draw [red] (2.934406, 5.000000) -- (4.416503, 5.000000);
      \draw [red] (4.416503, 5.000000) -- (4.658704, 5.000000);
      \draw [red] (4.658704, 5.000000) -- (4.913202, 5.000000);
      \draw [red] (4.913202, 5.000000) -- (5.000000, 5.000000);
      \draw [red] (5.000000, 3.598093) -- (5.000000, 5.000000);
      \draw [red] (5.000000, 3.598093) -- (5.000000, 2.436215);
      \draw [red] (2.835736, 3.005796) -- (4.795322, 5.000000);
      \draw [red] (4.913202, 5.000000) -- (4.795322, 5.000000);
      \draw [red] (2.835736, 3.005796) -- (0.008949, 0.049773);
      \draw [red] (2.835736, 3.005796) -- (4.597218, 5.000000);
      \draw [red] (4.416503, 5.000000) -- (4.597218, 5.000000);
      \draw [red] (5.000000, 2.436215) -- (5.000000, 2.094688);
      \draw [red] (5.000000, 2.094688) -- (5.000000, 1.769017);
      \draw [red] (5.000000, 1.769017) -- (5.000000, 1.701253);
      \draw [red] (5.000000, 1.701253) -- (5.000000, 1.634044);
      \draw [red] (4.916825, 1.724287) -- (5.000000, 1.753714);
      \draw [red] (5.000000, 1.769017) -- (5.000000, 1.753714);
      \draw [red] (5.000000, 1.634044) -- (5.000000, 0.829427);
      \draw [red] (5.000000, 0.829427) -- (5.000000, 0.064552);
      \draw [red] (5.000000, 0.064552) -- (5.000000, -0.503193);
      \draw [red] (5.000000, -0.503193) -- (5.000000, -1.084080);
      \draw [red] (0.344936, -0.050521) -- (5.000000, 0.009577);
      \draw [red] (5.000000, 0.064552) -- (5.000000, 0.009577);
      \draw [red] (-0.005662, 0.005193) -- (-0.005563, 0.003957);
      \draw [red] (4.916825, 1.724287) -- (0.372054, 0.141140);
      \draw [red] (4.916825, 1.724287) -- (5.000000, 1.751469);
      \draw [red] (5.000000, 1.634044) -- (5.000000, 1.751469);
      \draw [red] (0.003421, -0.004188) -- (-0.000686, -0.000686);
      \draw [red] (-0.002828, 0.001140) -- (-0.000686, -0.000686);
      \draw [red] (0.372054, 0.141140) -- (0.011297, 0.011297);
      \draw [red] (0.000000, 0.007231) -- (0.011297, 0.011297);
      \draw [red] (-0.005662, 0.005193) -- (-0.005317, 0.005317);
      \draw [red] (0.000000, 0.007231) -- (-0.005317, 0.005317);
      \draw [red] (0.372054, 0.141140) -- (5.000000, 2.396074);
      \draw [red] (5.000000, 2.436215) -- (5.000000, 2.396074);
      \draw [red] (5.000000, -1.084080) -- (5.000000, -2.251103);
      \draw [red] (5.000000, -2.251103) -- (5.000000, -3.665242);
      \draw [red] (5.000000, -3.665242) -- (5.000000, -4.735406);
      \draw [red] (5.000000, -4.735406) -- (5.000000, -5.000000);
      \draw [red] (4.100749, -5.000000) -- (5.000000, -5.000000);
      \draw [red] (0.071053, -0.071053) -- (5.000000, -3.684209);
      \draw [red] (5.000000, -3.665242) -- (5.000000, -3.684209);
      \draw [red] (0.023854, -0.036454) -- (0.071053, -0.071053);
      \draw [red] (4.100749, -5.000000) -- (3.236243, -5.000000);
      \draw [red] (3.236243, -5.000000) -- (2.489523, -5.000000);
      \draw [red] (2.489523, -5.000000) -- (1.901581, -5.000000);
      \draw [red] (1.901581, -5.000000) -- (1.356250, -5.000000);
      \draw [red] (0.136148, -0.342543) -- (2.455118, -5.000000);
      \draw [red] (2.489523, -5.000000) -- (2.455118, -5.000000);
      \draw [red] (0.024235, -0.037289) -- (0.023854, -0.036454);
      \draw [red] (0.023854, -0.036454) -- (0.022089, -0.033888);
      \draw [red] (0.024235, -0.037289) -- (4.094402, -5.000000);
      \draw [red] (4.100749, -5.000000) -- (4.094402, -5.000000);
      \draw [red] (1.356250, -5.000000) -- (0.793071, -5.000000);
      \draw [red] (0.793071, -5.000000) -- (0.248860, -5.000000);
      \draw [red] (0.248860, -5.000000) -- (-0.152522, -5.000000);
      \draw [red] (-0.152522, -5.000000) -- (-0.555878, -5.000000);
      \draw [red] (0.022089, -0.033888) -- (0.269263, -5.000000);
      \draw [red] (0.248860, -5.000000) -- (0.269263, -5.000000);
      \draw [red] (-0.555878, -5.000000) -- (-1.043646, -5.000000);
      \draw [red] (-1.043646, -5.000000) -- (-1.551222, -5.000000);
      \draw [red] (-1.551222, -5.000000) -- (-1.899479, -5.000000);
      \draw [red] (-1.899479, -5.000000) -- (-2.264623, -5.000000);
      \draw [red] (-0.025698, -0.025698) -- (-1.568948, -5.000000);
      \draw [red] (-1.551222, -5.000000) -- (-1.568948, -5.000000);
      \draw [red] (-0.020144, -0.007796) -- (-0.025698, -0.025698);
      \draw [red] (0.003421, -0.004188) -- (0.022089, -0.033888);
      \draw [red] (-0.002828, 0.001140) -- (-0.005563, 0.003957);
      \draw [red] (-0.003325, -0.003325) -- (-0.558833, -5.000000);
      \draw [red] (-0.555878, -5.000000) -- (-0.558833, -5.000000);
      \draw [red] (-0.002828, 0.001140) -- (-0.003325, -0.003325);
      \draw [red] (-0.007986, 0.014440) -- (-0.005820, 0.005820);
      \draw [red] (-0.005662, 0.005193) -- (-0.005820, 0.005820);
      \draw [red] (0.136148, -0.342543) -- (0.024235, -0.037289);
      \draw [red] (0.136148, -0.342543) -- (1.399483, -5.000000);
      \draw [red] (1.356250, -5.000000) -- (1.399483, -5.000000);
      \draw [red] (-0.018102, -0.006025) -- (-0.005563, 0.003957);
      \draw [red] (0.344936, -0.050521) -- (0.004308, -0.004308);
      \draw [red] (0.003421, -0.004188) -- (0.004308, -0.004308);
      \draw [red] (0.344936, -0.050521) -- (5.000000, -1.059814);
      \draw [red] (5.000000, -1.084080) -- (5.000000, -1.059814);
      \draw [red] (-2.264623, -5.000000) -- (-3.046845, -5.000000);
      \draw [red] (-3.046845, -5.000000) -- (-3.955388, -5.000000);
      \draw [red] (-3.955388, -5.000000) -- (-3.965031, -5.000000);
      \draw [red] (-3.965031, -5.000000) -- (-3.974692, -5.000000);
      \draw [red] (-3.974692, -5.000000) -- (-5.000000, -5.000000);
      \draw [red] (-5.000000, -4.289231) -- (-5.000000, -5.000000);
      \draw [red] (-5.000000, -4.289231) -- (-5.000000, -2.855603);
      \draw [red] (-3.974692, -5.000000) -- (-3.969978, -5.000000);
      \draw [red] (-0.127860, -0.127860) -- (-3.994447, -5.000000);
      \draw [red] (-3.969978, -5.000000) -- (-3.994447, -5.000000);
      \draw [red] (-0.048664, -0.028068) -- (-0.127860, -0.127860);
      \draw [red] (-3.955388, -5.000000) -- (-3.969978, -5.000000);
      \draw [red] (-5.000000, -2.855603) -- (-5.000000, -2.141128);
      \draw [red] (-5.000000, -2.141128) -- (-5.000000, -1.493618);
      \draw [red] (-5.000000, -1.493618) -- (-5.000000, -0.819248);
      \draw [red] (-5.000000, -0.819248) -- (-5.000000, -0.172708);
      \draw [red] (-0.351898, -0.154785) -- (-5.000000, -1.543283);
      \draw [red] (-5.000000, -1.493618) -- (-5.000000, -1.543283);
      \draw [red] (-5.000000, -0.172708) -- (-5.000000, 0.184485);
      \draw [red] (-5.000000, 0.184485) -- (-5.000000, 0.543567);
      \draw [red] (-5.000000, 0.543567) -- (-5.000000, 0.861869);
      \draw [red] (-5.000000, 0.861869) -- (-5.000000, 1.187103);
      \draw [red] (-0.134319, 0.065642) -- (-5.000000, 0.594607);
      \draw [red] (-5.000000, 0.543567) -- (-5.000000, 0.594607);
      \draw [red] (-0.020144, -0.007796) -- (-0.018102, -0.006025);
      \draw [red] (-0.035290, -0.018926) -- (-0.048664, -0.028068);
      \draw [red] (-0.018102, -0.006025) -- (-5.000000, -0.178107);
      \draw [red] (-5.000000, -0.172708) -- (-5.000000, -0.178107);
      \draw [red] (-0.134319, 0.065642) -- (-0.018838, 0.018838);
      \draw [red] (-0.007986, 0.014440) -- (-0.018838, 0.018838);
      \draw [red] (-0.351898, -0.154785) -- (-0.048664, -0.028068);
      \draw [red] (-0.351898, -0.154785) -- (-5.000000, -2.809412);
      \draw [red] (-5.000000, -2.855603) -- (-5.000000, -2.809412);
      \draw [red] (-5.000000, 1.187103) -- (-5.000000, 1.488092);
      \draw [red] (-5.000000, 1.488092) -- (-5.000000, 1.799326);
      \draw [red] (-5.000000, 1.799326) -- (-5.000000, 2.296825);
      \draw [red] (-5.000000, 2.296825) -- (-5.000000, 2.834943);
      \draw [red] (-0.554107, 0.211568) -- (-5.000000, 1.811490);
      \draw [red] (-5.000000, 1.799326) -- (-5.000000, 1.811490);
      \draw [red] (-5.000000, 2.834943) -- (-5.000000, 3.290101);
      \draw [red] (-5.000000, 3.290101) -- (-5.000000, 3.786783);
      \draw [red] (-5.000000, 3.786783) -- (-5.000000, 5.000000);
      \draw [red] (-4.814343, 5.000000) -- (-5.000000, 5.000000);
      \draw [red] (-4.814343, 5.000000) -- (-3.497300, 5.000000);
      \draw [red] (-0.153975, 0.074489) -- (-5.000000, 3.744658);
      \draw [red] (-5.000000, 3.786783) -- (-5.000000, 3.744658);
      \draw [red] (-0.134319, 0.065642) -- (-0.153975, 0.074489);
      \draw [red] (-0.265384, 0.118542) -- (-0.554107, 0.211568);
      \draw [red] (-0.265384, 0.118542) -- (-5.000000, 2.803016);
      \draw [red] (-5.000000, 2.834943) -- (-5.000000, 2.803016);
      \draw [red] (-3.497300, 5.000000) -- (-2.417186, 5.000000);
      \draw [red] (-2.417186, 5.000000) -- (-1.493459, 5.000000);
      \draw [red] (-1.493459, 5.000000) -- (-1.096172, 5.000000);
      \draw [red] (-1.096172, 5.000000) -- (-0.711668, 5.000000);
      \draw [red] (-0.135547, 0.562446) -- (-1.461007, 5.000000);
      \draw [red] (-1.493459, 5.000000) -- (-1.461007, 5.000000);
      \draw [red] (-0.711668, 5.000000) -- (-0.038572, 5.000000);
      \draw [red] (-0.038572, 5.000000) -- (0.633129, 5.000000);
      \draw [red] (0.633129, 5.000000) -- (1.180949, 5.000000);
      \draw [red] (1.180949, 5.000000) -- (1.757009, 5.000000);
      \draw [red] (0.025645, 0.109455) -- (0.644914, 5.000000);
      \draw [red] (0.633129, 5.000000) -- (0.644914, 5.000000);
      \draw [red] (-0.007032, 0.023019) -- (-0.007986, 0.014440);
      \draw [red] (-0.135547, 0.562446) -- (-0.034410, 0.088844);
      \draw [red] (-0.135547, 0.562446) -- (-0.767159, 5.000000);
      \draw [red] (-0.711668, 5.000000) -- (-0.767159, 5.000000);
      \draw [red] (-0.007032, 0.023019) -- (0.008949, 0.049773);
      \draw [red] (-0.034410, 0.088844) -- (-0.007032, 0.023019);
      \draw [red] (-0.034410, 0.088844) -- (-3.469568, 5.000000);
      \draw [red] (-3.497300, 5.000000) -- (-3.469568, 5.000000);
      \draw [red] (0.025645, 0.109455) -- (0.008949, 0.049773);
      \draw [red] (-0.265384, 0.118542) -- (-0.153975, 0.074489);
      \draw [red] (-0.554107, 0.211568) -- (-5.000000, 1.267115);
      \draw [red] (-5.000000, 1.187103) -- (-5.000000, 1.267115);
      \draw [red] (0.025645, 0.109455) -- (1.744191, 5.000000);
      \draw [red] (1.757009, 5.000000) -- (1.744191, 5.000000);
      \draw [red] (-0.035290, -0.018926) -- (-0.020144, -0.007796);
      \draw [red] (-0.048838, -0.048838) -- (-2.291340, -5.000000);
      \draw [red] (-2.264623, -5.000000) -- (-2.291340, -5.000000);
      \draw [red] (-0.035290, -0.018926) -- (-0.048838, -0.048838);

      \draw [fill = black] (-0.097570, -0.957952) circle [radius = 0.005];
      \draw [fill = black] (0.770609, -0.476450) circle [radius = 0.005];
      \draw [fill = black] (0.040970, -0.960957) circle [radius = 0.005];
      \draw [fill = black] (0.013376, 0.997841) circle [radius = 0.005];
      \draw [fill = black] (-0.971949, -0.100398) circle [radius = 0.005];
      \draw [fill = black] (-0.731669, -0.537918) circle [radius = 0.005];
      \draw [fill = black] (0.549356, 0.793807) circle [radius = 0.005];
      \draw [fill = black] (-0.718613, 0.546428) circle [radius = 0.005];
      \draw [fill = black] (-0.052334, 0.933458) circle [radius = 0.005];
      \draw [fill = black] (-0.797832, -0.535641) circle [radius = 0.005];
      \draw [fill = black] (-0.750289, 0.620360) circle [radius = 0.005];
      \draw [fill = black] (-0.904683, 0.026419) circle [radius = 0.005];
      \draw [fill = black] (-0.953114, -0.035441) circle [radius = 0.005];
      \draw [fill = black] (-0.903827, 0.421326) circle [radius = 0.005];
      \draw [fill = black] (0.228361, -0.881811) circle [radius = 0.005];
      \draw [fill = black] (-0.628953, 0.672674) circle [radius = 0.005];
      \draw [fill = black] (0.810743, -0.480038) circle [radius = 0.005];
      \draw [fill = black] (-0.594638, -0.718847) circle [radius = 0.005];
      \draw [fill = black] (0.979708, -0.098334) circle [radius = 0.005];
      \draw [fill = black] (0.955319, -0.201058) circle [radius = 0.005];
      \draw [fill = black] (0.748860, 0.638487) circle [radius = 0.005];
      \draw [fill = black] (0.782118, 0.566145) circle [radius = 0.005];
      \draw [fill = black] (-0.455680, -0.834109) circle [radius = 0.005];
      \draw [fill = black] (0.005741, -0.970491) circle [radius = 0.005];
      \draw [fill = black] (0.355582, 0.928879) circle [radius = 0.005];
      \draw [fill = black] (-0.367804, -0.824608) circle [radius = 0.005];
      \draw [fill = black] (0.228125, 0.919265) circle [radius = 0.005];
      \draw [fill = black] (-0.789962, 0.517379) circle [radius = 0.005];
      \draw [fill = black] (-0.956410, 0.136210) circle [radius = 0.005];
      \draw [fill = black] (0.916544, -0.204020) circle [radius = 0.005];
      \draw [fill = black] (0.072020, -0.929603) circle [radius = 0.005];
      \draw [fill = black] (-0.890157, -0.373949) circle [radius = 0.005];
      \draw [fill = black] (-0.966259, -0.021449) circle [radius = 0.005];
      \draw [fill = black] (0.025994, 0.930529) circle [radius = 0.005];
      \draw [fill = black] (0.890636, 0.390245) circle [radius = 0.005];
      \draw [fill = black] (-0.957284, 0.168913) circle [radius = 0.005];
      \draw [fill = black] (-0.568908, 0.811315) circle [radius = 0.005];
      \draw [fill = black] (0.322438, -0.908073) circle [radius = 0.005];
      \draw [fill = black] (0.930215, 0.204870) circle [radius = 0.005];
      \draw [fill = black] (0.354909, 0.891552) circle [radius = 0.005];
      \draw [fill = black] (-0.984404, -0.172595) circle [radius = 0.005];
      \draw [fill = black] (-0.216508, -0.891559) circle [radius = 0.005];
      \draw [fill = black] (-0.447476, -0.801233) circle [radius = 0.005];
      \draw [fill = black] (0.928514, -0.274879) circle [radius = 0.005];
      \draw [fill = black] (0.791318, -0.574760) circle [radius = 0.005];
      \draw [fill = black] (0.970447, -0.011294) circle [radius = 0.005];
      \draw [fill = black] (0.268248, 0.911400) circle [radius = 0.005];
      \draw [fill = black] (-0.170437, 0.930701) circle [radius = 0.005];
      \draw [fill = black] (-0.987113, 0.037193) circle [radius = 0.005];
      \draw [fill = black] (-0.984601, -0.082280) circle [radius = 0.005];
      \draw [fill = black] (-0.905033, -0.060704) circle [radius = 0.005];
      \draw [fill = black] (0.718902, 0.563448) circle [radius = 0.005];
      \draw [fill = black] (0.807363, 0.490243) circle [radius = 0.005];
      \draw [fill = black] (-0.002601, 0.982632) circle [radius = 0.005];
      \draw [fill = black] (0.962872, -0.090119) circle [radius = 0.005];
      \draw [fill = black] (-0.336185, 0.898247) circle [radius = 0.005];
      \draw [fill = black] (-0.758603, -0.557996) circle [radius = 0.005];
      \draw [fill = black] (-0.885755, -0.455790) circle [radius = 0.005];
      \draw [fill = black] (-0.554502, 0.754878) circle [radius = 0.005];
      \draw [fill = black] (-0.965689, -0.229863) circle [radius = 0.005];
      \draw [fill = black] (0.360399, 0.835738) circle [radius = 0.005];
      \draw [fill = black] (0.906648, -0.236301) circle [radius = 0.005];
      \draw [fill = black] (0.930937, -0.024858) circle [radius = 0.005];
      \draw [fill = black] (-0.613462, -0.706316) circle [radius = 0.005];
      \draw [fill = black] (-0.359685, 0.915488) circle [radius = 0.005];
      \draw [fill = black] (0.419329, 0.900006) circle [radius = 0.005];
      \draw [fill = black] (-0.710893, -0.554022) circle [radius = 0.005];
      \draw [fill = black] (0.658631, 0.658750) circle [radius = 0.005];
      \draw [fill = black] (0.727138, -0.659503) circle [radius = 0.005];
      \draw [fill = black] (-0.173249, 0.930610) circle [radius = 0.005];
      \draw [fill = black] (0.661218, 0.659689) circle [radius = 0.005];
      \draw [fill = black] (-0.676973, 0.629119) circle [radius = 0.005];
      \draw [fill = black] (0.988919, -0.013602) circle [radius = 0.005];
      \draw [fill = black] (-0.482419, -0.855241) circle [radius = 0.005];
      \draw [fill = black] (-0.071689, 0.993607) circle [radius = 0.005];
      \draw [fill = black] (-0.604652, -0.672503) circle [radius = 0.005];
      \draw [fill = black] (0.532575, 0.810420) circle [radius = 0.005];
      \draw [fill = black] (0.203333, -0.944114) circle [radius = 0.005];
      \draw [fill = black] (-0.040616, -0.978836) circle [radius = 0.005];
      \draw [fill = black] (-0.873211, -0.481535) circle [radius = 0.005];
      \draw [fill = black] (0.929992, 0.155452) circle [radius = 0.005];
      \draw [fill = black] (-0.913715, -0.117828) circle [radius = 0.005];
      \draw [fill = black] (0.776932, 0.506580) circle [radius = 0.005];
      \draw [fill = black] (-0.665012, -0.609110) circle [radius = 0.005];
      \draw [fill = black] (-0.691874, 0.710465) circle [radius = 0.005];
      \draw [fill = black] (0.407365, -0.883171) circle [radius = 0.005];
      \draw [fill = black] (-0.521920, 0.740489) circle [radius = 0.005];
      \draw [fill = black] (-0.861895, 0.409471) circle [radius = 0.005];
      \draw [fill = black] (-0.754758, 0.602736) circle [radius = 0.005];
      \draw [fill = black] (-0.202606, -0.929125) circle [radius = 0.005];
      \draw [fill = black] (-0.586115, -0.763646) circle [radius = 0.005];
      \draw [fill = black] (-0.918867, 0.197965) circle [radius = 0.005];
      \draw [fill = black] (-0.876981, 0.266458) circle [radius = 0.005];
      \draw [fill = black] (0.738586, -0.592758) circle [radius = 0.005];
      \draw [fill = black] (0.909196, -0.022605) circle [radius = 0.005];
      \draw [fill = black] (-0.723595, 0.643480) circle [radius = 0.005];
      \draw [fill = black] (-0.181448, -0.961003) circle [radius = 0.005];
      \draw [fill = black] (0.872122, -0.432246) circle [radius = 0.005];
      \draw [fill = black] (0.200275, 0.977062) circle [radius = 0.005];
      \draw [fill = black] (0.985996, 0.071019) circle [radius = 0.005];
    \end{scope}
  \end{tikzpicture}

%% file: OptimalCompressionOfAPolylineWithSegmentsAndArcsAppendices.tex

\clearpage

\section
{
  \label{appendix:FittingArcByTolerance}
  Fitting an Arc by Tolerance
}

\begin{figure} [ht]
  \input{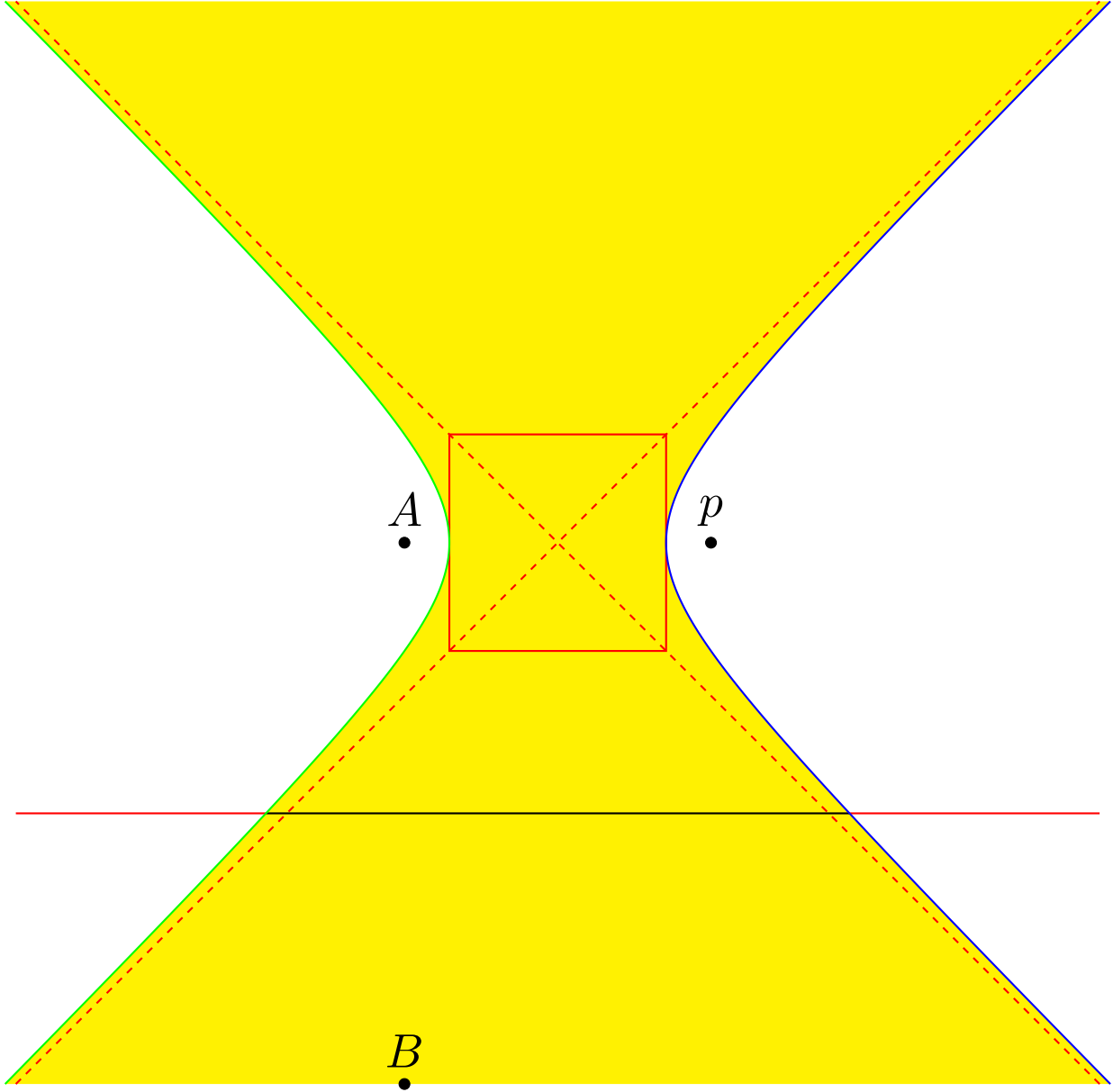}
  \caption
  {
    The hyperbola for focal points $ A $ and $ p $ is shown as green and blue curved lines. The area between hyperbola branches is shown in yellow. The line that is equidistant from $ A $ and $ B $ is shown as a red-black-red line.
  }
  \label{fig:Hyperbola}
\end{figure}

The center of an arc, passing through points $ A $ and $ B $, should lie on the line that is equidistant from these points, see Figure~\ref{fig:Hyperbola}. The radius of the arc is the distance between its center and point $ A $ or $ B $. Only if such an arc has its center lying on the red-black-red line between the hyperbola branches (black part in yellow area) will point $ p $ be within the specified tolerance from the arc. Points that are no farther from $ A $ or $ B $ than the specified tolerance always satisfy the tolerance requirement.

Note that there is a possibility that the equidistant line will intersect only one hyperbola branch. In the case where the equidistant line intersects both hyperbola branches, two open intervals will satisfy the tolerance requirement. The number of possible intersections between a line and a hyperbola is $ 0 $, $ 1 $, and $ 2 $.

The intersection of all intervals constructed for all points will produce a set of intervals that satisfy the specified tolerance. The complexity of the algorithm is $ O{\left( N \log{ \left( N \right) } \right)} $, where $ N $ is the number of points; however, if we take into account that the most likely intervals are the one shown in Figure~\ref{fig:Hyperbola}, then the complexity of the algorithm is $ O{\left( N \right)} $. This paper will only consider worst-case complexity. If the set of intervals is empty, then no arc satisfies the specified tolerance; otherwise, of all the possible solutions, the one that corresponds to an arc with the minimum approximate sum of squared deviations \cite{EfficientFittingOfCircularArcs} is taken. Following are the steps of the algorithm for finding an arc within the specified tolerance:
\begin{enumerate}
  \item Fit an arc when two points are known by the approximate solution of fitting an arc by the least squares approach \cite{EfficientFittingOfCircularArcs}.
  \item Check if this arc satisfies the tolerance requirement. If it does, then take it as the solution.
  \item \label{enum:ArcFittingByToleranceIntervals} Use the approach described in this appendix. If this algorithm does not find any solution (the set of intervals is empty), then no arc can satisfy the specified tolerance.
  \item Evaluate the approximate sum of squared deviations for each end of the set of intervals found in step~\ref{enum:ArcFittingByToleranceIntervals} for corresponding arcs.
  \item Among the evaluated solutions, find one with the minimum value.
\end{enumerate}

\begin{figure} [ht]
  \centering
  \includegraphics[width = 17 cm, keepaspectratio]{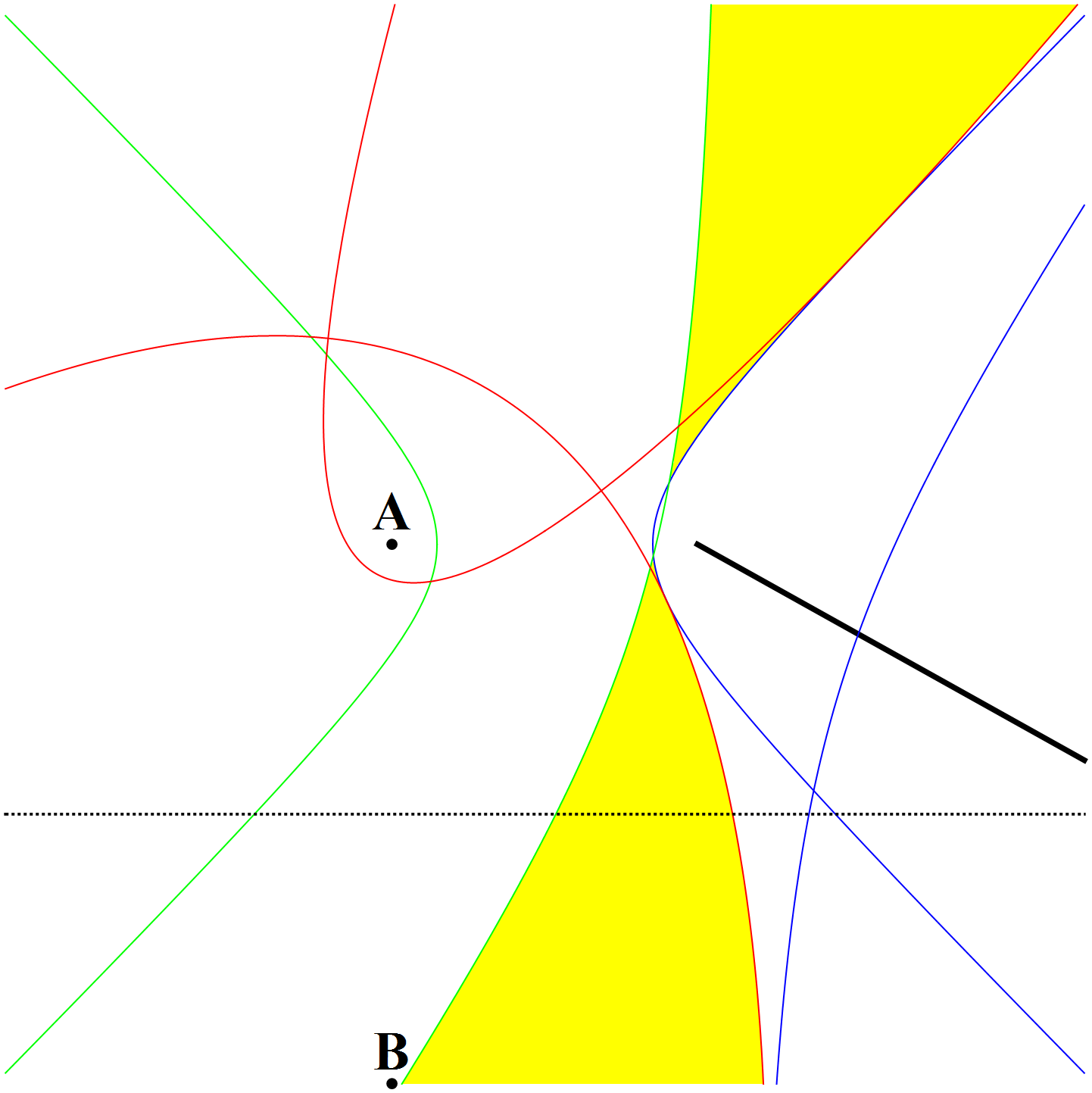}
  \caption
  {
    The hyperbolas for focal point $ A $ and end points of the black segment are shown as green and blue lines. Parabolas equidistant from focal point $ A $ and the black segment extended infinitely in both directions and shifted orthogonally in both directions by the specified tolerance are shown as red lines. The line that is equidistant from $ A $ and $ B $ is shown as a dotted black line. The solution to where the center of an arc can be located (the dotted line inside the yellow area) passes through points $ A $ and $ B $, while the black segment is within the specified tolerance from the arc.
  }
  \label{fig:SegmentArea}
\end{figure}

In the case where the tolerance requirement should be satisfied for the source polyline segments, the solution is more complicated; see the example in Figure~\ref{fig:SegmentArea}. The yellow area is limited by the combination of hyperbolas and parabolas.

\clearpage

\section
{
  \label{appendix:ConvexHullTree}
  Efficient Tolerance Checking of a Segment
}

A test to determine if a segment from a start point to an end point has all vertices of the source polyline from $ i $ to $ j $ within the specified tolerance is performed by using the convex hull. The same test is performed for the end points to be within the tolerance, see \cite[section II.C]{OptimalCompression}. The number of possible combinations of indices $ i $ and $ j $, $ i < j $, is $ O{ \left( N^2 \right) } $, where $ N $ is the number of vertices in the source polyline, and therefore the complexity to construct all convex hulls is $ O{ \left( N^3 \log{ \left( N \right) } \right) } $. However, not all of them are used. This paper will describe a different approach. Construct all convex hulls for subsets from vertex $ k \cdot 2^q $ to vertex $ \left( k + 1 \right) 2^q - 1 $, $ \forall k, q \in \mathbb{N}_0 $ and $ \left( k + 1 \right) 2^q \leq N $, see Figure~\ref{fig:ConvexHullTree}. Note that it is possible to delay construction of some of the convex hulls until they are needed. Complexity to construct all necessary convex hulls is $ O{ \left( N \log{ \left( N \right) } \right) } $. To check the segment for the tolerance requirement: First, convex hulls corresponding to the part of the source polyline from vertex $ i $ to vertex $ j $ need to be found with the preferences for the longest; see the green cells in Figure~\ref{fig:ConvexHullTree}.  There is no need to merge convex hulls to perform a tolerance check. Second, the segment is tested against each convex hull. The complexity of one test is $ O{ \left( \log^2{ \left( K \right) } \right) } $, where $ K $ is the number of vertices in the part of the source polyline.

\begin{figure} [ht]
  \centering
  \begin{tabular}{|c|c|c|c|c|c|c|c|c|c|c|c|c|c|c|c|ccccc}
  \hline
  0 & 1 & 2 & \cellcolor{green}3 & 4 & 5 & 6 & 7 & 8 & 9 & 10 & 11 & 12 & 13 & 14 & 15 & \multicolumn{1}{c|}{16} & \multicolumn{1}{c|}{17} & \multicolumn{1}{c|}{18} & \multicolumn{1}{c|}{19} & \multicolumn{1}{c|}{20} \\ \hline
  \multicolumn{2}{|c|}{0, 1} & \multicolumn{2}{c|}{2, 3} & \multicolumn{2}{c|}{4, 5} & \multicolumn{2}{c|}{6, 7} & \multicolumn{2}{c|}{8, 9} & \multicolumn{2}{c|}{10, 11} & \multicolumn{2}{c|}{12, 13} & \multicolumn{2}{c|}{14, 15} & \multicolumn{2}{c|}{\cellcolor{green}16, 17} & \multicolumn{2}{c|}{18, 19} &  \\ \cline{1-20}
  \multicolumn{4}{|c|}{0, 1, 2, 3} & \multicolumn{4}{c|}{\cellcolor{green}4, 5, 6, 7} & \multicolumn{4}{c|}{8, 9, 10, 11} & \multicolumn{4}{c|}{12, 13, 14, 15} & \multicolumn{4}{c|}{16, 17, 18, 19} &  \\ \cline{1-20}
  \multicolumn{8}{|c|}{0, 1, 2, 3, 4, 5, 6, 7} & \multicolumn{8}{c|}{\cellcolor{green}8, 9, 10, 11, 12, 13, 14, 15} & \multicolumn{5}{c}{} \\ \cline{1-16}
  \multicolumn{16}{|c|}{0, 1, 2, 3, 4, 5, 6, 7, 8, 9, 10, 11, 12, 13, 14, 15} & \multicolumn{5}{c}{} \\ \cline{1-16}
  \end{tabular}
  \caption
  {
    Each number represents an index of the vertex in the source polyline with $ N = 21 $ vertices. For each cell in the table, the convex hull is constructed for the list of indices. This is done iteratively, with the first iteration constructing a convex hull for each pair of points, then at each successive iteration, merging each pair of convex hulls found in the previous iteration. Four green cells represent four convex hulls for the part of the source polyline from vertex~$ 3 $ to vertex~$ 17 $.
  }
  \label{fig:ConvexHullTree}
\end{figure}

\clearpage

\section
{
  \label{appendix:AppendixTestingArcs}
  Testing Parts of The Source Polyline for Arc Fitting
}

Function {\bf{Test}}, see Figure~\ref{fig:FunctionTest}, will return a sorted list of parts of the source polyline that cannot be fitted with any arc within the specified tolerance. If the subset of points includes any of these parts, then it is not possible to fit an arc to this set of points within the specified tolerance. For example, see the source polyline with $ N = 65 $ vertices in Figure~\ref{fig:ExampleArcTest} with the list of pairs
$
  \left(
    \begin{matrix}
      10\\
      17
    \end{matrix}
  \right)
$,
$
  \left(
    \begin{matrix}
      12\\
      19
    \end{matrix}
  \right)
$,
$
  \left(
    \begin{matrix}
      14\\
      21
    \end{matrix}
  \right)
$,
$
  \left(
    \begin{matrix}
      26\\
      33
    \end{matrix}
  \right)
$,
$
  \left(
    \begin{matrix}
      28\\
      35
    \end{matrix}
  \right)
$,
$
  \left(
    \begin{matrix}
      30\\
      37
    \end{matrix}
  \right)
$,
$
  \left(
    \begin{matrix}
      42\\
      49
    \end{matrix}
  \right)
$,
$
  \left(
    \begin{matrix}
      44\\
      51
    \end{matrix}
  \right)
$,
and
$
  \left(
    \begin{matrix}
      46\\
      53
    \end{matrix}
  \right)
$.
Adding
$
  \left(
    \begin{matrix}
      -1\\
      0
    \end{matrix}
  \right)
$
and
$
  \left(
    \begin{matrix}
      N - 1\\
      N
    \end{matrix}
  \right)
$
at the front and at the end of this list correspondingly will make it simpler to use.
This list is in the sorted order. Note that there are no pairs completely inside any other pair. Therefore, in the list, both indices of the pairs are increasing.

\begin{figure} [ht]
  \centering
\begin{algorithmic}
  \Function{Test}{}
    \Function{TestPart}{$ q $, $ i $, $ j $}
      \If{$ j < i^\star + 4 \cdot 2^q $}
        \Return
      \EndIf
      \For{$ i^\star$ from $ i $ while $ i^\star + 4 \cdot 2^q \leq j $ step $ 2^q $}
        \State $ j^\star = i^\star + 4 \cdot 2^q - 1 $
        \If{no arc within the specified tolerance can be fitted to $ P_{ i^\star, j^\star } $}
          \State
          \Call{TestPart}{$ q + 1 $, $ i $, $ j^\star $}
          \State Report pair $ \left( i^\star, j^\star \right) $
          \State $ i = i^\star + 2^q $
        \EndIf
      \EndFor
      \State
      \Call{TestPart}{$ q + 1 $, $ i $, $ j $}
    \EndFunction
    \State
    \Call{TestPart}{$ 0 $, $ 0 $, $ N $}
    \Comment{$ N $ is the number of vertices in the source polyline}
  \EndFunction
\end{algorithmic}
  \caption
  {
    Function {\bf{Test}} finds a sorted list of parts of the source polyline (pair of start and end indices) that cannot be fitted with any arc within the specified tolerance.
  }
  \label{fig:FunctionTest}
\end{figure}

\begin{figure} [htb]
  \input{FigureArcFitting.tex}
  \caption
  {
    The red polyline with black vertices is the source polyline. The black vertices above are their indices. The green polyline is the optimal solution.
  }
  \label{fig:ExampleArcTest}
\end{figure}

Determining if part of the source polyline $ P_{i, j} $ (all vertices from $ i $ to $ j $) can be fitted with an arc within the specified tolerance is performed by checking if pair
$
  \left(
    \begin{matrix}
      i\\
      j
    \end{matrix}
  \right)
$
contains any pair from the sorted list of pairs. This is done by locating the position of index $ i $ in the upper array
$
  a
  =
  \left(
    -1,
    10,
    12,
    14,
    26,
    28,
    30,
    42,
    44,
    46,
    64
  \right)
$
(find the smallest $ q_a $ satisfying $ i \leq a_{q_a} $)
and the position of index $ j $ in the lower array
$
  b
  =
  \left(
    0,
    17,
    19,
    21,
    33,
    35,
    37,
    49,
    51,
    53,
    65
  \right)
$
(find the smallest $ q_b $ satisfying $ j < b_{q_b} $).
If $ q_a < q_b $ then pair
$
  \left(
    \begin{matrix}
      i\\
      j
    \end{matrix}
  \right)
$
contains a pair from the list, and therefore, part of the source polyline $ P_{i, j} $ cannot be fitted with any arc within the specified tolerance. This test is more than is needed in this paper.

In order to reduce the number of combination in the dynamic search, it is necessary to find for index $ j $ the first index where an arc can be fitted. This first index equals
$
  a_{b_j - 1} + 1
$.
If searching for the last index where an arc can be fitted for the index $ i $, the index is equal to 
$
  b_{a_i} - 1
$.

If, instead of the arrays $ a $ and $ b $, the array with the first indices where an arc can be fitted for each index is prepared
($ 0 $,
$ 0 $,
$ 0 $,
$ 0 $,
$ 0 $,
$ 0 $,
$ 0 $,
$ 0 $,
$ 0 $,
$ 0 $,
$ 0 $,
$ 0 $,
$ 0 $,
$ 0 $,
$ 0 $,
$ 0 $,
$ 0 $,
$ 11 $,
$ 11 $,
$ 13 $,
$ 13 $,
$ 15 $,
$ 15 $,
$ 15 $,
$ 15 $,
$ 15 $,
$ 15 $,
$ 15 $,
$ 15 $,
$ 15 $,
$ 15 $,
$ 15 $,
$ 15 $,
$ 27 $,
$ 27 $,
$ 29 $,
$ 29 $,
$ 31 $,
$ 31 $,
$ 31 $,
$ 31 $,
$ 31 $,
$ 31 $,
$ 31 $,
$ 31 $,
$ 31 $,
$ 31 $,
$ 31 $,
$ 31 $,
$ 43 $,
$ 43 $,
$ 45 $,
$ 45 $,
$ 47 $,
$ 47 $,
$ 47 $,
$ 47 $,
$ 47 $,
$ 47 $,
$ 47 $,
$ 47 $,
$ 47 $,
$ 47 $,
$ 47 $,
$ 47 $),
then the binary search can be avoided. The complexity to create this array is $ O{ \left( N \right) } $.
For the last index, the array can also be prepared
($ 16 $,
$ 16 $,
$ 16 $,
$ 16 $,
$ 16 $,
$ 16 $,
$ 16 $,
$ 16 $,
$ 16 $,
$ 16 $,
$ 16 $,
$ 18 $,
$ 18 $,
$ 20 $,
$ 20 $,
$ 32 $,
$ 32 $,
$ 32 $,
$ 32 $,
$ 32 $,
$ 32 $,
$ 32 $,
$ 32 $,
$ 32 $,
$ 32 $,
$ 32 $,
$ 32 $,
$ 34 $,
$ 34 $,
$ 36 $,
$ 36 $,
$ 48 $,
$ 48 $,
$ 48 $,
$ 48 $,
$ 48 $,
$ 48 $,
$ 48 $,
$ 48 $,
$ 48 $,
$ 48 $,
$ 48 $,
$ 48 $,
$ 50 $,
$ 50 $,
$ 52 $,
$ 52 $,
$ 64 $,
$ 64 $,
$ 64 $,
$ 64 $,
$ 64 $,
$ 64 $,
$ 64 $,
$ 64 $,
$ 64 $,
$ 64 $,
$ 64 $,
$ 64 $,
$ 64 $,
$ 64 $,
$ 64 $,
$ 64 $,
$ 64 $,
$ 64 $).

\clearpage

\section
{
  \label{appendix:DualityDelaunayDiagramsInversiveGeometry}
  Duality of the Farthest and the Closest Delaunay Diagrams by Inversive Geometry
}

This appendix shows duality of the farthest Delaunay diagram with the closest Delaunay diagram by using inversive geometry \cite{InversionTheoryAndConformalMapping}. The farthest Delaunay diagram only includes points on the convex hull. For some choice of the reference circle center, the closest Delaunay diagram constructed on an inverted set of points is dual to the farthest Delaunay diagram. This gives the possibility to construct the farthest Delaunay diagram using algorithms for the closest Delaunay diagram.

The closest and the farthest Voronoi diagrams have opposite definitions. The closest Voronoi cell is a set of points closest to the same point. The farthest Voronoi cell is a set of points farthest from the same point. They are both dual to the closest and the farthest Delaunay diagrams, respectively. When there are no coincident points, they are both uniquely defined. The closest Voronoi diagram for each point has a cell; however, that is not the case for the farthest Voronoi diagram. Only points of the convex hull have cells. The link between the closest and the farthest Delaunay diagrams with the convex hull can be established by adding an extra dimension and by mapping each vertex to the paraboloid $ \left( x, y \right) \rightarrow \left( x, y, x^2 + y^2 \right) $ (see \cite{FarthestDelaunayTriangulation} and \cite{ComputationalGeometry1}) or by inversion to sphere \cite[section 6.3.2]{ComputationalGeometry2}. Therefore, the closest and the farthest Delaunay diagrams can be calculated by constructing convex hull in higher dimensions.


This appendix establishes another link between the farthest and the closest Delaunay diagrams without adding an extra dimension~--- by using inversive geometry in the same space. The link is from the farthest Delaunay diagram to the closest Delaunay diagram and is not unique. The farthest Delaunay diagram is dual to the closest Delaunay diagram for an inverted set of points inside an \textit{inverted convex hull}. Therefore, all algorithms to construct the closest Delaunay diagram can be used to construct the farthest Delaunay diagram.

\subsection
{
  \label{sec:Duality}
  Duality of the Farthest and the Closest Delaunay Diagrams
}

To establish duality between the farthest Delaunay diagram and the closest Delaunay diagram, the following definitions will be used
\begin{itemize} [\textbullet]
  \item $ conv{ \left( S \right) } $ is a set of vertices of the convex hull $ S $.

  \item $ CH{ \left( S \right) } $ is a convex hull of $ S $.

  \item $ vol{ \left( P \right) } $ is an area of $ P $.

  \item Circles are represented as their borders.

  \item $ \Inside{ c } $ is the area inside the circle $ c $.

  \item $ \Outside{ c } $ is the area outside the circle $ c $.

  \item $ CDC{ \left( S \right) } $ is the set of closest Delaunay circles for the set of points $ S $ \cite{ComputationalGeometry1}, i.e., satisfying that all points are outside or on the border of each circle and each circle has points on the border forming nonempty geometry, $ \left\{ c | S \subset \OutsideAndBorder{ c } \right. $ $ \wedge $ $ \left. vol{ \left( CH{ \left( S \cap c \right) } \right) } \neq 0 \right\} $.

  \item $ FDC{ \left( S \right) } $ is the set of farthest Delaunay circles for the set of points $ S $ \cite{ComputationalGeometry1}, i.e., satisfying that all points are inside or on the border of each circle and each circle has points on the border forming nonempty geometry, $ \left\{ c | S \subset \InsideAndBorder{ c } \right. $ $ \wedge $ $ \left. vol{ \left( CH{ \left( S \cap c \right) } \right) } \neq 0 \right\} $. This is the opposite of the definition of $ CDC{ \left( S \right) } $.
\end{itemize}

Define circle inversion for a reference unit circle\footnote{Reference circles with the same center but having different radiuses are equivalent up to the scale.} with center $ O $
\begin{equation*}
  \InversiveGeometryTransformation{ O }{ x },
\end{equation*}
where $ x $ is an object or a set of objects that is inverted (point, circle, set of points, set of circles, etc.).
The inverse of a point $ p $ is equal to
$
  p'
  =
  \dfrac{p}{\norm{p}^2}
$,
where
$
  \norm{p}
$ is Euclidean norm \cite{InversionTheoryAndConformalMapping}.

\begin{figure} [p]
  \centering
  \begin{tikzpicture} [scale = 2.75] 
    \draw [thick,->, >=stealth'] (-2.5, 0) -- (2.5, 0) node [anchor = south] {\scalebox{2}{$X$}};
    \draw [thick,->, >=stealth'] (0, -1.8) -- (0, 1.8) node [anchor = west ] {\scalebox{2}{$Y$}};

    \draw (0, 0) circle (1);

    \draw [red] (-1/2, 2/2) -- (1/2, -1/2);
    \draw [red] (-1/2, 2/2) -- (0.5/2, 1/2);
    \draw [red] (0.5/2, 1/2) -- (1/2, -1/2);
    \draw [red] (-1/2, 2/2) -- (-2/2, 1/2);
    \draw [red] (-2/2, 1/2) -- (1/2, -1/2);

    \node at (0, 0) [ label = south west : {\scalebox{2}{$O$}} ] {};

    \draw [thick, green] (-1.65/2, -0.6/2) circle (2.6800186566514793573839164063832/2);
    \draw [thick, blue] (-0.3/2, 0.3/2) circle (1.8384776310850235634421953414726/2);

    \draw [fill = black] (-1/2, 2/2) circle [radius = 0.03];
    \node at (-1/2, 2/2) [ label = north : {\scalebox{2}{$A$}} ] {};
    \draw [fill = black] (1/2, -1/2) circle [radius = 0.03];
    \node at (1/2, -1/2) [ label = east : {\scalebox{2}{$B$}} ] {};
    \draw [fill = black] (0.5/2, 1/2) circle [radius = 0.03];
    \node at (0.5/2, 1/2) [ label = north east : {\scalebox{2}{$C$}} ] {};
    \draw [fill = black] (-2/2, 1/2) circle [radius = 0.03];
    \node at (-2/2, 1/2) [ label = west : {\scalebox{2}{$D$}} ] {};
  \end{tikzpicture}
  \caption
  {
    The reference circle with center $ O $ is shown as a black circle. Triangles $ ABC $ and $ ABD $ are shown with their circumscribed green and blue circles, respectively. $ C $ is inside the blue circle, and $ D $ is inside the green circle.
  }
  \label{fig:InversiveGeometryTwoCircles}

  \begin{tikzpicture} [scale = 2.75] 
    \draw [thick,->, >=stealth'] (-2.5, 0) -- (2.5, 0) node [anchor = south] {\scalebox{2}{$X$}};
    \draw [thick,->, >=stealth'] (0, -1.8) -- (0, 1.8) node [anchor = west ] {\scalebox{2}{$Y$}};

    \draw (0, 0) circle (1);

    \draw [red] (-0.2*2, 0.4*2) -- (0.5*2, -0.5*2);
    \draw [red] (-0.2*2, 0.4*2) -- (0.4*2, 0.8*2);
    \draw [red] (0.4*2, 0.8*2) -- (0.5*2, -0.5*2);
    \draw [red] (-0.2*2, 0.4*2) -- (-0.4*2, 0.2*2);
    \draw [red] (-0.4*2, 0.2*2) -- (0.5*2, -0.5*2);

    \node at (0, 0) [ label = south west : {\scalebox{2}{$O$}} ] {};

    \draw [thick, green] (0.40243902439024390243902439024390*2, 0.14634146341463414634146341463415*2) circle (0.65366308698816569692290644058126*2);
    \draw [thick, blue] (0.09375*2, -0.09375*2) circle (0.57452425971406986357568604421019*2);

    \draw [fill = black] (-0.2*2, 0.4*2) circle [radius = 0.03];
    \node at (-0.2*2, 0.4*2) [ label = north west : {\scalebox{2}{$A'$}} ] {};
    \draw [fill = black] (0.5*2, -0.5*2) circle [radius = 0.03];
    \node at (0.5*2, -0.5*2) [ label = south : {\scalebox{2}{$B'$}} ] {};
    \draw [fill = black] (0.4*2, 0.8*2) circle [radius = 0.03];
    \node at (0.4*2, 0.8*2) [ label = north : {\scalebox{2}{$C'$}} ] {};
    \draw [fill = black] (-0.4*2, 0.2*2) circle [radius = 0.03];
    \node at (-0.4*2, 0.2*2) [ label = north west : {\scalebox{2}{$D'$}} ] {};
  \end{tikzpicture}
  \caption
  {
    The reference circle with center $ O $ is shown as a black circle. Points $ A $, $ B $, $ C $, and $ D $, shown in Figure~\ref{fig:InversiveGeometryTwoCircles}, are inverted ($ A' = \InversiveGeometryTransformation{ O }{ A } $, $ B' = \InversiveGeometryTransformation{ O }{ B } $, $ C' = \InversiveGeometryTransformation{ O }{ C } $, and $ D' = \InversiveGeometryTransformation{ O }{ D } $). Triangles $ A'B'C' $ and $ A'B'D' $ are shown with their circumscribed green and blue circles, respectively. $ C' $ is outside the blue circle, and $ D' $ is outside the green circle.
  }
  \label{fig:InversiveGeometryInvertedTwoCircles}
\end{figure}

From inversive geometry theory \cite{InversionTheoryAndConformalMapping}, the circle inversion has the following properties:
\begin{enumerate} [label = (\emph{\alph*})]
  \item
    $ \forall x, \forall O \Rightarrow \InversiveGeometryTransformation{ O }{ \InversiveGeometryTransformation{ O }{ x } } = x $.

  \item
    $ \forall \text{ circle } c, \forall O \notin c \Rightarrow \InversiveGeometryTransformation{ O }{ c } $ is a circle.

  \item
    \label{ig:OInsideCircle}
    $ \forall \text{ circle } c, \forall O \in \Inside{ c } \Rightarrow O \in \Inside{ \InversiveGeometryTransformation{ O }{ c } } $.

  \item
    $ \forall \text{ circle } c, \forall O \in \Outside{ c } \Rightarrow O \in \Outside{ \InversiveGeometryTransformation{ O }{ c } } $.

  \item
    \label{ig:PointsOnCircle}
    $ \forall \text{ circle } c, \forall O \notin c \wedge \forall p \in c \Rightarrow p' \in \InversiveGeometryTransformation{ O }{ c } $.
    For the circle containing $ O $, the inverted points on the circle will maintain their relative order and orientation. If $ O $ is outside of the circle, their relative order will be maintained but orientation will be reversed.

  \item
    \label{ig:OInsideCirclePointsInsideCircle}
    $ \forall \text{ circle } c, \forall O \in \Inside{ c } \wedge \forall p \in \Inside{ c } \Rightarrow \InversiveGeometryTransformation{ O }{ p } \in \Outside{ \InversiveGeometryTransformation{ O }{ c } } $.

  \item
    \label{ig:OInsideCirclePointsOutsideCircle}
    $ \forall \text{ circle } c, \forall O \in \Inside{ c } \wedge \forall p \in \Outside{ c } \Rightarrow \InversiveGeometryTransformation{ O }{ p } \in \Inside{ \InversiveGeometryTransformation{ O }{ c } } $.

  \item
    $ \forall \text{ circle } c, \forall O \in \Outside{ c } \wedge \forall p \in \Inside{ c } \Rightarrow \InversiveGeometryTransformation{ O }{ p } \in \Inside{ \InversiveGeometryTransformation{ O }{ c } } $.

  \item
    $ \forall \text{ circle } c, \forall O \in \Outside{ c } \wedge \forall p \in \Outside{ c } \Rightarrow \InversiveGeometryTransformation{ O }{ p } \in \Outside{ \InversiveGeometryTransformation{ O }{ c } } $.
\end{enumerate}

From the properties of inversive geometry and definitions of $ FDC{ \left( S \right) } $ and $ CDC{ \left( S \right) } $, it follows that
\begin{enumerate} [label = (\emph{\alph*}), resume]
  \item
    \label{ig:InverseTwoTriangles}
    Any two triangles sharing the edge and satisfying the farthest or the closest Delaunay triangulation property (no points outside or no points inside, respectively) and having circumscribed circles containing $ O $ will be inverted to nonoverlapping triangles with opposite properties, see example in Figures~\ref{fig:InversiveGeometryTwoCircles} and \ref{fig:InversiveGeometryInvertedTwoCircles}. The proof follows from \ref{ig:PointsOnCircle}, \ref{ig:OInsideCirclePointsInsideCircle}, and \ref{ig:OInsideCirclePointsOutsideCircle} that the part of the circle on one side of $ AB $ containing $ C $ is inside the blue circle (see Figure~\ref{fig:InversiveGeometryTwoCircles}) and will be inverted to the part of the circle that is outside of the blue circle (see Figure~\ref{fig:InversiveGeometryInvertedTwoCircles}). Therefore, the test to decide if point $ D $ lies inside the circumscribed circle of $ A $, $ B $, and $ C $ is equivalent if point $ D' $ lies outside of the circumscribed circle of $ A' $, $ B' $, and $ C' $.
\end{enumerate}

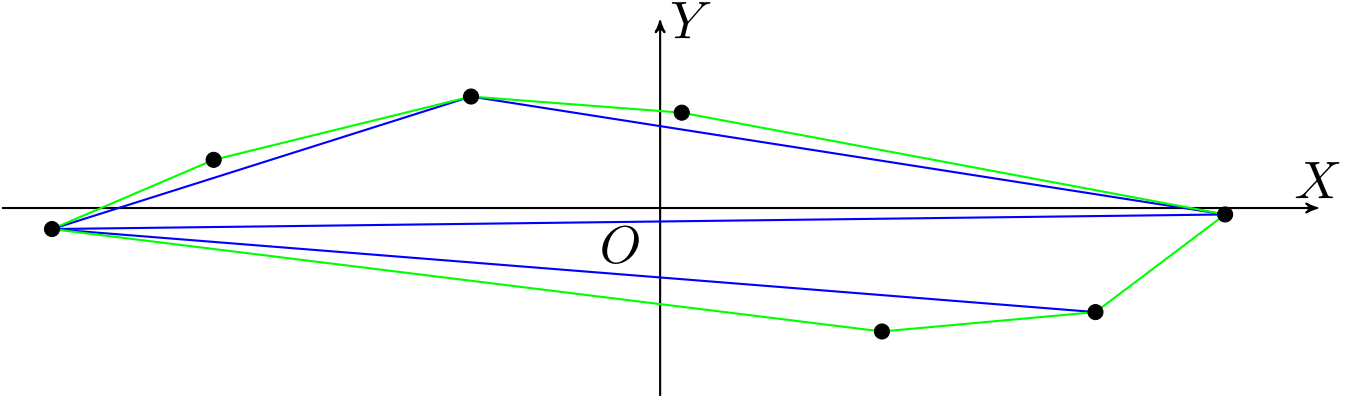
\begin{figure} [htb]
  \centering
  \begin{tikzpicture} [scale = 5]
    \draw [thick,->, >=stealth'] (-1.75, 0) -- (1.75, 0) node [anchor = south] {\scalebox{2}{$X$}};
    \draw [thick,->, >=stealth'] (0, -0.5) -- (0, 0.5) node [anchor = west ] {\scalebox{2}{$Y$}};

    \draw [thick, blue] (-1.616946, -0.055959) -- (-0.502804, 0.296642);
    \draw [thick, blue] (-1.616946, -0.055959) -- (1.502235, -0.017142);
    \draw [thick, blue] (-1.616946, -0.055959) -- (1.157448, -0.276951);
    \draw [thick, blue] (-0.502804, 0.296642) -- (1.502235, -0.017142);

    \draw [thick, green] (0.589759, -0.328516) -- (-1.616946, -0.055959);
    \draw [thick, green] (-1.616946, -0.055959) -- (-1.187119, 0.128211);
    \draw [thick, green] (-1.187119, 0.128211) -- (-0.502804, 0.296642);
    \draw [thick, green] (-0.502804, 0.296642) -- (0.057428, 0.253716);
    \draw [thick, green] (0.057428, 0.253716) -- (1.502235, -0.017142);
    \draw [thick, green] (1.502235, -0.017142) -- (1.157448, -0.276951);
    \draw [thick, green] (1.157448, -0.276951) -- (0.589759, -0.328516);

    \draw [fill = black] (0.589759, -0.328516) circle [radius = 0.02];
    \draw [fill = black] (-1.616946, -0.055959) circle [radius = 0.02];
    \draw [fill = black] (-1.187119, 0.128211) circle [radius = 0.02];
    \draw [fill = black] (-0.502804, 0.296642) circle [radius = 0.02];
    \draw [fill = black] (0.057428, 0.253716) circle [radius = 0.02];
    \draw [fill = black] (1.502235, -0.017142) circle [radius = 0.02];
    \draw [fill = black] (1.157448, -0.276951) circle [radius = 0.02];

    \node at (0, 0) [ label = south west : {\scalebox{2}{$O$}} ] {};
  \end{tikzpicture}
  \caption
  {
    The farthest Delaunay triangulation (green and blue lines) for the points (black points) of the convex hull. The convex hull border is shown as a green polyline.
  }
  \label{fig:ConvexHull}
\end{figure}

\begin{figure} [p]
  \centering
  \begin{tikzpicture} [scale = 5]
    \draw [thick,->, >=stealth'] (-1.75, 0) -- (1.75, 0) node [anchor = south] {\scalebox{2}{$X$}};
    \draw [thick,->, >=stealth'] (0, -0.75) -- (0, 3.75) node [anchor = west ] {\scalebox{2}{$Y$}};

    \draw [thick, blue] (-0.617710, -0.021378) -- (-1.475328, 0.870406);
    \draw [thick, blue] (-0.617710, -0.021378) -- (0.665588, -0.007595);
    \draw [thick, blue] (-0.617710, -0.021378) -- (0.817183, -0.195534);
    \draw [thick, blue] (-1.475328, 0.870406) -- (0.665588, -0.007595);

    \draw [thick, red] (1.294074, -0.720843) -- (0.848648, 3.749331);
    \draw [thick, red] (0.848648, 3.749331) -- (0.817183, -0.195534);

    \draw [thick, green] (1.294074, -0.720843) -- (-0.617710, -0.021378);
    \draw [thick, green] (-0.617710, -0.021378) -- (-0.832663, 0.089929);
    \draw [thick, green] (-0.832663, 0.089929) -- (-1.475328, 0.870406);
    \draw [thick, green] (-1.475328, 0.870406) -- (0.848648, 3.749331);
    \draw [thick, green] (0.848648, 3.749331) -- (0.665588, -0.007595);
    \draw [thick, green] (0.665588, -0.007595) -- (0.817183, -0.195534);
    \draw [thick, green] (0.817183, -0.195534) -- (1.294074, -0.720843);

    \draw [fill = black] (1.294074, -0.720843) circle [radius = 0.02];
    \draw [fill = black] (-0.617710, -0.021378) circle [radius = 0.02];
    \draw [fill = black] (-0.832663, 0.089929) circle [radius = 0.02];
    \draw [fill = black] (-1.475328, 0.870406) circle [radius = 0.02];
    \draw [fill = black] (0.848648, 3.749331) circle [radius = 0.02];
    \draw [fill = black] (0.665588, -0.007595) circle [radius = 0.02];
    \draw [fill = black] (0.817183, -0.195534) circle [radius = 0.02];

    \node at (0, 0) [ label = south west : {\scalebox{2}{$O$}} ] {};
  \end{tikzpicture}
  \caption
  {
    The closest Delaunay triangulation (green, blue, and red lines) for the inverted set of points (black points), see Figure~\ref{fig:ConvexHull}. The \textit{inverted convex hull} border is shown as a green polyline. Red lines are outside of the \textit{inverted convex hull}.
  }
  \label{fig:InvertedConvexHull}
\end{figure}
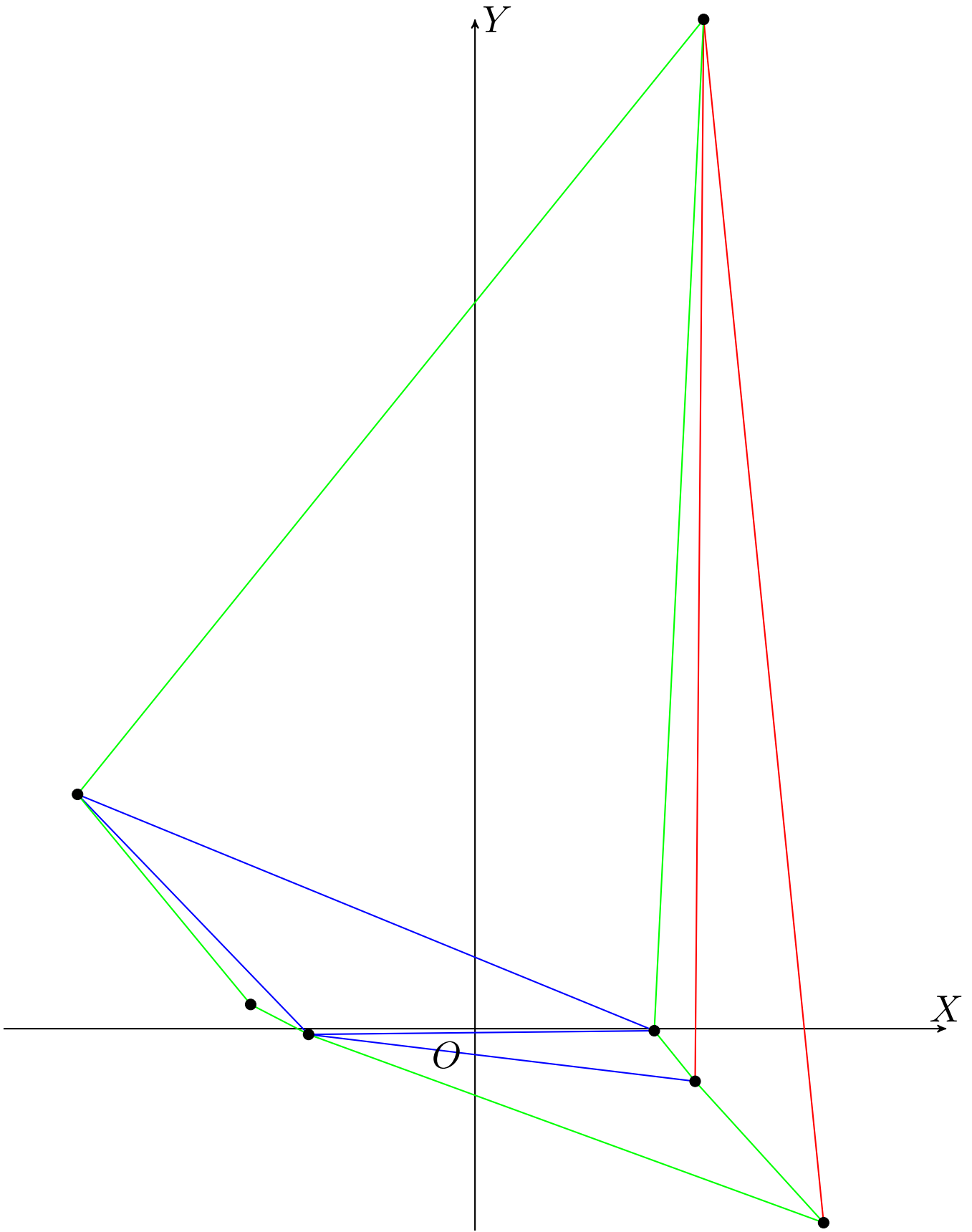

\begin{figure} [htb]
  \centering
  \begin{tikzpicture} [scale = 1.2]
    \draw [thick,->, >=stealth'] (-7.25, 0) -- (7.25, 0) node [anchor = south] {\scalebox{2}{$X$}};
    \draw [thick,->, >=stealth'] (0, -1.5) -- (0, 0.75) node [anchor = west ] {\scalebox{2}{$Y$}};

    \draw [thick, blue] (-7.098226, 0.543878) -- (1.127102, -0.191009);
    \draw [thick, blue] (-1.357863, -0.322494) -- (-7.098226, 0.543878);
    \draw [thick, blue] (-1.357863, -0.322494) -- (1.127102, -0.191009);
    \draw [thick, blue] (-1.357863, -0.322494) -- (1.249205, -0.499751);

    \draw [thick, red] (1.586010, -1.178495) -- (7.098226, -0.543878);
    \draw [thick, red] (1.586010, -1.178495) -- (-7.098226, 0.543878);
    \draw [thick, red] (1.586010, -1.178495) -- (-2.026698, -0.308845);
    \draw [thick, red] (7.098226, -0.543878) -- (1.249205, -0.499751);

    \draw [thick, green] (1.586010, -1.178495) -- (-1.357863, -0.322494);
    \draw [thick, green] (-1.357863, -0.322494) -- (-2.026698, -0.308845);
    \draw [thick, green] (-2.026698, -0.308845) -- (-7.098226, 0.543878);
    \draw [thick, green] (-7.098226, 0.543878) -- (7.098226, -0.543878);
    \draw [thick, green] (7.098226, -0.543878) -- (1.127102, -0.191009);
    \draw [thick, green] (1.127102, -0.191009) -- (1.249205, -0.499751);
    \draw [thick, green] (1.249205, -0.499751) -- (1.586010, -1.178495);

    \draw [fill = black] (1.586010, -1.178495) circle [radius = 0.08];
    \draw [fill = black] (-1.357863, -0.322494) circle [radius = 0.08];
    \draw [fill = black] (-2.026698, -0.308845) circle [radius = 0.08];
    \draw [fill = black] (-7.098226, 0.543878) circle [radius = 0.08];
    \draw [fill = black] (7.098226, -0.543878) circle [radius = 0.08];
    \draw [fill = black] (1.127102, -0.191009) circle [radius = 0.08];
    \draw [fill = black] (1.249205, -0.499751) circle [radius = 0.08];

    \node at (0, 0) [ label = south west : {\scalebox{2}{$O$}} ] {};
  \end{tikzpicture}
  \caption
  {
    Same as Figure~\ref{fig:InvertedConvexHull}, with the difference that $ O $ is shifted to the border of the convex hull, see point $ O $ in the middle of the green segment. The image was resized to fit the page.
  }
  \label{fig:InvertedConvexHullOnBorderOfTheConvexHull}
\end{figure}
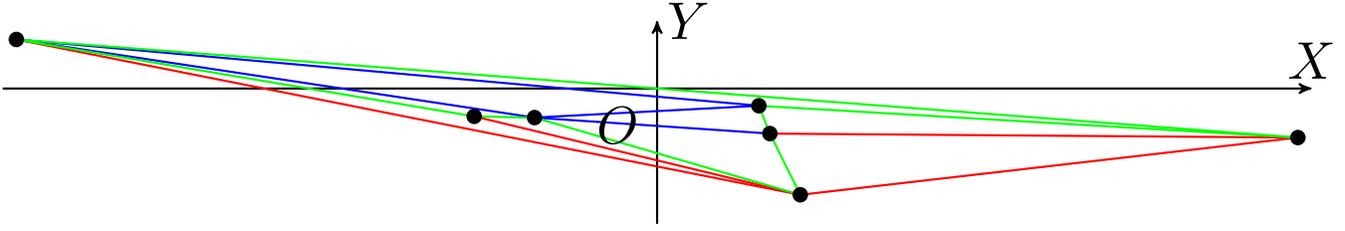

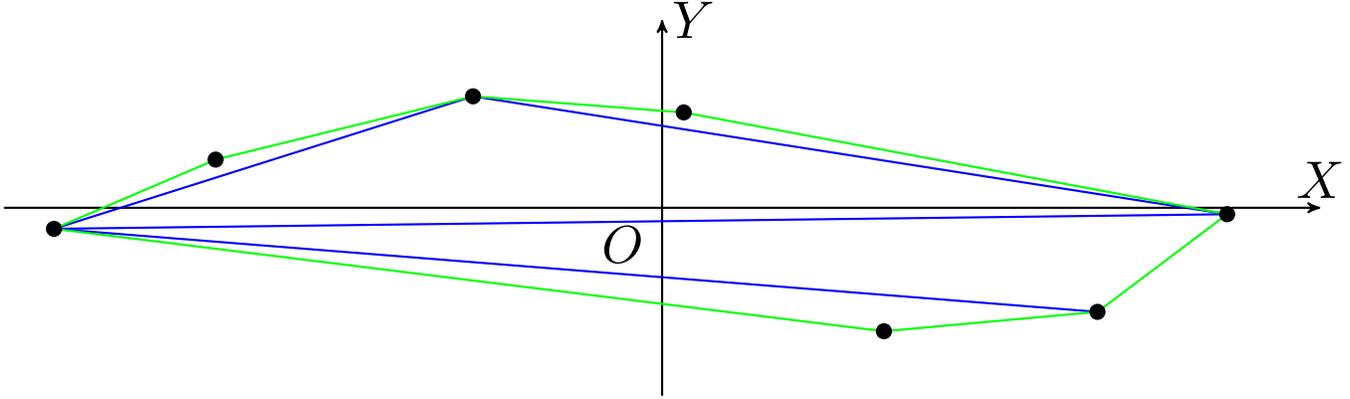
\begin{figure} [htb]
  \centering
  \begin{tikzpicture} [scale = 5]
    \draw [thick,->, >=stealth'] (-1.75, 0) -- (1.75, 0) node [anchor = south] {\scalebox{2}{$X$}};
    \draw [thick,->, >=stealth'] (0, -0.5) -- (0, 0.5) node [anchor = west ] {\scalebox{2}{$Y$}};

    \draw [thick, blue] (-1.616946, -0.055959) -- (-0.502804, 0.296642);
    \draw [thick, blue] (-1.616946, -0.055959) -- (1.502235, -0.017142);
    \draw [thick, blue] (-1.616946, -0.055959) -- (1.157448, -0.276951);
    \draw [thick, blue] (-0.502804, 0.296642) -- (1.502235, -0.017142);

    \draw [thick, green] (0.589759, -0.328516) -- (-1.616946, -0.055959);
    \draw [thick, green] (-1.616946, -0.055959) -- (-1.187119, 0.128211);
    \draw [thick, green] (-1.187119, 0.128211) -- (-0.502804, 0.296642);
    \draw [thick, green] (-0.502804, 0.296642) -- (0.057428, 0.253716);
    \draw [thick, green] (0.057428, 0.253716) -- (1.502235, -0.017142);
    \draw [thick, green] (1.502235, -0.017142) -- (1.157448, -0.276951);
    \draw [thick, green] (1.157448, -0.276951) -- (0.589759, -0.328516);

    \draw [fill = black] (0.589759, -0.328516) circle [radius = 0.02];
    \draw [fill = black] (-1.616946, -0.055959) circle [radius = 0.02];
    \draw [fill = black] (-1.187119, 0.128211) circle [radius = 0.02];
    \draw [fill = black] (-0.502804, 0.296642) circle [radius = 0.02];
    \draw [fill = black] (0.057428, 0.253716) circle [radius = 0.02];
    \draw [fill = black] (1.502235, -0.017142) circle [radius = 0.02];
    \draw [fill = black] (1.157448, -0.276951) circle [radius = 0.02];

    \node at (0, 0) [ label = south west : {\scalebox{2}{$O$}} ] {};
  \end{tikzpicture}
  \caption
  {
    Inverted closest Delaunay triangulation (green and blue lines) for Delaunay circles containing $ O $ or inside of the \textit{inverted convex hull}, see Figures~\ref{fig:InvertedConvexHull} or \ref{fig:InvertedConvexHullOnBorderOfTheConvexHull}. This figure is exactly the same as Figure~\ref{fig:ConvexHull}.
  }
  \label{fig:FinalConvexHull}
\end{figure}

For the set of points $ S $ that are not on the line (otherwise, the Delaunay diagrams are trivial) this appendix will show the duality of the farthest Delaunay diagram with the closest Delaunay diagram by circle inversion. For any number of dimensions, the requirement is that the set of points $ S $ is not in the hyperplane; otherwise, the Delaunay diagram can be constructed in the hyperplane.
\begin{enumerate} [label = (\emph{\Roman*})]
  \item
    \label{item:S}
      The farthest Voronoi diagram only includes points of the convex hull \cite{ComputationalGeometry1}.
      Therefore, $ S $ is a set of points that satisfy $ S = conv{ \left( S \right) } \wedge vol{ \left( CH{ \left( S \right) } \right) } \neq 0 $.
      From $ S = conv{ \left( S \right) } $, it follows that $ S $ does not have any coincident and does not have more than two collinear points.

  \item
    \label{item:Inside}
    The convex hull of $ S $ is inside of all circles of the farthest Delaunay diagram with the exception that points of $ S $ are on the border of some circles. From the definition for the farthest Delaunay circles, stating that any circle of the farthest Delaunay diagram contains all points inside or on the border, it follows that any circle contains a convex hull of these points,
    $ \forall c \in FDC{ \left( S \right) } \Rightarrow S \subset \InsideAndBorder{ c } \Rightarrow CH{ \left( S \right) } \subset \InsideAndBorder{ c } \Rightarrow CH{ \left( S \right) } \setminus S \subset \Inside{ c } $.

  \item
    \label{item:O}
    Choose any point $ O $ inside or on the border of $ CH{ \left( S \right) } $ but not a point from $ S $, $ O \in CH{ \left( S \right) } \setminus S $.

  \item
    \label{item:OInsideCircle}
    From \ref{item:Inside} and \ref{item:O}, it follows that $ O $ is inside of all the circles of the farthest Delaunay diagram,
    $ \forall c \in FDC{ \left( S \right) } \Rightarrow O \in \Inside{ c } $. Because $ O $ is inside of all the circles of the farthest Delaunay diagram, from~\ref{ig:OInsideCircle}, it follows that $ O $ is inside of all inverted circles of the farthest Delaunay diagram, $ \forall c' \in \InversiveGeometryTransformation{ O }{ FDC{ \left( S \right) } } \Rightarrow O \in \Inside{ c' } $.

  \item
    Invert set of points $ S $ by unit circle with center $ O $,
    $ S' = \InversiveGeometryTransformation{ O }{ S } $.

  \item
    \label{item:FDCpartCDC}
    From \ref{item:OInsideCircle}, \ref{ig:PointsOnCircle}, \ref{ig:OInsideCirclePointsInsideCircle}, and definitions of $ FDC{ \left( S \right) } $ and $ CDC{ \left( S \right) } $, it follows that any inverted circle of the farthest Delaunay diagram is also present in the closest Delaunay diagram,
    $ \InversiveGeometryTransformation{ O }{ FDC{ \left( S \right) } } \subset CDC{ \left( S' \right) } $.

  \item
    \label{item:CDCMinusFCDOutside}
    From \ref{item:FDCpartCDC}, and definitions of $ FDC{ \left( S \right) } $ and $ CDC{ \left( S \right) } $, it follows that
    $
      \forall c' \in CDC{ \left( S' \right) } \setminus \InversiveGeometryTransformation{ O }{ FDC{ \left( S \right) } }
      \Rightarrow
      O \in \OutsideAndBorder{ c' }
    $.
    From the inversion of a circle that does not contain any point of $ S' $ inside but has inversion center $ O $ inside, it follows that the inverted circle contains all the points of $ S $ inside or on the border. However, that circle is part of the farthest Delaunay circles, which is a contradiction. Suppose $ O \in \Inside{ c' } $, then $ \InversiveGeometryTransformation{ O }{ c' } \in FDC{ \left( S \right) } \Rightarrow c' \in \InversiveGeometryTransformation{ O }{ FDC{ \left( S \right) } } $.
    Note that in two dimensions, $ O \in \Outside{ c' } $.

  \item
    \label{item:CDCEqualFDC}
    From \ref{item:OInsideCircle}, \ref{item:FDCpartCDC}, and \ref{item:CDCMinusFCDOutside}, it follows that
    $ \InversiveGeometryTransformation{ O }{ \left\{ c' | CDC{ \left( S' \right) } \wedge O \in \Inside{ c' } \right\} } = FDC{ \left( S \right) } $.

  \item
    \textit{Inverted convex hull} means the polygon connecting inverted convex hull vertices in the order of the original convex hull vertices, see Figures~\ref{fig:ConvexHull} and \ref{fig:InvertedConvexHull}.

  \item
    From \ref{ig:PointsOnCircle}, \ref{ig:InverseTwoTriangles}, and \ref{item:CDCEqualFDC}, it follows that inversion of the farthest Delaunay diagram will be the closest Delaunay diagram inside and on the border of the \textit{inverted convex hull}. Note that this does not depend on the choice of point $ O $ in step~\ref{item:O}; however, the closest Delaunay diagram outside of the \textit{inverted convex hull} will depend on the choice of $ O $, see two-dimensional example in Figures~\ref{fig:ConvexHull}, \ref{fig:InvertedConvexHull}, \ref{fig:InvertedConvexHullOnBorderOfTheConvexHull}, and \ref{fig:FinalConvexHull}. Cases where more than $ n + 1 $ points are on the $n$ dimensional circle of the Delaunay diagram correspond to nonunique Delaunay triangulation. These circles will have one-to-one correspondence in original and inverted spaces; therefore, the farthest and the closest Delaunay triangulations have a one-to-one relationship.
\end{enumerate}

These statements prove the duality of the farthest Delaunay diagram with the closest Delaunay diagram in any number of dimensions.

\subsection{Properties of the Closest Delaunay Triangulation of \textit{Inverted Convex Hull}}

Because points of the convex hull were inverted, the closest Delaunay triangulation of the \textit{inverted convex hull} has the following properties:
\begin{itemize}
  \item
    In two dimensions, for any triangulation of $ S' $ that does not intersect the border of the \textit{inverted convex hull}, for triangles inside of the \textit{inverted convex hull}, only one side is visible unless it contains point $ O $, and for triangles outside of the \textit{inverted convex hull}, two sides are visible from point $ O $.

  \item
    In three dimensions, for any tetrahedration of $ S' $ that does not intersect the border of the \textit{inverted convex hull}, for tetrahedrons inside of the \textit{inverted convex hull}, only one or two sides are visible unless it contains point $ O $, and for tetrahedron outside of the \textit{inverted convex hull}, two or three sides are visible from point $ O $.

  \item
    In two dimensions, any triangulation inside the \textit{inverted convex hull} will be inverted to the proper triangulation of the convex hull.
\end{itemize}

\subsection{Calculation of the Farthest Delaunay triangulation by the Closest Delaunay Triangulation}

From section~``\nameref{sec:Duality}'', the farthest Delaunay triangulation can be constructed from the closest Delaunay triangulation.\footnote{If floating-point arithmetic is used, due to roundoff error the \textit{inverted convex hull} might not correspond to the convex hull in the original space. In such a case, the connection between the farthest Delaunay diagram and the closest Delaunay diagram might be broken.} The test to decide if point $ D' $ lies inside the circumscribed circle of $ A' $, $ B' $, $ C' $, and $ D' $ is evaluated by the sign of determinant
\begin{equation}
  \left|
    \begin{matrix}
      x'_{A} & y'_{A} & x'^{2}_{A} + y'^{2}_{A} & 1\\
      x'_{B} & y'_{B} & x'^{2}_{B} + y'^{2}_{B} & 1\\
      x'_{C} & y'_{C} & x'^{2}_{C} + y'^{2}_{C} & 1\\
      x'_{D} & y'_{D} & x'^{2}_{D} + y'^{2}_{D} & 1
    \end{matrix}
  \right|
  ,
  \label{eq:InvertedDeterminant}
\end{equation}
where
$
  \left(
    x',
    y'
  \right)
  =
  \InversiveGeometryTransformation
  {
    O
  }
  {
    \left(
      x,
      y
    \right)
  }
  =
  \left(
    x' = \dfrac{x}{x^2 + y^2},
    y' = \dfrac{y}{x^2 + y^2}
  \right)
$
are inverted coordinates of
$
  A = \left( x_{A}, y_{A} \right)
$,
$
  B = \left( x_{B}, y_{B} \right)
$,
$
  C = \left( x_{C}, y_{C} \right)
$,
and
$
  D = \left( x_{D}, y_{D} \right)
$.

Multiplying each row of \eqref{eq:InvertedDeterminant} by $ x^{2}_{A} + y^{2}_{A} $, $ x^{2}_{B} + y^{2}_{B} $, $ x^{2}_{C} + y^{2}_{C} $, and $ x^{2}_{D} + y^{2}_{D} $, respectively, will not change the sign of the determinant (note that from \ref{item:O}, it follows that there are no points coincident with point $ O $).
\begin{equation}
  \left|
    \begin{matrix}
      x_{A} & y_{A} & 1 & x^{2}_{A} + y^{2}_{A}\\
      x_{B} & y_{B} & 1 & x^{2}_{B} + y^{2}_{B}\\
      x_{C} & y_{C} & 1 & x^{2}_{C} + y^{2}_{C}\\
      x_{D} & y_{D} & 1 & x^{2}_{D} + y^{2}_{D}
    \end{matrix}
  \right|
  .
  \label{eq:DeterminantSign}
\end{equation}

Note that \eqref{eq:DeterminantSign} is equal to minus
\begin{equation}
  \left|
    \begin{matrix}
      x_{A} & y_{A} & x^{2}_{A} + y^{2}_{A} & 1\\
      x_{B} & y_{B} & x^{2}_{B} + y^{2}_{B} & 1\\
      x_{C} & y_{C} & x^{2}_{C} + y^{2}_{C} & 1\\
      x_{D} & y_{D} & x^{2}_{D} + y^{2}_{D} & 1
    \end{matrix}
  \right|
  .
  \label{eq:Determinant}
\end{equation}

Therefore, the test to decide if point $ D' $ lies inside the circumscribed circle of $ A' $, $ B' $, $ C' $, and $ D' $ can be performed in the original space with the inversion of the sign of the determinant \eqref{eq:Determinant}. Note that if coordinates of the points, in the original space, are integer numbers, then they will be represented in the inverted space as rational numbers.

\subsection{Special Case where Points of the Convex Hull are Inverted to Points of the Convex Hull}

From \cite[chapter 7]{InversionTheoryAndConformalMapping} (\textit{When Does Inversion Preserve Convexity?}), there is a case where points of the convex hull can be inverted to the set of points of the convex hull. This happens when the intersection of the interiors of all circles constructed on neighboring vertices of the convex hull is not empty and the center of inversion is located in that intersection area. For such a special set of points, it is possible to construct the closest Delaunay triangulation using the farthest Delaunay triangulation.

\subsection{Remarks}

This appendix shows the duality of the farthest Delaunay triangulation with the closest Delaunay triangulation by inversive geometry. The choice of $ O $ is arbitrary, with the only requirement that it is inside of all farthest Delaunay hyperspheres. The new result of this appendix is that all algorithms for calculation of the closest Delaunay triangulation are also applicable to calculation of the farthest Delaunay triangulation.

\clearpage

\section
{
  \label{appendix:RandomConvexHull}
  Algorithms to Generate Random Convex Hulls
}

The ability to simulate a random convex hull is very important for testing and performance evaluation of different implementations for which the convex hull is an input. The distribution of the random convex hulls depends on the algorithm used to produce one. Therefore, several algorithms will be described in this appendix and, for three of them, examples will be shown.

\subsection{Random Set of Points}

The algorithm is based on construction of a convex hull for a set of random points uniformly distributed in a square, a circle, or other shapes \cite{ConvexHullsComplexity}. The main disadvantage of this algorithm is its tendency to reproduce a convex hull of the original shape.

\subsection{Random Distribution}

This is similar to the algorithm described in the previous section, with the difference that the random points are distributed with density having radial form \cite{OnTheComputerGenerationOfRandomConvexHulls}, random walk, or other distributions \cite{RandomConvexHullsAndExtremeValueStatistics}. From \cite{RandomConvexHullsAndExtremeValueStatistics}, the expected number of vertices in the convex hull of a random walk of length $ n $ is approximately
\begin{equation}
  2 \log{ \left( n \right) }
  .
  \label{eq:ApproximateNumberOfVerticesInConvexHullForRandomWalk}
\end{equation}
Examples for convex hulls of random walks are shown in Figure~\ref{fig:ExamplesRandomConvexHullRandomWalk}.

\input{ConvexHullRandomWalk}

\subsection{Random Modifications of Edges}

The algorithm is based on selecting randomly an edge and adding a vertex so that the new polygon is convex \cite{WebSiteAnswer1}.

\subsection{Random Modifications of Adjacent Edges}

The algorithm is based on selecting randomly two neighboring edges of the convex hull, removing their common vertex, and randomly placing one point on each of them \cite{WebSiteAnswer2}. This approach has the property that, if it starts from any triangle, then it cannot generate a square.

\subsection
{
  \label{sec:RandomSetOfDirections}
  Random Set of Directions
}

This is a new approach to generate a convex hull from a random set of directions by simulating random weights to rescale directions, which satisfies two restrictions:
\begin{enumerate} [label = {(\alph*)}]
  \item \label{enum:RandomConvexHullConditionSumOfWeights} The sum of squares of weights is equal to one.
  \item \label{enum:RandomConvexHullConditionClosedConvexHull} The sum of weighted directions is equal to a vector of zero length.
\end{enumerate}

The steps of the algorithm are as follows:
\begin{enumerate}
  \item
    \label{enum:RandomConvexHullStart}
    Simulate $ n $ random unit vectors $ \left( x_i, y_i \right) $, $ i = \overline{1, n} $. To simulate each random unit vector, the well-known solution is to simulate a random angle $ \alpha \in \left[ -\pi, \pi \right) $ and obtain the unit vector as $ \left( \cos{\left( \alpha \right)}, \sin{\left( \alpha \right)} \right) $. Another well-known solution is to simulate two random variables $ \left( u, v \right) $ from uniform distribution in $ \left[ -1, 1 \right] $ and obtain the unit vector as $ \dfrac{ \left( u, v \right) }{\sqrt{u^2 + v^2}} $, if $ u^2 + v^2 < 1 $ and $ u^2 + v^2 $ is not too small; otherwise, try again.

  \item
    Find orthogonal complement $ X $ of the subspace formed by two vectors $ \left( x_1, x_2, ..., x_n \right) $ and $ \left( y_1, y_2, ..., y_n \right) $, unless these two vectors are close to collinear (absolute value of their correlation is close to one). In such a case, return to step \ref{enum:RandomConvexHullStart}. Matrix $ X $ has dimensions $ n \times \left( n - 2 \right) $ and satisfies
    \begin{equation*}
      \mytranspose{X}
      \cdot
      \begin{bmatrix}
        x_1 & y_1 \\
        x_2 & y_2 \\
        \vdots\\
        x_n & y_n
      \end{bmatrix}
      =
      0
      .
    \end{equation*}

  \item
    Generate random unit vector $ z $ of dimension $ n - 2 $, $ \mytranspose{z} \cdot z = 1 $. The well-known solution is to simulate a vector with $ n - 2 $ random variables from standard normal distribution and divide it by its length if the length is not too small; otherwise, try again.

  \item
    Let
    $
      w
      =
      X
      \cdot
      z
    $.

    Note that this satisfies conditions \ref{enum:RandomConvexHullConditionSumOfWeights} and \ref{enum:RandomConvexHullConditionClosedConvexHull} because
    \begin{equation*}
      \mytranspose{w}
      \cdot
      w
      =
      \mytranspose{z}
      \cdot
      \mytranspose{X}
      \cdot
      X
      \cdot
      z
      =
      \left[
        \mytranspose{X}
        \cdot
        X
        =
        I
      \right]
      =
      \mytranspose{z}
      \cdot
      z
      =
      1
    \end{equation*}
    and
    \begin{equation*}
      \begin{bmatrix}
        x_1 & x_2 & \cdots & x_n \\
        y_1 & y_2 & \cdots & y_n
      \end{bmatrix}
      \cdot
      w
      =
      \begin{bmatrix}
        x_1 & x_2 & \cdots & x_n \\
        y_1 & y_2 & \cdots & y_n
      \end{bmatrix}
      \cdot
      X
      \cdot
      z
      =
      0
      \cdot
      z
      =
      0
      .
    \end{equation*}
    
    Resize all vectors $ \left( x_i, y_i \right) $ by $ w_i $,
    \begin{equation*}
      \left( x_i, y_i \right)
      =
      w_i
      \cdot
      \left( x_i, y_i \right)
      .
    \end{equation*}

  \item
    Sort all vectors $ \left( x_i, y_i \right) $, $ i = \overline{1, n} $ in a clockwise (or counterclockwise) direction\footnote{For example, using $ \arctan2{ \left( y_i, x_i \right) } = -\ImaginaryUnit \log{ \left( \dfrac{x_i + \ImaginaryUnit y_i}{\sqrt{ x^2_i + y^2_i}} \right)} $ and requiring all angles to be in $ \left( -\pi, \pi \right] $.}. Vectors of zero length can be ignored.

  \item
    Construct the convex hull:
    \begin{itemize} [label = {}]
      \item
        Set $ p_0 = \left( 0, 0 \right) $
      \item
        For $ i = \overline{1, n - 1} $
        \begin{itemize} [label = {}]
          \item $ p_i = p_{i - 1} + \left( x_i, y_i \right) $
        \end{itemize}
    \end{itemize}
    Note that $ p_n = p_{n - 1} + \left( x_n, y_n \right) $ is approximately equal to $ p_0 $ due to inexact floating-point arithmetic.
\end{enumerate}

The complexity of this algorithm is $ O{ \left( N \log{ \left( N \right) } \right) } $. See examples of generated convex hulls in Figures~\ref{fig:ExamplesRandomConvexHull}, \ref{fig:ExamplesRandomConvexHullGeometrical1}, and \ref{fig:ExamplesRandomConvexHullGeometrical2}. While this approach efficiently generates convex hulls with a large number of vertices, as the number of vertices increases, the shape of the convex hull becomes more circular, see Figures~\ref{fig:ExamplesRandomConvexHullGeometrical1} and \ref{fig:ExamplesRandomConvexHullGeometrical2}.


\input{ConvexHullOne}

\subsection
{
  \label{sec:RandomConvexHullFromTheFarthestDelaunayTriangulation}
  From the Farthest Delaunay Triangulation
}

The algorithm described in this section is based on simulation of the farthest Delaunay triangulation \cite{ComputationalGeometry1}. Because the farthest Delaunay triangulation only includes vertices of the convex hull, the result of the simulated triangulation is the convex hull. From the property of the farthest Delaunay triangulation that the circumscribed circle of each triangle contains all vertices, the simulation starts by generating a random segment on the unit circle, which cut the unit circle into two circular segments. Each consequent simulation consists of placing a point inside the circular segment, finding the circumscribed circle for the triangle formed by the segment and the point, and replacing the circular segment with two new circular segments formed by the end points of the segment and the point. This process guarantees that each iteration will not break the consistency of the farthest Delaunay triangulation.

The steps of the algorithm are as follows:
\begin{enumerate}
  \item
    Generate two points on the unit circle. The segment connecting these two points divides the circle into two circular segments. Put two circular segments into the list.

  \item
    Randomly select a circular segment from the list with probabilities proportional to the areas of circular segments.

  \item
    Simulate a random point inside the selected circular segment so that the distance from the point to the segment divided by the height of the circular segment (this is similar to using the area of the triangle formed by the segment and the random point) follows the Beta distribution. In this appendix, Beta distribution, with parameters $ \left( 3, 1 \right) $, was used, which gives preference to a larger area.

  \item
    Find the circumscribed circle for the ends of the segment and the simulated random point.

  \item
    Replace the selected circular segment with two circular segments between the end points of the segment and the simulated point.
\end{enumerate}

The complexity of this algorithm is $ O{ \left( N \log{ \left( N \right) } \right) } $. See examples of generated convex hulls in Figures~\ref{fig:ExamplesRandomConvexHullDelaunayTriangulation}, \ref{fig:ExamplesRandomConvexHullDelaunayTriangulationGeometrical1}, and \ref{fig:ExamplesRandomConvexHullDelaunayTriangulationGeometrical2}. Unlike the algorithm described in section~``\nameref{sec:RandomSetOfDirections}'', there is no tendency to produce convex hulls like a circle, see Figures~\ref{fig:ExamplesRandomConvexHullDelaunayTriangulationGeometrical1} and \ref{fig:ExamplesRandomConvexHullDelaunayTriangulationGeometrical2}; however, the circular segments tend to become too narrow. Practically, this algorithm will not be able to generate convex hulls with more than a few hundred thousand vertices, as the new points will lie on the existing segments due to the finite precision of floating-point arithmetic.

\input{ConvexHullTwo}

\subsection{Remarks}

Due to the use of rounded arithmetic, some of the generated vertices might not be vertices of the convex hull. Therefore, an additional step is needed to remove such vertices and obtain the final convex hull.

Let's reiterate the importance of using different algorithms to generate a random convex hull to test and evaluate the performance of different implementations for which the convex hull is an input. Using different algorithms to generate a random convex hull will improve the quality of testing, as the algorithms will cover convex hulls with different properties: small angles, small segments, clustered vertices, etc..

\clearpage

\section
{
  \label{appendix:ClippingBySquare}
  Clipping Segments by Square
}

This section will describe a robust algorithm (proper despite inexact floating-point arithmetic) to clip all segments to a square centered at the origin of a coordinate system. Because all segments are part of some polygons, it is necessary to clip them in a way that preserves the topology of the polygons. The main ideas are to project points that are outside the square to the square outline by finding the intersection of the square outline and the line from the center of the coordinate system to the projected point, and for all segments that have any part outside the square and are crossing any diagonal line, to divide them by the diagonal lines. Dividing segments by the diagonal lines solves inexact floating-point arithmetic issues and simplifies the algorithm to clip all segments by the square.

The area is divided into eight zones, see Figure~\ref{fig:DiagonalQuadrants}.
\begin{figure} [ht]
  \centering
  \begin{tikzpicture} [scale = 0.9]
    \tkzInit[xmin = -3, ymin = -3, xmax = 3, ymax = 3]

    \draw [fill, red] (0, 0) -- (3, -3) -- (3, 3) -- cycle;
    \draw [fill, yellow] (0, 0) -- (3, 3) -- (-3, 3) -- cycle;
    \draw [fill, green] (0, 0) -- (-3, 3) -- (-3, -3) -- cycle;
    \draw [fill, blue] (0, 0) -- (-3, -3) -- (3, -3) -- cycle;
    \draw [thick, black] (0, 0) -- (3, 3);
    \draw [thick, black] (0, 0) -- (-3, 3);
    \draw [thick, black] (0, 0) -- (-3, -3);
    \draw [thick, black] (0, 0) -- (3, -3);

    \tkzGrid

    \draw [thick,->, >=stealth'] (-3.5, 0) -- (3.5, 0) node [anchor = south] {X};
    \draw [thick,->, >=stealth'] (0, -3.5) -- (0, 3.5) node [anchor = west ] {Y};

    \node at ( 2,  0) { \contour{white}{\large{\textbf{0}}} };
    \node at ( 2,  2) { \contour{white}{\large{\textbf{1}}} };
    \node at ( 0,  2) { \contour{white}{\large{\textbf{2}}} };
    \node at (-2,  2) { \contour{white}{\large{\textbf{3}}} };
    \node at (-2,  0) { \contour{white}{\large{\textbf{4}}} };
    \node at (-2, -2) { \contour{white}{\large{\textbf{5}}} };
    \node at ( 0, -2) { \contour{white}{\large{\textbf{6}}} };
    \node at ( 2, -2) { \contour{white}{\large{\textbf{7}}} };
  \end{tikzpicture}
  \caption
  {
    The division of a coordinate system into four areas (red, yellow, green, and blue) and four lines (black) with the exception of point $ \left( 0, 0 \right) $. They are numbered counterclockwise from $ 0 $ to $ 7 $.
  }
  \label{fig:DiagonalQuadrants}
\end{figure}

The algorithm to do robust clipping by square is as follows; see the examples shown in Figures~\ref{fig:ClippingBySquare}, \ref{fig:Projection}, and \ref{fig:FinalResult}. For each segment in the list
\begin{enumerate}
  \item If the segment does not have any points outside the square, return the segment.
  \item If the segment has one point in the origin of the coordinate system $ \left( 0, 0 \right) $, then return the intersection of this segment and the square.
  \item If the points are in zones $ 0 $ and $ 4 $ or $ 2 $ and $ 6 $, the segment crosses two diagonal lines. Intersect this segment with OY or OX axes, respectively (see red points in Figure~\ref{fig:ClippingBySquare}), and add two segments to the list.
  \item If both points of the segment are in the odd zone, zones $ 1 $ and $ 5 $ or zones $ 3 $ and $ 7 $, then return the intersection of this segment and the square.
  \item If the segment is crossing a diagonal line, then divide it by the diagonal line and add both segments to the list.
  \item \label{en:OutsideEvenZone} Both points of the segment are in the even zone extended by the zone's border (see red, yellow, green, or blue areas in Figure~\ref{fig:DiagonalQuadrants}). If there are no points inside the square (excluding the square outline), project both points to the square outline (see Figure~\ref{fig:Projection}) and return the resultant segment.
  \item Both points of the segment are in the even zone extended by the zone's border; one of the points is inside the square, and another is outside. Therefore, the segment intersects the outline of the square only at one point. Divide this segment by the intersection point. The intersection point must be in the same zone extended by the zone's border. If the intersection point falls within another zone due to inexact calculations, it has to be adjusted. Add the segment that is outside the square to the list (process it by step \ref{en:OutsideEvenZone}) and return the segment that is inside the square.
\end{enumerate}

\begin{figure} [p]
  \centering
  \begin{tikzpicture} [scale = 0.9]
    \tkzInit[xmin = -9, ymin = -9, xmax = 9, ymax = 9]

    \draw [fill, yellow] (-3, -3) -- (-3, 3) -- (3, 3) -- (3, -3) -- cycle;

    \tkzGrid

    \draw [thick, red] (-3, -3) -- (-3, 3) -- (3, 3) -- (3, -3) -- cycle;
    \draw [dashed, red] (-9, -9) -- (9, 9);
    \draw [dashed, red] (-9, 9) -- (9, -9);

    \draw [thick, black] (0, 1) -- (4, 9);
    \draw [fill = black, black] (0, 1) circle [radius = 0.08];
    \draw [fill = black, black] (4, 9) circle [radius = 0.08];

    \draw [thick, black] (4, 4) -- (9, -1);
    \draw [fill = black, black] (4, 4) circle [radius = 0.08];
    \draw [fill = black, black] (9, -1) circle [radius = 0.08];

    \draw [thick, black] (7, -3) -- (5, -7);
    \draw [fill = black, black] (7, -3) circle [radius = 0.08];
    \draw [fill = black, black] (5, -7) circle [radius = 0.08];

    \draw [thick, black] (-4, 9) -- (2, -9) -- cycle;
    \draw [fill = black, black] (-4, 9) circle [radius = 0.08];
    \draw [fill = black, black] (2, -9) circle [radius = 0.08];

    \draw [thick, black] (-7, 9) -- (-8, -9);
    \draw [fill = black, black] (-7, 9) circle [radius = 0.08];
    \draw [fill = black, black] (-8, -9) circle [radius = 0.08];

    \tkzAxeXY

    \draw [fill = red, red] (-1, 0) circle [radius = 0.08];
    \draw [fill = red, red] (-7.5, 0) circle [radius = 0.08];

    \draw [fill = green, green] (17/3, -17/3) circle [radius = 0.08];
    \draw [fill = green, green] (-1.5, 1.5) circle [radius = 0.08];
    \draw [fill = green, green] (-0.75, -0.75) circle [radius = 0.08];
    \draw [fill = green, green] (-135/17, -135/17) circle [radius = 0.08];
    \draw [fill = green, green] (-135/19, 135/19) circle [radius = 0.08];

    \draw [fill = blue, blue] (-2, 3) circle [radius = 0.08];
    \draw [fill = blue, blue] (1, 3) circle [radius = 0.08];
    \draw [fill = blue, blue] (0, -3) circle [radius = 0.08];
  \end{tikzpicture}
  \caption
  {
    Examples of segments are shown as black lines having ends with black circles. The points where they intersect the coordinate system axis are shown as red circles. The points where segments intersect diagonal lines (dashed red lines) are shown as green circles. The points where segments intersect the square outline are shown as blue circles.
  }
  \label{fig:ClippingBySquare}
\end{figure}

\begin{figure} [p]
  \centering
  \begin{tikzpicture} [scale = 0.9]
    \tkzInit[xmin = -9, ymin = -9, xmax = 9, ymax = 9]

    \draw [fill, yellow] (-3, -3) -- (-3, 3) -- (3, 3) -- (3, -3) -- cycle;

    \tkzGrid

    \draw [thick, red] (-3, -3) -- (-3, 3) -- (3, 3) -- (3, -3) -- cycle;
    \draw [dashed, red] (-9, -9) -- (9, 9);
    \draw [dashed, red] (-9, 9) -- (9, -9);

    \draw [thick, black] (0, 1) -- (4, 9);
    \draw [fill = black, black] (0, 1) circle [radius = 0.08];
    \draw [fill = black, black] (4, 9) circle [radius = 0.08];

    \draw [thick, black] (4, 4) -- (9, -1);
    \draw [fill = black, black] (4, 4) circle [radius = 0.08];
    \draw [fill = black, black] (9, -1) circle [radius = 0.08];

    \draw [thick, black] (7, -3) -- (5, -7);
    \draw [fill = black, black] (7, -3) circle [radius = 0.08];
    \draw [fill = black, black] (5, -7) circle [radius = 0.08];

    \draw [thick, black] (-4, 9) -- (2, -9) -- cycle;
    \draw [fill = black, black] (-4, 9) circle [radius = 0.08];
    \draw [fill = black, black] (2, -9) circle [radius = 0.08];

    \draw [thick, black] (-7, 9) -- (-8, -9);
    \draw [fill = black, black] (-7, 9) circle [radius = 0.08];
    \draw [fill = black, black] (-8, -9) circle [radius = 0.08];

    \tkzAxeXY

    \draw [fill = black, black] (-1, 0) circle [radius = 0.08];
    \draw [fill = black, black] (-7.5, 0) circle [radius = 0.08];

    \draw [fill = black, black] (17/3, -17/3) circle [radius = 0.08];
    \draw [fill = black, black] (-1.5, 1.5) circle [radius = 0.08];
    \draw [fill = black, black] (-0.75, -0.75) circle [radius = 0.08];
    \draw [fill = black, black] (-135/17, -135/17) circle [radius = 0.08];
    \draw [fill = black, black] (-135/19, 135/19) circle [radius = 0.08];

    \draw [fill = black, black] (-2, 3) circle [radius = 0.08];
    \draw [fill = black, black] (1, 3) circle [radius = 0.08];
    \draw [fill = black, black] (0, -3) circle [radius = 0.08];

    \draw [dashed, blue] (0, 0) -- (4, 9) circle [radius = 0.08];
    \draw [dashed, blue] (0, 0) -- (9, -1) circle [radius = 0.08];
    \draw [dashed, blue] (0, 0) -- (7, -3) circle [radius = 0.08];
    \draw [dashed, blue] (0, 0) -- (5, -7) circle [radius = 0.08];
    \draw [dashed, blue] (0, 0) -- (-4, 9) circle [radius = 0.08];
    \draw [dashed, blue] (0, 0) -- (2, -9) circle [radius = 0.08];
    \draw [dashed, blue] (0, 0) -- (-7, 9) circle [radius = 0.08];
    \draw [dashed, blue] (0, 0) -- (-8, -9) circle [radius = 0.08];

    \draw [fill = blue, blue] (-3, -3) circle [radius = 0.08];
    \draw [fill = blue, blue] (-3, 3) circle [radius = 0.08];
    \draw [fill = blue, blue] (3, 3) circle [radius = 0.08];
    \draw [fill = blue, blue] (3, -3) circle [radius = 0.08];

    \draw [fill = blue, blue] (4/3, 3) circle [radius = 0.08];
    \draw [fill = blue, blue] (3, -1/3) circle [radius = 0.08];
    \draw [fill = blue, blue] (3, -9/7) circle [radius = 0.08];
    \draw [fill = blue, blue] (15/7, -3) circle [radius = 0.08];
    \draw [fill = blue, blue] (-4/3, 3) circle [radius = 0.08];
    \draw [fill = blue, blue] (2/3, -3) circle [radius = 0.08];
    \draw [fill = blue, blue] (-7/3, 3) circle [radius = 0.08];
    \draw [fill = blue, blue] (-8/3, -3) circle [radius = 0.08];
    \draw [fill = blue, blue] (-3, 0) circle [radius = 0.08];
  \end{tikzpicture}
  \caption
  {
    All vertices (black circles) outside the square (yellow square) are projected to the outline of the square (blue circles) by intersecting the line from the origin of the coordinate system to the vertex (dashed blue line) with the square outline (red line).
  }
  \label{fig:Projection}
\end{figure}

The result of performing this algorithm is shown in Figure~\ref{fig:FinalResult}.

\begin{figure} [ht]
  \centering
  \begin{tikzpicture} [scale = 2 * 0.9]
    \tkzInit[xmin = -4, ymin = -4, xmax = 4, ymax = 4]
    \tkzGrid
    \tkzAxeXY

    \draw [thick, blue] (0, 1) -- (1, 3) -- (4/3, 3);
    \draw [fill = blue, blue] (0, 1) circle [radius = 0.08];
    \draw [fill = blue, blue] (1, 3) circle [radius = 0.08];
    \draw [fill = blue, blue] (4/3, 3) circle [radius = 0.08];

    \draw [thick, blue] (3, 3) -- (3, -1/3);
    \draw [fill = blue, blue] (3, 3) circle [radius = 0.08];
    \draw [fill = blue, blue] (3, -1/3) circle [radius = 0.08];

    \draw [thick, blue] (3, -9/7) -- (3, -3) -- (15/7, -3);
    \draw [fill = blue, blue] (3, -9/7) circle [radius = 0.08];
    \draw [fill = blue, blue] (3, -3) circle [radius = 0.08];
    \draw [fill = blue, blue] (15/7, -3) circle [radius = 0.08];

    \draw [thick, blue] (2/3, -3) -- (0, -3) -- (-2, 3) -- (-4/3, 3);
    \draw [fill = blue, blue] (2/3, -3) circle [radius = 0.08];
    \draw [fill = blue, blue] (0, -3) circle [radius = 0.08];
    \draw [fill = blue, blue] (-0.75, -0.75) circle [radius = 0.08];
    \draw [fill = blue, blue] (-1, 0) circle [radius = 0.08];
    \draw [fill = blue, blue] (-1.5, 1.5) circle [radius = 0.08];
    \draw [fill = blue, blue] (-2, 3) circle [radius = 0.08];
    \draw [fill = blue, blue] (-4/3, 3) circle [radius = 0.08];

    \draw [thick, blue] (-8/3, -3) -- (-3, -3) -- (-3, 3) -- (-7/3, 3);
    \draw [fill = blue, blue] (-8/3, -3) circle [radius = 0.08];
    \draw [fill = blue, blue] (-3, -3) circle [radius = 0.08];
    \draw [fill = blue, blue] (-3, 0) circle [radius = 0.08];
    \draw [fill = blue, blue] (-3, 3) circle [radius = 0.08];
    \draw [fill = blue, blue] (-7/3, 3) circle [radius = 0.08];
  \end{tikzpicture}
  \caption
  {
    The result of clipping all the segments (shown in Figure~\ref{fig:ClippingBySquare}) by the square.
  }
  \label{fig:FinalResult}
\end{figure}

\clearpage

\section
{
  \label{appendix:RemoveOverlappingSegments}
  Algorithm to Remove Overlapping Segments
}

All segments are specified with integer coordinates. Each segment has a list of indices. The indices of each segment are indices of the polygons for which the segment is a border segment. The next algorithm will detect all overlapping segments. In some parts of the algorithm, it would be necessary to merge lists of indices. This operation applies the ``XOR'' rule, which means that only indices occurring an odd number of times in both lists will be present in the final list, but only once.
\begin{enumerate}
  \item
    Discard segments of zero length.
  \item
    Assign each segment an integer vector:
    \begin{equation*}
      \left(
        \begin{matrix}
          x\\
          y
        \end{matrix}
      \right)
      =
      \dfrac
      {
        \left(
          \begin{matrix}
            x_1\\
            y_1
          \end{matrix}
        \right)
        -
        \left(
          \begin{matrix}
            x_0\\
            y_0
          \end{matrix}
        \right)
      }
      {
        \gcd{ \left( x_1 - x_0, y_1 - y_0 \right) }
      }
      ,
    \end{equation*}
    where
    $
      \left(
        \begin{matrix}
          x_0\\
          y_0
        \end{matrix}
      \right)
    $
    and
    $
      \left(
        \begin{matrix}
          x_1\\
          y_1
        \end{matrix}
      \right)
    $
    are starting and ending vertices of the segment, and
    $ \gcd $
    is the greatest common divisor.

    Make adjustments for opposite directions
    \begin{equation*}
      \left(
        \begin{matrix}
          x\\
          y
        \end{matrix}
      \right)
      =
      \left\{
        \begin{aligned}
          -
          &
          \left(
            \begin{matrix}
              x\\
              y
            \end{matrix}
          \right)
          ,
          \text{ if }
          x < 0 \lor x = 0 \land y < 0,
          \\
          &
          \left(
            \begin{matrix}
              x\\
              y
            \end{matrix}
          \right)
          ,
          \text{ otherwise}.
        \end{aligned}
      \right.
    \end{equation*}
  \item
    Using a hash table or sorting all integer vectors, group all segments having the same integer vector.
  \item
    For each group of segments sharing an integer vector
    $
      \left(
        \begin{matrix}
          x\\
          y
        \end{matrix}
      \right)
    $
    \begin{enumerate}
      \item
        Assign each segment an integer value equal to the vector product of the integer vector to any end point of the segment (the result is the same because the integer vector is parallel to the segment).\footnote{This operation will approximately double the number of bits and might require the use of extended precision.}
      \item
        Using a hash table or sorting all integer values, group all segments having the same integer value.
      \item
        For each group of segments sharing an integer value
        \begin{enumerate}
          \item
            Put all end points of segments with the list of indices into an array. Each element of the array is a pair of a point and a list of indices.
          \item
            Sort this array by $ x $-coordinate if the $ x $-coordinates are different; otherwise, sort them by $ y $-coordinate.
          \item
            Merge elements of the array with equal points by merging their lists of indices.
          \item
            Remove elements of the array with an empty list of indices.
          \item
            For each element in the array, merge indices with indices of the previous element from beginning to end (the last element should have an empty list of indices).
          \item
            For each element in the array with a nonempty list of indices, create a segment starting with the element point, ending with the next element point, and having indices of that element.
        \end{enumerate}
    \end{enumerate}
\end{enumerate}

This algorithm creates a new list of segments without any overlaps.

\clearpage

\section
{
  \label{appendix:MinimumSubElement}
  Efficient Extraction of Elements in Sorted Order from any Subarray
}

Efficient extraction of elements in sorted order from any subarray is performed by preprocessing the array using a method very similar to \textbf{mergesort} \cite[chapter 5]{Knuth}, see Figure~\ref{fig:SortedSubArray}. The only difference is that the previous step of the \textbf{mergesort} algorithm is kept in memory. The complexity of this step is the same as the complexity of \textbf{mergesort}, which is $ O{ \left( N \log{ \left( N \right) } \right) } $, where $ N $ is the number of elements in the array. Then, extraction of elements in sorted order from any subarray is performed by finding corresponding sorted arrays with the preferences for the longest (see the green cells in Figure~\ref{fig:SortedSubArray}) and merging, starting with the shortest (similar to steps of the \textbf{mergesort} algorithm). Because there are no more than two sorted arrays for each size of $ 2^i $, $ i \in \mathbb{N}_0 \wedge 2^i \leq K $, where $ K $ is the number of elements in the subarray, the complexity of merging sorted arrays is $ O{ \left( K \right) } $; however, not all elements are needed. To extract a few elements in sorted order, put all sorted arrays in the tree structure, starting with the shortest, and rearrange them to make all child nodes of the tree no smaller than their parent nodes, as shown in Figure~\ref{fig:TreeA}. This is similar to merging of \textbf{leftist trees} \cite[part 5]{HandbookOfDataStructuresAndApplications}. The complexity of this step is $ O{ \left( \log{ \left( K \right) } \right) } $. This forms a priority tree with some nodes referring to the position in the sorted arrays shown in green in Figures~\ref{fig:SortedSubArray}, \ref{fig:TreeA}, \ref{fig:TreeB}, and \ref{fig:TreeC}. Modifications of the tree with the first and second elements removed are shown in Figures~\ref{fig:TreeB} and \ref{fig:TreeC}. While the worst-case complexity of each request to remove the minimum element is $ O{ \left( \log{ \left( K \right) } \right) } $, the total complexity to extract all elements of the subarray in sorted order is still $ O{ \left( K \right) } $. Therefore, the amortized time complexity to remove an element is $ O{ \left( 1 \right) } $.

\begin{figure} [ht]
  \centering
  \begin{tabular}{|c|c|c|c|c|c|c|c|c|c|c|c|c|c|c|c|ccccc}
    \hline
    8 & \cellcolor{green}5 & 13 & 16 & 6 & 1 & 14 & 3 & 15 & 4 & 17 & 0 & 12 & 10 & 19 & 7 & \multicolumn{1}{c|}{20} & \multicolumn{1}{c|}{10} & \multicolumn{1}{c|}{\cellcolor{green}18} & \multicolumn{1}{c|}{11} & \multicolumn{1}{c|}{9} \\ \hline
    \multicolumn{2}{|c|}{5, 8} & \multicolumn{2}{c|}{\cellcolor{green}13, 16} & \multicolumn{2}{c|}{1, 6} & \multicolumn{2}{c|}{3, 14} & \multicolumn{2}{c|}{4, 15} & \multicolumn{2}{c|}{0, 17} & \multicolumn{2}{c|}{10, 12} & \multicolumn{2}{c|}{7, 19} & \multicolumn{2}{c|}{\cellcolor{green}10, 20} & \multicolumn{2}{c|}{11, 18} & \\ \cline{1-20}
    \multicolumn{4}{|c|}{5, 8, 13, 16} & \multicolumn{4}{c|}{\cellcolor{green}1, 3, 6, 14} & \multicolumn{4}{c|}{0, 4, 15, 17} & \multicolumn{4}{c|}{7, 10, 12, 19} & \multicolumn{4}{c|}{10, 11, 18, 20} & \\ \cline{1-20}
    \multicolumn{8}{|c|}{1, 3, 5, 6, 8, 13, 14, 16} & \multicolumn{8}{c|}{\cellcolor{green}0, 4, 7, 10, 12, 15, 17, 19} & \multicolumn{5}{c}{} \\ \cline{1-16}
    \multicolumn{16}{|c|}{0, 1, 3, 4, 5, 6, 7, 8, 10, 12, 13, 14, 15, 16, 17, 19} & \multicolumn{5}{c}{} \\ \cline{1-16}
  \end{tabular}
  \caption
  {
    The first row is the array of size $ N = 21 $. Each element of the next row groups two elements of the previous row and sorts all elements in them (similar to a step of the \textbf{mergesort} algorithm). The process is continued for all rows until there is nothing left to group. In the result, any subarray can be represented as several sorted arrays (green cells for the subarray of size $ K = 18 $). Note that it is possible to delay construction of the sorted arrays until they are needed.
  }
  \label{fig:SortedSubArray}
\end{figure}

\begin{figure} [ht]
  \centering
    \begin{tikzpicture}
      \tikzset{level distance = 1.5 cm}
      \Tree[.\addcircle{\color{red}0}
             [.\addcircle{\color{red}1}
               [.\addcircle{\color{red}5}
                 [.\addcircle{\color{red}13}
                   [.\addsquare{18} ]
                   [.\addsquare{{\color{red}\sout{13}}, 16} ]
                 ]
                 [.\addsquare{10, 20} ]
               ]
               [.\addsquare{{\color{red}\sout{1}}, 3, 6, 14} ]
             ]
             [.\addsquare{{\color{red}\sout{0}}, 4, 7, 10, 12, 15, 17, 19} ]
           ]
    \end{tikzpicture}
  \caption
  {
    Priority queue based on the tree with all child nodes no smaller than their parent nodes (changes are shown in red).
  }
  \label{fig:TreeA}
\end{figure}

\begin{figure} [htb]
  \centering
    \begin{tikzpicture}
      \tikzset{level distance = 1.5 cm}
      \Tree[.\addcircle{\color{red}1}
             [.\addcircle{\color{red}3}
               [.\addcircle{5}
                 [.\addcircle{13}
                   [.\addsquare{18} ]
                   [.\addsquare{\sout{13}, 16} ]
                 ]
                 [.\addsquare{10, 20} ]
               ]
               [.\addsquare{\sout{1}, {\color{red}\sout{3}}, 6, 14} ]
             ]
             [.\addsquare{\sout{0}, 4, 7, 10, 12, 15, 17, 19} ]
      ]
    \end{tikzpicture}
  \caption
  {
    Modified priority queue after removing the first element (changes are shown in red).
  }
  \label{fig:TreeB}
\end{figure}

\begin{figure} [htb]
  \centering
    \begin{tikzpicture}
      \tikzset{level distance = 1.5 cm}
      \Tree[.\addcircle{\color{red}3}
             [.\addcircle{\color{red}5}
               [.\addcircle{\color{red}10}
                 [.\addcircle{13}
                   [.\addsquare{18} ]
                   [.\addsquare{\sout{13}, 16} ]
                 ]
                 [.\addsquare{{\color{red}\sout{10}}, 20} ]
               ]
               [.\addsquare{\sout{1}, {\sout{3}}, 6, 14} ]
             ]
             [.\addsquare{\sout{0}, 4, 7, 10, 12, 15, 17, 19} ]
      ]
    \end{tikzpicture}
  \caption
  {
    Modified priority queue after removing the second element (changes are shown in red).
  }
  \label{fig:TreeC}
\end{figure}

Note that in a dynamic programming approach, see step~\ref{enum:SubArrayStep} in section~\myref{sec:DynamicProgramming}, elements of the array are added one by one. 

%% file: FigureArcFitting.tex

  \centering
  \begin{tikzpicture} [scale = 0.18]
    \draw [thick, red]
    (-10.000000, 0.000000) --
    (-9.826016, 1.954516) --
    (-9.238432, 3.826684) --
    (-8.626057, 5.763747) --
    (-7.302913, 7.302913) --
    (-5.408999, 8.095138) --
    (-3.642285, 8.793255) --
    (-2.028990, 10.200423) --
    (-0.000000, 9.646971) --
    (2.025162, 10.181179) --
    (3.643924, 8.797212) --
    (5.445700, 8.150066) --
    (7.089360, 7.089360) --
    (8.191286, 5.473242) --
    (8.974809, 3.717488) --
    (9.665902, 1.922667) --
    (10.000000, 0.000000) --
    (11.875000, -0.179909) --
    (13.750000, 0.100368) --
    (15.625000, 0.406337) --
    (17.500000, -0.030850) --
    (19.375000, -0.056515) --
    (21.250000, -0.102612) --
    (23.125000, 0.059574) --
    (25.000000, 0.025121) --
    (26.875000, -0.318717) --
    (28.750000, -0.353412) --
    (30.625000, 0.004383) --
    (32.500000, -0.458428) --
    (34.375000, -0.337692) --
    (36.250000, -0.145258) --
    (38.125000, 0.086489) --
    (40.000000, 0.000000) --
    (40.250610, 1.939274) --
    (40.847778, 3.790975) --
    (41.991879, 5.350855) --
    (42.730882, 7.269118) --
    (44.401162, 8.379253) --
    (46.350539, 8.810579) --
    (48.082292, 9.640970) --
    (50.000000, 9.686431) --
    (51.873440, 9.418419) --
    (53.829417, 9.245031) --
    (55.732531, 8.579339) --
    (57.388207, 7.388207) --
    (58.116020, 5.422951) --
    (59.286081, 3.846421) --
    (60.169792, 2.022897) --
    (60.000000, 0.000000) --
    (61.875000, -0.399138) --
    (63.750000, -0.168001) --
    (65.625000, 0.064184) --
    (67.500000, 0.182067) --
    (69.375000, -0.247732) --
    (71.250000, -0.415730) --
    (73.125000, 0.025796) --
    (75.000000, -0.303001) --
    (76.875000, 0.344346) --
    (78.750000, 0.321963) --
    (80.625000, -0.498818) --
    (82.500000, -0.299388) --
    (84.375000, 0.251906) --
    (86.250000, 0.293652) --
    (88.125000, 0.487014) --
    (90.000000, 0.000000);

    \draw [thick, MyGreenColor]
    (-10.000000, 0.000000) --
    (-9.998561, 0.122269) --
    (-9.995627, 0.244512) --
    (-9.991199, 0.366710) --
    (-9.985277, 0.488844) --
    (-9.977862, 0.610897) --
    (-9.968955, 0.732850) --
    (-9.958557, 0.854685) --
    (-9.946671, 0.976384) --
    (-9.933297, 1.097929) --
    (-9.918438, 1.219300) --
    (-9.902097, 1.340481) --
    (-9.884275, 1.461453) --
    (-9.864975, 1.582199) --
    (-9.844200, 1.702699) --
    (-9.821953, 1.822936) --
    (-9.798237, 1.942892) --
    (-9.773057, 2.062549) --
    (-9.746416, 2.181890) --
    (-9.718317, 2.300895) --
    (-9.688765, 2.419548) --
    (-9.657765, 2.537831) --
    (-9.625320, 2.655727) --
    (-9.591437, 2.773216) --
    (-9.556119, 2.890283) --
    (-9.519373, 3.006909) --
    (-9.481203, 3.123076) --
    (-9.441616, 3.238769) --
    (-9.400617, 3.353969) --
    (-9.358213, 3.468658) --
    (-9.314410, 3.582821) --
    (-9.269213, 3.696440) --
    (-9.222631, 3.809497) --
    (-9.174670, 3.921977) --
    (-9.125338, 4.033862) --
    (-9.074641, 4.145135) --
    (-9.022587, 4.255780) --
    (-8.969184, 4.365780) --
    (-8.914440, 4.475119) --
    (-8.858363, 4.583780) --
    (-8.800962, 4.691747) --
    (-8.742245, 4.799005) --
    (-8.682221, 4.905537) --
    (-8.620899, 5.011327) --
    (-8.558288, 5.116359) --
    (-8.494397, 5.220617) --
    (-8.429236, 5.324087) --
    (-8.362815, 5.426752) --
    (-8.295144, 5.528597) --
    (-8.226232, 5.629607) --
    (-8.156090, 5.729768) --
    (-8.084729, 5.829063) --
    (-8.012159, 5.927477) --
    (-7.938392, 6.024998) --
    (-7.863437, 6.121609) --
    (-7.787307, 6.217296) --
    (-7.710012, 6.312045) --
    (-7.631564, 6.405842) --
    (-7.551976, 6.498673) --
    (-7.471258, 6.590524) --
    (-7.389424, 6.681381) --
    (-7.306484, 6.771230) --
    (-7.222452, 6.860059) --
    (-7.137340, 6.947853) --
    (-7.051162, 7.034601) --
    (-6.963928, 7.120288) --
    (-6.875654, 7.204902) --
    (-6.786352, 7.288430) --
    (-6.696035, 7.370860) --
    (-6.604717, 7.452180) --
    (-6.512411, 7.532377) --
    (-6.419132, 7.611439) --
    (-6.324892, 7.689355) --
    (-6.229708, 7.766113) --
    (-6.133591, 7.841701) --
    (-6.036558, 7.916108) --
    (-5.938622, 7.989323) --
    (-5.839799, 8.061336) --
    (-5.740102, 8.132134) --
    (-5.639547, 8.201708) --
    (-5.538148, 8.270048) --
    (-5.435922, 8.337142) --
    (-5.332883, 8.402982) --
    (-5.229047, 8.467556) --
    (-5.124429, 8.530857) --
    (-5.019044, 8.592873) --
    (-4.912909, 8.653596) --
    (-4.806040, 8.713017) --
    (-4.698452, 8.771126) --
    (-4.590162, 8.827916) --
    (-4.481185, 8.883378) --
    (-4.371538, 8.937502) --
    (-4.261238, 8.990282) --
    (-4.150301, 9.041710) --
    (-4.038743, 9.091777) --
    (-3.926581, 9.140476) --
    (-3.813832, 9.187800) --
    (-3.700513, 9.233743) --
    (-3.586640, 9.278296) --
    (-3.472232, 9.321453) --
    (-3.357304, 9.363208) --
    (-3.241874, 9.403555) --
    (-3.125960, 9.442487) --
    (-3.009578, 9.479999) --
    (-2.892746, 9.516086) --
    (-2.775482, 9.550741) --
    (-2.657803, 9.583959) --
    (-2.539726, 9.615737) --
    (-2.421270, 9.646068) --
    (-2.302451, 9.674948) --
    (-2.183289, 9.702374) --
    (-2.063799, 9.728340) --
    (-1.944002, 9.752843) --
    (-1.823914, 9.775880) --
    (-1.703553, 9.797447) --
    (-1.582937, 9.817540) --
    (-1.462084, 9.836157) --
    (-1.341013, 9.853295) --
    (-1.219742, 9.868951) --
    (-1.098288, 9.883123) --
    (-0.976670, 9.895809) --
    (-0.854906, 9.907007) --
    (-0.733014, 9.916716) --
    (-0.611013, 9.924933) --
    (-0.488920, 9.931658) --
    (-0.366754, 9.936889) --
    (-0.244533, 9.940627) --
    (-0.122276, 9.942869) --
    (-0.000000, 9.943617) --
    (0.122276, 9.942869) --
    (0.244533, 9.940627) --
    (0.366754, 9.936889) --
    (0.488920, 9.931658) --
    (0.611013, 9.924933) --
    (0.733014, 9.916716) --
    (0.854906, 9.907007) --
    (0.976670, 9.895809) --
    (1.098288, 9.883123) --
    (1.219742, 9.868951) --
    (1.341013, 9.853295) --
    (1.462084, 9.836157) --
    (1.582937, 9.817540) --
    (1.703553, 9.797447) --
    (1.823914, 9.775880) --
    (1.944002, 9.752843) --
    (2.063799, 9.728340) --
    (2.183289, 9.702374) --
    (2.302451, 9.674948) --
    (2.421270, 9.646068) --
    (2.539726, 9.615737) --
    (2.657803, 9.583959) --
    (2.775482, 9.550741) --
    (2.892746, 9.516086) --
    (3.009578, 9.479999) --
    (3.125960, 9.442487) --
    (3.241874, 9.403555) --
    (3.357304, 9.363208) --
    (3.472232, 9.321453) --
    (3.586640, 9.278296) --
    (3.700513, 9.233743) --
    (3.813832, 9.187800) --
    (3.926581, 9.140476) --
    (4.038743, 9.091777) --
    (4.150301, 9.041710) --
    (4.261238, 8.990282) --
    (4.371538, 8.937502) --
    (4.481185, 8.883378) --
    (4.590162, 8.827916) --
    (4.698452, 8.771126) --
    (4.806040, 8.713017) --
    (4.912909, 8.653596) --
    (5.019044, 8.592873) --
    (5.124429, 8.530857) --
    (5.229047, 8.467556) --
    (5.332883, 8.402982) --
    (5.435922, 8.337142) --
    (5.538148, 8.270048) --
    (5.639547, 8.201708) --
    (5.740102, 8.132134) --
    (5.839799, 8.061336) --
    (5.938622, 7.989323) --
    (6.036558, 7.916108) --
    (6.133591, 7.841701) --
    (6.229708, 7.766113) --
    (6.324892, 7.689355) --
    (6.419132, 7.611439) --
    (6.512411, 7.532377) --
    (6.604717, 7.452180) --
    (6.696035, 7.370860) --
    (6.786352, 7.288430) --
    (6.875654, 7.204902) --
    (6.963928, 7.120288) --
    (7.051162, 7.034601) --
    (7.137340, 6.947853) --
    (7.222452, 6.860059) --
    (7.306484, 6.771230) --
    (7.389424, 6.681381) --
    (7.471258, 6.590524) --
    (7.551976, 6.498673) --
    (7.631564, 6.405842) --
    (7.710012, 6.312045) --
    (7.787307, 6.217296) --
    (7.863437, 6.121609) --
    (7.938392, 6.024998) --
    (8.012159, 5.927477) --
    (8.084729, 5.829063) --
    (8.156090, 5.729768) --
    (8.226232, 5.629607) --
    (8.295144, 5.528597) --
    (8.362815, 5.426752) --
    (8.429236, 5.324087) --
    (8.494397, 5.220617) --
    (8.558288, 5.116359) --
    (8.620899, 5.011327) --
    (8.682221, 4.905537) --
    (8.742245, 4.799005) --
    (8.800962, 4.691747) --
    (8.858363, 4.583780) --
    (8.914440, 4.475119) --
    (8.969184, 4.365780) --
    (9.022587, 4.255780) --
    (9.074641, 4.145135) --
    (9.125338, 4.033862) --
    (9.174670, 3.921977) --
    (9.222631, 3.809497) --
    (9.269213, 3.696440) --
    (9.314410, 3.582821) --
    (9.358213, 3.468658) --
    (9.400617, 3.353969) --
    (9.441616, 3.238769) --
    (9.481203, 3.123076) --
    (9.519373, 3.006909) --
    (9.556119, 2.890283) --
    (9.591437, 2.773216) --
    (9.625320, 2.655727) --
    (9.657765, 2.537831) --
    (9.688765, 2.419548) --
    (9.718317, 2.300895) --
    (9.746416, 2.181890) --
    (9.773057, 2.062549) --
    (9.798237, 1.942892) --
    (9.821953, 1.822936) --
    (9.844200, 1.702699) --
    (9.864975, 1.582199) --
    (9.884275, 1.461453) --
    (9.902097, 1.340481) --
    (9.918438, 1.219300) --
    (9.933297, 1.097929) --
    (9.946671, 0.976384) --
    (9.958557, 0.854685) --
    (9.968955, 0.732850) --
    (9.977862, 0.610897) --
    (9.985277, 0.488844) --
    (9.991199, 0.366710) --
    (9.995627, 0.244512) --
    (9.998561, 0.122269) --
    (10.000000, 0.000000) --
    (40.000000, 0.000000) --
    (40.001638, 0.122139) --
    (40.004768, 0.244250) --
    (40.009389, 0.366313) --
    (40.015500, 0.488310) --
    (40.023102, 0.610224) --
    (40.032191, 0.732036) --
    (40.042768, 0.853727) --
    (40.054831, 0.975281) --
    (40.068377, 1.096678) --
    (40.083406, 1.217900) --
    (40.099914, 1.338930) --
    (40.117898, 1.459749) --
    (40.137358, 1.580340) --
    (40.158288, 1.700684) --
    (40.180688, 1.820763) --
    (40.204552, 1.940560) --
    (40.229877, 2.060056) --
    (40.256661, 2.179234) --
    (40.284898, 2.298076) --
    (40.314584, 2.416564) --
    (40.345716, 2.534680) --
    (40.378288, 2.652408) --
    (40.412296, 2.769729) --
    (40.447734, 2.886626) --
    (40.484597, 3.003081) --
    (40.522880, 3.119078) --
    (40.562577, 3.234598) --
    (40.603682, 3.349624) --
    (40.646189, 3.464140) --
    (40.690091, 3.578128) --
    (40.735383, 3.691572) --
    (40.782056, 3.804454) --
    (40.830105, 3.916757) --
    (40.879523, 4.028465) --
    (40.930301, 4.139561) --
    (40.982432, 4.250028) --
    (41.035909, 4.359851) --
    (41.090723, 4.469012) --
    (41.146866, 4.577495) --
    (41.204330, 4.685285) --
    (41.263107, 4.792365) --
    (41.323187, 4.898718) --
    (41.384561, 5.004330) --
    (41.447221, 5.109185) --
    (41.511157, 5.213266) --
    (41.576360, 5.316559) --
    (41.642819, 5.419047) --
    (41.710526, 5.520716) --
    (41.779469, 5.621551) --
    (41.849639, 5.721535) --
    (41.921024, 5.820656) --
    (41.993615, 5.918897) --
    (42.067401, 6.016243) --
    (42.142370, 6.112682) --
    (42.218511, 6.208197) --
    (42.295814, 6.302776) --
    (42.374266, 6.396403) --
    (42.453855, 6.489065) --
    (42.534571, 6.580747) --
    (42.616400, 6.671438) --
    (42.699331, 6.761121) --
    (42.783352, 6.849785) --
    (42.868449, 6.937417) --
    (42.954610, 7.024002) --
    (43.041822, 7.109528) --
    (43.130072, 7.193983) --
    (43.219347, 7.277354) --
    (43.309634, 7.359628) --
    (43.400920, 7.440793) --
    (43.493189, 7.520837) --
    (43.586430, 7.599747) --
    (43.680627, 7.677514) --
    (43.775768, 7.754123) --
    (43.871837, 7.829565) --
    (43.968820, 7.903828) --
    (44.066703, 7.976901) --
    (44.165471, 8.048772) --
    (44.265110, 8.119432) --
    (44.365605, 8.188870) --
    (44.466940, 8.257075) --
    (44.569101, 8.324037) --
    (44.672072, 8.389746) --
    (44.775838, 8.454193) --
    (44.880383, 8.517367) --
    (44.985692, 8.579260) --
    (45.091750, 8.639862) --
    (45.198539, 8.699164) --
    (45.306045, 8.757157) --
    (45.414252, 8.813832) --
    (45.523142, 8.869182) --
    (45.632701, 8.923197) --
    (45.742911, 8.975870) --
    (45.853756, 9.027194) --
    (45.965220, 9.077159) --
    (46.077286, 9.125759) --
    (46.189937, 9.172987) --
    (46.303157, 9.218835) --
    (46.416928, 9.263297) --
    (46.531233, 9.306366) --
    (46.646057, 9.348035) --
    (46.761380, 9.388299) --
    (46.877187, 9.427152) --
    (46.993460, 9.464586) --
    (47.110181, 9.500598) --
    (47.227334, 9.535182) --
    (47.344900, 9.568332) --
    (47.462863, 9.600043) --
    (47.581203, 9.630311) --
    (47.699905, 9.659132) --
    (47.818950, 9.686500) --
    (47.938321, 9.712413) --
    (48.057999, 9.736865) --
    (48.177966, 9.759854) --
    (48.298206, 9.781376) --
    (48.418699, 9.801427) --
    (48.539429, 9.820005) --
    (48.660376, 9.837107) --
    (48.781523, 9.852731) --
    (48.902852, 9.866873) --
    (49.024345, 9.879533) --
    (49.145983, 9.890708) --
    (49.267749, 9.900396) --
    (49.389624, 9.908596) --
    (49.511590, 9.915307) --
    (49.633628, 9.920527) --
    (49.755722, 9.924257) --
    (49.877852, 9.926495) --
    (50.000000, 9.927241) --
    (50.122148, 9.926495) --
    (50.244278, 9.924257) --
    (50.366372, 9.920527) --
    (50.488410, 9.915307) --
    (50.610376, 9.908596) --
    (50.732251, 9.900396) --
    (50.854017, 9.890708) --
    (50.975655, 9.879533) --
    (51.097148, 9.866873) --
    (51.218477, 9.852731) --
    (51.339624, 9.837107) --
    (51.460571, 9.820005) --
    (51.581301, 9.801427) --
    (51.701794, 9.781376) --
    (51.822034, 9.759854) --
    (51.942001, 9.736865) --
    (52.061679, 9.712413) --
    (52.181050, 9.686500) --
    (52.300095, 9.659132) --
    (52.418797, 9.630311) --
    (52.537137, 9.600043) --
    (52.655100, 9.568332) --
    (52.772666, 9.535182) --
    (52.889819, 9.500598) --
    (53.006540, 9.464586) --
    (53.122813, 9.427152) --
    (53.238620, 9.388299) --
    (53.353943, 9.348035) --
    (53.468767, 9.306366) --
    (53.583072, 9.263297) --
    (53.696843, 9.218835) --
    (53.810063, 9.172987) --
    (53.922714, 9.125759) --
    (54.034780, 9.077159) --
    (54.146244, 9.027194) --
    (54.257089, 8.975870) --
    (54.367299, 8.923197) --
    (54.476858, 8.869182) --
    (54.585748, 8.813832) --
    (54.693955, 8.757157) --
    (54.801461, 8.699164) --
    (54.908250, 8.639862) --
    (55.014308, 8.579260) --
    (55.119617, 8.517367) --
    (55.224162, 8.454193) --
    (55.327928, 8.389746) --
    (55.430899, 8.324037) --
    (55.533060, 8.257075) --
    (55.634395, 8.188870) --
    (55.734890, 8.119432) --
    (55.834529, 8.048772) --
    (55.933297, 7.976901) --
    (56.031180, 7.903828) --
    (56.128163, 7.829565) --
    (56.224232, 7.754123) --
    (56.319373, 7.677514) --
    (56.413570, 7.599747) --
    (56.506811, 7.520837) --
    (56.599080, 7.440793) --
    (56.690366, 7.359628) --
    (56.780653, 7.277354) --
    (56.869928, 7.193983) --
    (56.958178, 7.109528) --
    (57.045390, 7.024002) --
    (57.131551, 6.937417) --
    (57.216648, 6.849785) --
    (57.300669, 6.761121) --
    (57.383600, 6.671438) --
    (57.465429, 6.580747) --
    (57.546145, 6.489065) --
    (57.625734, 6.396403) --
    (57.704186, 6.302776) --
    (57.781489, 6.208197) --
    (57.857630, 6.112682) --
    (57.932599, 6.016243) --
    (58.006385, 5.918897) --
    (58.078976, 5.820656) --
    (58.150361, 5.721535) --
    (58.220531, 5.621551) --
    (58.289474, 5.520716) --
    (58.357181, 5.419047) --
    (58.423640, 5.316559) --
    (58.488843, 5.213266) --
    (58.552779, 5.109185) --
    (58.615439, 5.004330) --
    (58.676813, 4.898718) --
    (58.736893, 4.792365) --
    (58.795670, 4.685285) --
    (58.853134, 4.577495) --
    (58.909277, 4.469012) --
    (58.964091, 4.359851) --
    (59.017568, 4.250028) --
    (59.069699, 4.139561) --
    (59.120477, 4.028465) --
    (59.169895, 3.916757) --
    (59.217944, 3.804454) --
    (59.264617, 3.691572) --
    (59.309909, 3.578128) --
    (59.353811, 3.464140) --
    (59.396318, 3.349624) --
    (59.437423, 3.234598) --
    (59.477120, 3.119078) --
    (59.515403, 3.003081) --
    (59.552266, 2.886626) --
    (59.587704, 2.769729) --
    (59.621712, 2.652408) --
    (59.654284, 2.534680) --
    (59.685416, 2.416564) --
    (59.715102, 2.298076) --
    (59.743339, 2.179234) --
    (59.770123, 2.060056) --
    (59.795448, 1.940560) --
    (59.819312, 1.820763) --
    (59.841712, 1.700684) --
    (59.862642, 1.580340) --
    (59.882102, 1.459749) --
    (59.900086, 1.338930) --
    (59.916594, 1.217900) --
    (59.931623, 1.096678) --
    (59.945169, 0.975281) --
    (59.957232, 0.853727) --
    (59.967809, 0.732036) --
    (59.976898, 0.610224) --
    (59.984500, 0.488310) --
    (59.990611, 0.366313) --
    (59.995232, 0.244250) --
    (59.998362, 0.122139) --
    (60.000000, 0.000000) --
    (90.000000, 0.000000);

    \draw [fill = black] (-10.000000, 0.000000) circle [radius = 0.15] node [above] {\tiny{$ 0 $}};
    \draw [fill = black] (-9.826016, 1.954516) circle [radius = 0.15] node [above] {\tiny{$ 1 $}};
    \draw [fill = black] (-9.238432, 3.826684) circle [radius = 0.15] node [above] {\tiny{$ 2 $}};
    \draw [fill = black] (-8.626057, 5.763747) circle [radius = 0.15] node [above] {\tiny{$ 3 $}};
    \draw [fill = black] (-7.302913, 7.302913) circle [radius = 0.15] node [above] {\tiny{$ 4 $}};
    \draw [fill = black] (-5.408999, 8.095138) circle [radius = 0.15] node [above] {\tiny{$ 5 $}};
    \draw [fill = black] (-3.642285, 8.793255) circle [radius = 0.15] node [above] {\tiny{$ 6 $}};
    \draw [fill = black] (-2.028990, 10.200423) circle [radius = 0.15] node [above] {\tiny{$ 7 $}};
    \draw [fill = black] (-0.000000, 9.646971) circle [radius = 0.15] node [above] {\tiny{$ 8 $}};
    \draw [fill = black] (2.025162, 10.181179) circle [radius = 0.15] node [above] {\tiny{$ 9 $}};
    \draw [fill = black] (3.643924, 8.797212) circle [radius = 0.15] node [above] {\tiny{$ 10 $}};
    \draw [fill = black] (5.445700, 8.150066) circle [radius = 0.15] node [above] {\tiny{$ 11 $}};
    \draw [fill = black] (7.089360, 7.089360) circle [radius = 0.15] node [above] {\tiny{$ 12 $}};
    \draw [fill = black] (8.191286, 5.473242) circle [radius = 0.15] node [above] {\tiny{$ 13 $}};
    \draw [fill = black] (8.974809, 3.717488) circle [radius = 0.15] node [above] {\tiny{$ 14 $}};
    \draw [fill = black] (9.665902, 1.922667) circle [radius = 0.15] node [above] {\tiny{$ 15 $}};
    \draw [fill = black] (10.000000, 0.000000) circle [radius = 0.15] node [above] {\tiny{$ 16 $}};
    \draw [fill = black] (11.875000, -0.179909) circle [radius = 0.15] node [above] {\tiny{$ 17 $}};
    \draw [fill = black] (13.750000, 0.100368) circle [radius = 0.15] node [above] {\tiny{$ 18 $}};
    \draw [fill = black] (15.625000, 0.406337) circle [radius = 0.15] node [above] {\tiny{$ 19 $}};
    \draw [fill = black] (17.500000, -0.030850) circle [radius = 0.15] node [above] {\tiny{$ 20 $}};
    \draw [fill = black] (19.375000, -0.056515) circle [radius = 0.15] node [above] {\tiny{$ 21 $}};
    \draw [fill = black] (21.250000, -0.102612) circle [radius = 0.15] node [above] {\tiny{$ 22 $}};
    \draw [fill = black] (23.125000, 0.059574) circle [radius = 0.15] node [above] {\tiny{$ 23 $}};
    \draw [fill = black] (25.000000, 0.025121) circle [radius = 0.15] node [above] {\tiny{$ 24 $}};
    \draw [fill = black] (26.875000, -0.318717) circle [radius = 0.15] node [above] {\tiny{$ 25 $}};
    \draw [fill = black] (28.750000, -0.353412) circle [radius = 0.15] node [above] {\tiny{$ 26 $}};
    \draw [fill = black] (30.625000, 0.004383) circle [radius = 0.15] node [above] {\tiny{$ 27 $}};
    \draw [fill = black] (32.500000, -0.458428) circle [radius = 0.15] node [above] {\tiny{$ 28 $}};
    \draw [fill = black] (34.375000, -0.337692) circle [radius = 0.15] node [above] {\tiny{$ 29 $}};
    \draw [fill = black] (36.250000, -0.145258) circle [radius = 0.15] node [above] {\tiny{$ 30 $}};
    \draw [fill = black] (38.125000, 0.086489) circle [radius = 0.15] node [above] {\tiny{$ 31 $}};
    \draw [fill = black] (40.000000, 0.000000) circle [radius = 0.15] node [above] {\tiny{$ 32 $}};
    \draw [fill = black] (40.250610, 1.939274) circle [radius = 0.15] node [above] {\tiny{$ 33 $}};
    \draw [fill = black] (40.847778, 3.790975) circle [radius = 0.15] node [above] {\tiny{$ 34 $}};
    \draw [fill = black] (41.991879, 5.350855) circle [radius = 0.15] node [above] {\tiny{$ 35 $}};
    \draw [fill = black] (42.730882, 7.269118) circle [radius = 0.15] node [above] {\tiny{$ 36 $}};
    \draw [fill = black] (44.401162, 8.379253) circle [radius = 0.15] node [above] {\tiny{$ 37 $}};
    \draw [fill = black] (46.350539, 8.810579) circle [radius = 0.15] node [above] {\tiny{$ 38 $}};
    \draw [fill = black] (48.082292, 9.640970) circle [radius = 0.15] node [above] {\tiny{$ 39 $}};
    \draw [fill = black] (50.000000, 9.686431) circle [radius = 0.15] node [above] {\tiny{$ 40 $}};
    \draw [fill = black] (51.873440, 9.418419) circle [radius = 0.15] node [above] {\tiny{$ 41 $}};
    \draw [fill = black] (53.829417, 9.245031) circle [radius = 0.15] node [above] {\tiny{$ 42 $}};
    \draw [fill = black] (55.732531, 8.579339) circle [radius = 0.15] node [above] {\tiny{$ 43 $}};
    \draw [fill = black] (57.388207, 7.388207) circle [radius = 0.15] node [above] {\tiny{$ 44 $}};
    \draw [fill = black] (58.116020, 5.422951) circle [radius = 0.15] node [above] {\tiny{$ 45 $}};
    \draw [fill = black] (59.286081, 3.846421) circle [radius = 0.15] node [above] {\tiny{$ 46 $}};
    \draw [fill = black] (60.169792, 2.022897) circle [radius = 0.15] node [above] {\tiny{$ 47 $}};
    \draw [fill = black] (60.000000, 0.000000) circle [radius = 0.15] node [above] {\tiny{$ 48 $}};
    \draw [fill = black] (61.875000, -0.399138) circle [radius = 0.15] node [above] {\tiny{$ 49 $}};
    \draw [fill = black] (63.750000, -0.168001) circle [radius = 0.15] node [above] {\tiny{$ 50 $}};
    \draw [fill = black] (65.625000, 0.064184) circle [radius = 0.15] node [above] {\tiny{$ 51 $}};
    \draw [fill = black] (67.500000, 0.182067) circle [radius = 0.15] node [above] {\tiny{$ 52 $}};
    \draw [fill = black] (69.375000, -0.247732) circle [radius = 0.15] node [above] {\tiny{$ 53 $}};
    \draw [fill = black] (71.250000, -0.415730) circle [radius = 0.15] node [above] {\tiny{$ 54 $}};
    \draw [fill = black] (73.125000, 0.025796) circle [radius = 0.15] node [above] {\tiny{$ 55 $}};
    \draw [fill = black] (75.000000, -0.303001) circle [radius = 0.15] node [above] {\tiny{$ 56 $}};
    \draw [fill = black] (76.875000, 0.344346) circle [radius = 0.15] node [above] {\tiny{$ 57 $}};
    \draw [fill = black] (78.750000, 0.321963) circle [radius = 0.15] node [above] {\tiny{$ 58 $}};
    \draw [fill = black] (80.625000, -0.498818) circle [radius = 0.15] node [above] {\tiny{$ 59 $}};
    \draw [fill = black] (82.500000, -0.299388) circle [radius = 0.15] node [above] {\tiny{$ 60 $}};
    \draw [fill = black] (84.375000, 0.251906) circle [radius = 0.15] node [above] {\tiny{$ 61 $}};
    \draw [fill = black] (86.250000, 0.293652) circle [radius = 0.15] node [above] {\tiny{$ 62 $}};
    \draw [fill = black] (88.125000, 0.487014) circle [radius = 0.15] node [above] {\tiny{$ 63 $}};
    \draw [fill = black] (90.000000, 0.000000) circle [radius = 0.15] node [above] {\tiny{$ 64 $}};
  \end{tikzpicture}

%% file: ConvexHullRandomWalk.tex


\begin{figure} [p]
  \ifnum\MyGraphicsMode=1
    \includegraphics[width = \columnwidth, keepaspectratio]{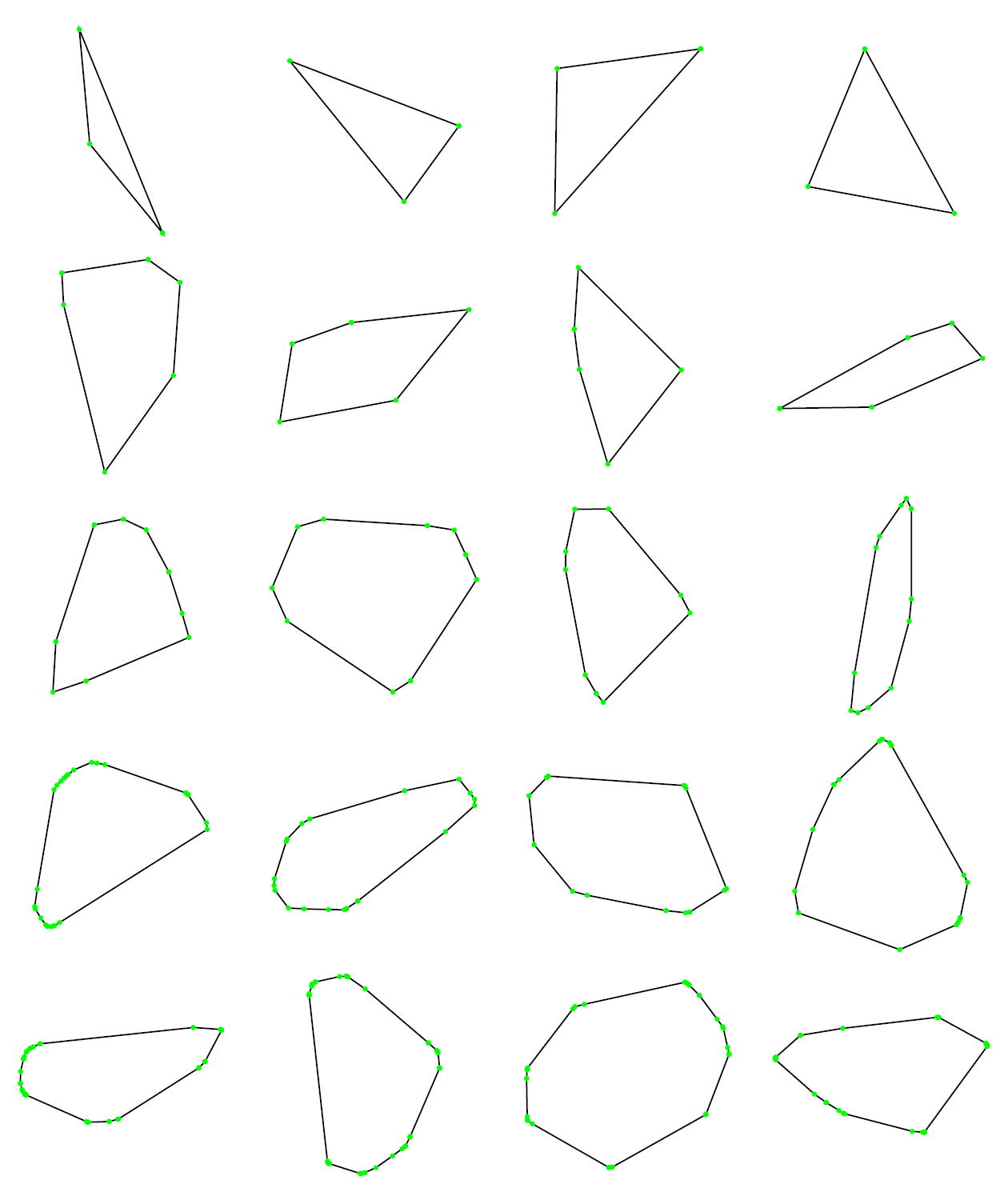}
  \fi
  \ifnum\MyGraphicsMode=2
    \input{ExamplesRandomConvexHullRandomWalk2481632.tex}
  \fi
  \caption
  {
    Examples of random convex hulls for a number of vertices in a random walk, $ 3 $, $ 7 $, $ 55 $, $ 2,981 $, and $ 8,886,111 $, for each row from top to bottom. The number of vertices in a random walk is chosen so that in an approximate average, see \eqref{eq:ApproximateNumberOfVerticesInConvexHullForRandomWalk}, the expected number of vertices in the convex hull will be $ 2 $ (a~convex hull of three random points from a random walk will almost surely have three vertices), $ 4 $, $ 8 $, $ 16 $, and $ 32 $, correspondingly.
  }
  \label{fig:ExamplesRandomConvexHullRandomWalk}
\end{figure}

%% file: ConvexHullOne.tex

\begin{figure} [p]
  \ifnum\MyGraphicsMode=1
    \includegraphics[width = \columnwidth, keepaspectratio]{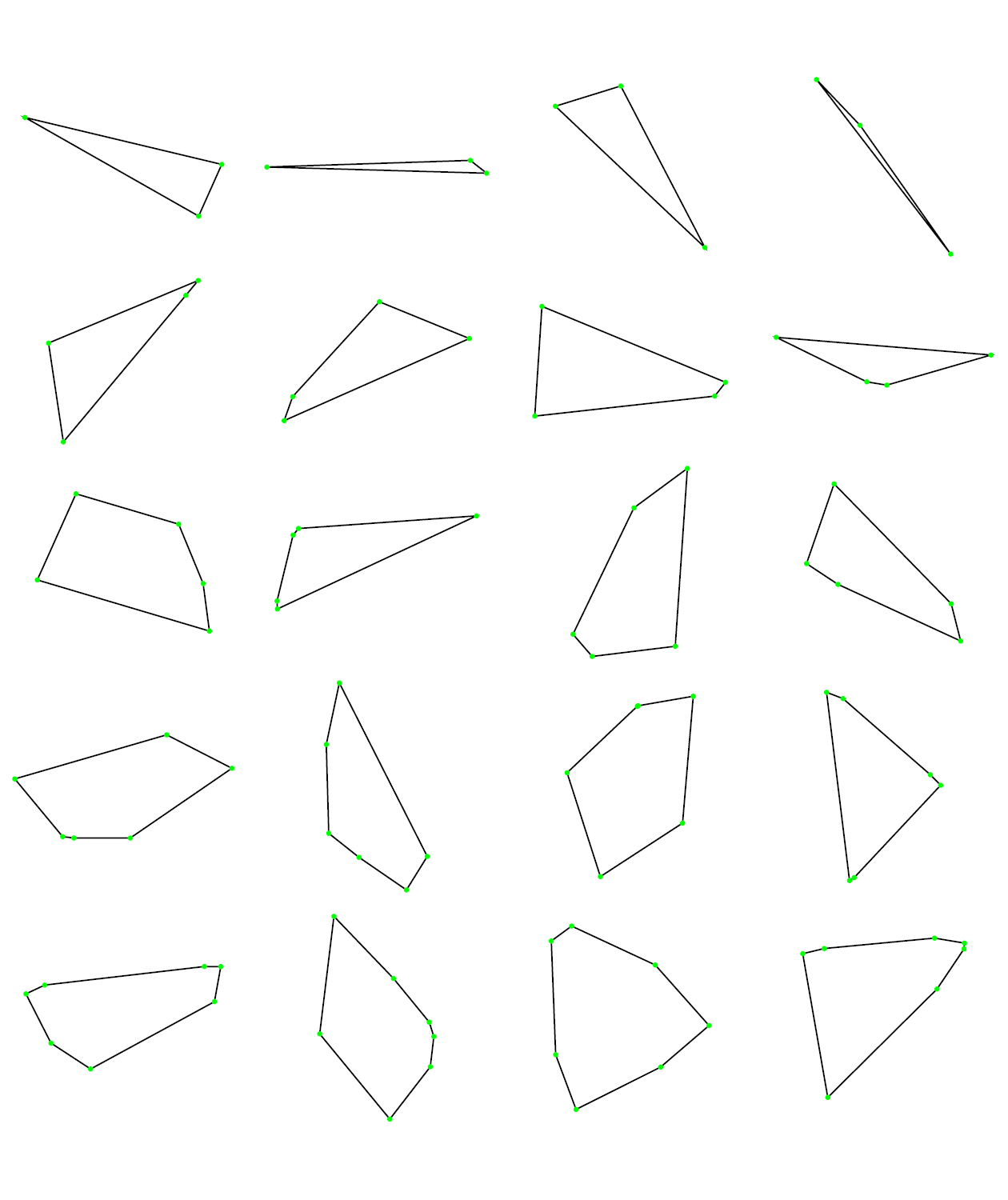}
  \fi
  \ifnum\MyGraphicsMode=2
    \input{ExamplesRandomConvexHull34567.tex}
  \fi
  \caption
  {
    Examples of random convex hulls for a number of vertices, $ 3 $, $ 4 $, $ 5 $, $ 6 $, and $ 7 $, for each row from top to bottom.
  }
  \label{fig:ExamplesRandomConvexHull}
\end{figure}

\begin{figure} [p]
  \ifnum\MyGraphicsMode=1
    \includegraphics[width = \columnwidth, keepaspectratio]{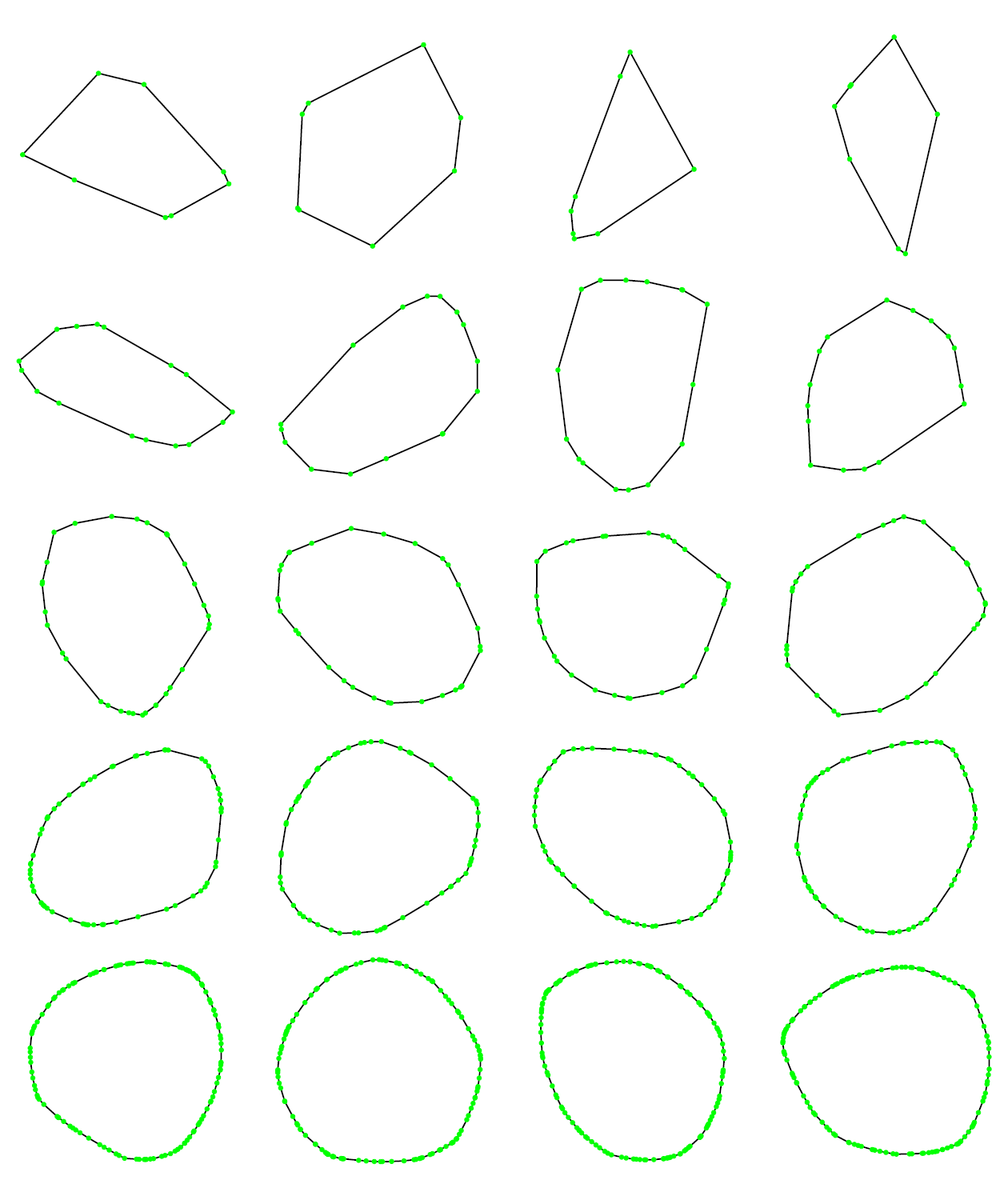}
  \fi
  \ifnum\MyGraphicsMode=2
    \input{ExamplesRandomConvexHull8163264128.tex}
  \fi
  \caption
  {
    Examples of random convex hulls for a number of vertices, $ 8 $, $ 16 $, $ 32 $, $ 64 $, and $ 128 $, for each row from top to bottom.
  }
  \label{fig:ExamplesRandomConvexHullGeometrical1}
\end{figure}

\begin{figure} [p]
  \ifnum\MyGraphicsMode=1
    \includegraphics[width = \columnwidth, keepaspectratio]{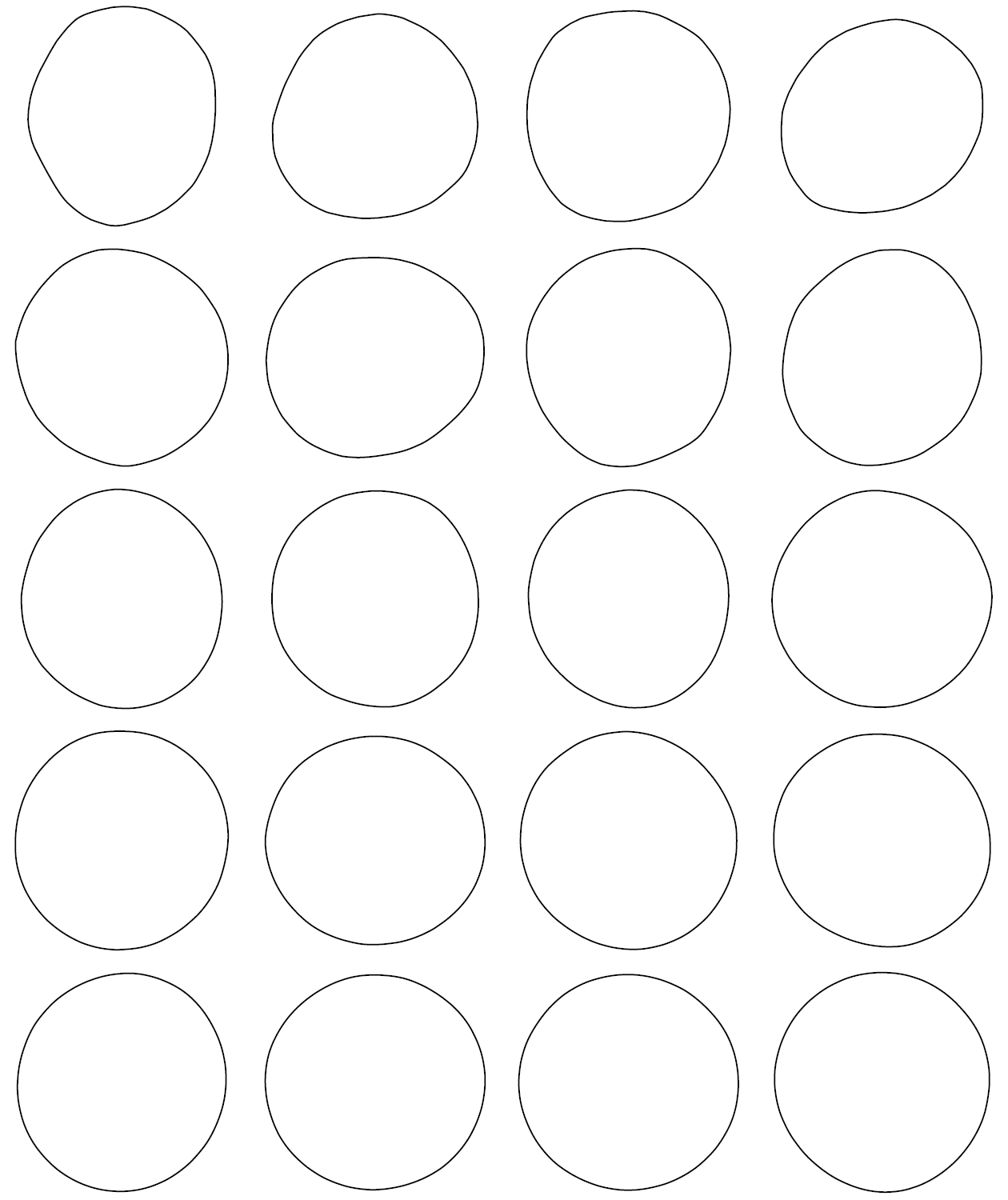}
  \fi
  \ifnum\MyGraphicsMode=2
    \input{ExamplesRandomConvexHull256512102420484096.tex}
  \fi
  \caption
  {
    Examples of random convex hulls for a number of vertices, $ 256 $, $ 512 $, $ 1,024 $, $ 2,048 $, and $ 4,096 $, for each row from top to bottom.
  }
  \label{fig:ExamplesRandomConvexHullGeometrical2}
\end{figure}

%% file: ConvexHullTwo.tex

\begin{figure} [p]
  \ifnum\MyGraphicsMode=1
    \includegraphics[width = \columnwidth, keepaspectratio]{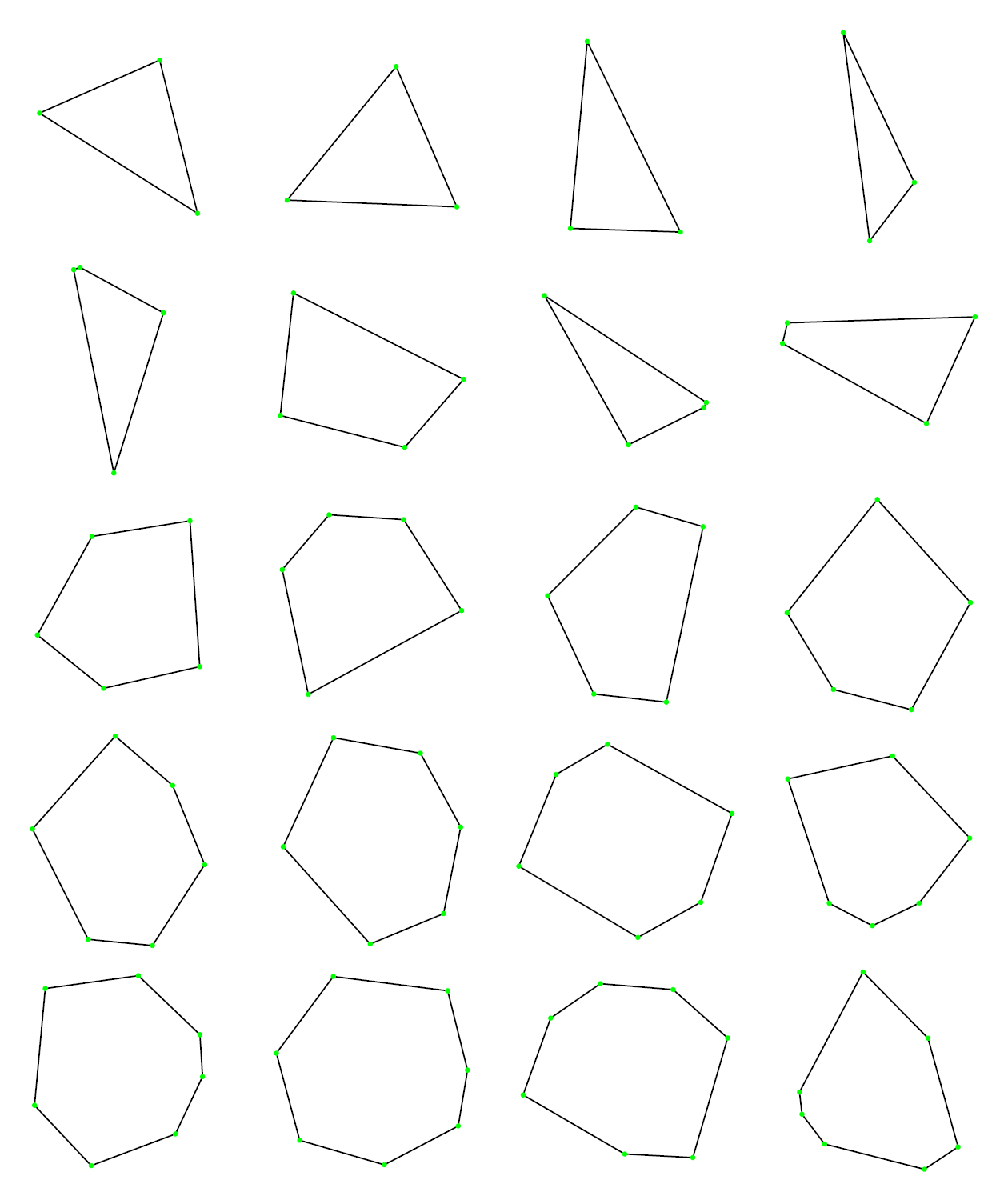}
  \fi
  \ifnum\MyGraphicsMode=2
    \input{ExamplesRandomConvexHullDelaunayTriangulation34567.tex}
  \fi
  \caption
  {
    Examples of random convex hulls for a number of vertices, $ 3 $, $ 4 $, $ 5 $, $ 6 $, and $ 7 $, for each row from top to bottom.
  }
  \label{fig:ExamplesRandomConvexHullDelaunayTriangulation}
\end{figure}

\begin{figure} [p]
  \ifnum\MyGraphicsMode=1
    \includegraphics[width = \columnwidth, keepaspectratio]{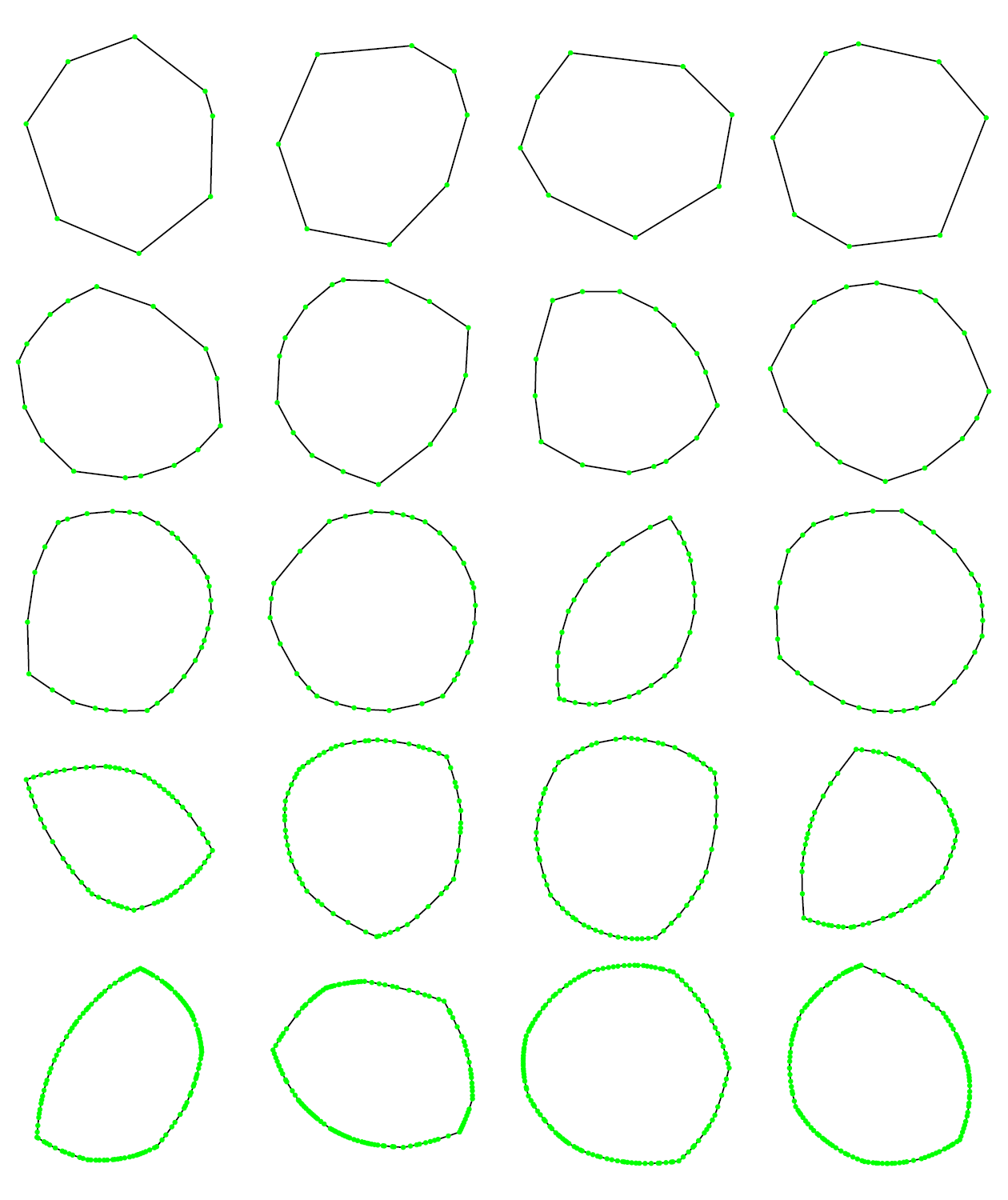}
  \fi
  \ifnum\MyGraphicsMode=2
    \input{ExamplesRandomConvexHullDelaunayTriangulation8163264128.tex}
  \fi
  \caption
  {
    Examples of random convex hulls for a number of vertices, $ 8 $, $ 16 $, $ 32 $, $ 64 $, and $ 128 $, for each row from top to bottom.
  }
  \label{fig:ExamplesRandomConvexHullDelaunayTriangulationGeometrical1}
\end{figure}

\begin{figure} [p]
  \ifnum\MyGraphicsMode=1
    \includegraphics[width = \columnwidth, keepaspectratio]{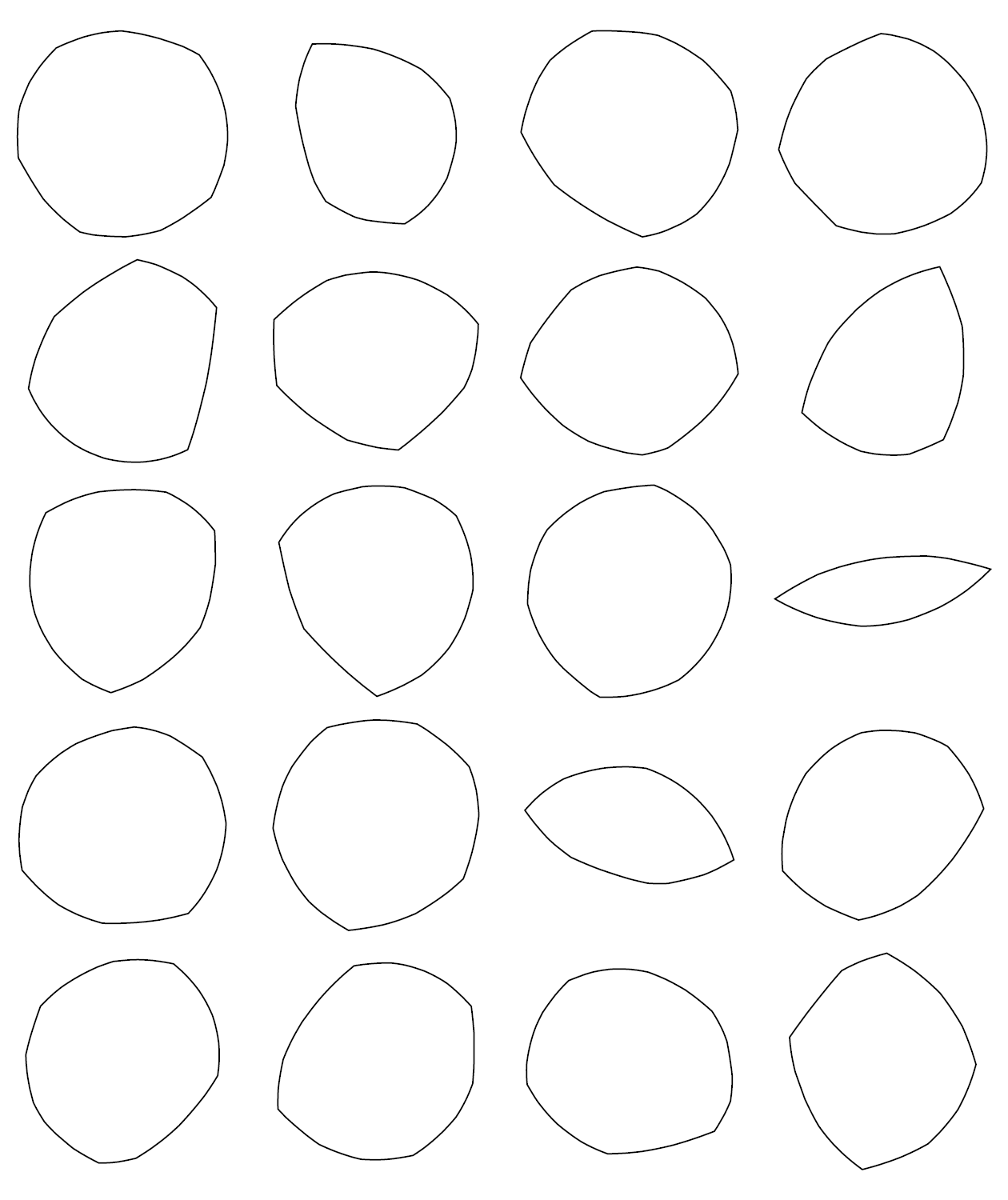}
  \fi
  \ifnum\MyGraphicsMode=2
    \input{ExamplesRandomConvexHullDelaunayTriangulation256512102420484096.tex}
  \fi
  \caption
  {
    Examples of random convex hulls for a number of vertices, $ 256 $, $ 512 $, $ 1,024 $, $ 2,048 $, and $ 4,096 $, for each row from top to bottom.
  }
  \label{fig:ExamplesRandomConvexHullDelaunayTriangulationGeometrical2}
\end{figure}